\def\lpmb#1{\mbox{\boldmath $#1$}}

\def\etal{{\it et al.}}

\def\half{{\textstyle{1\over2}}}
\def\thalf{{\textstyle{3\over2}}}

\def\>{\rangle}
\def\<{\langle}

\def\rmb#1{{\bf #1}}

\def\beq{\begin{equation}}
\def\eeq{\end{equation}}

\def\ma{m_{\Lambda_Q}}
\def\mb{m_{\Lambda_q}}
\def\jp{{J^P}}
\def\uv{u(v)}
\def\vp{v^\prime}
\def\vvp{v\cdot v^\prime}

\def\g5{\gamma_5}
\def\ll{\Lambda} 
\def\lls{\Lambda^*}
\def\lb{{\Lambda_b}}
\def\llcs{{\Lambda_c^*}}
\def\llc{{\Lambda_c^+}}

\def\beqy{\begin{eqnarray}}
\def\eeqy{\end{eqnarray}}

\def\gursp{u^{\mu_1\dots\mu_j}(\vp)}
\def\gbursp{\bar u^{\mu_1\dots\mu_j}(\vp)}
\def\gpurs{u^{\mu_1\dots\mu_n}(p)}
\def\gpburs{\bar u^{\mu_1\dots\mu_n}(p)}
\def\mdm{{\mu_1\dots\mu_j}}
\def\mdn{{\mu_1\dots\mu_n}}
\def\ep{\varepsilon}
\def\slash#1{#1 \hskip -0.5em / } 

\documentclass[aps,showpacs,floatfix]{revtex4}
\usepackage{graphicx,epsfig}
\newlength{\dinwidth}
\newlength{\dinmargin}
\begin{document}
\thispagestyle{empty}
\title{Semileptonic Decays of Heavy Lambda Baryons in a Quark Model}
\author{Muslema Pervin$^1$, Winston Roberts$^{2.3}$ and Simon Capstick$^1$}
\affiliation{$^1$ Department of Physics, Florida State University, Tallahassee,
 FL 32306\\
$^2$ Department of Physics, Old Dominion University, Norfolk, VA
23529, USA\\
and\\
Thomas Jefferson National Accelerator Facility,
12000 Jefferson Avenue, Newport News, VA 23606, USA\\
$^3$ On leave at the Office of Nuclear Physics, 
Department of Energy, 19901 Germantown Road, Germantown, MD 20874}
\begin{abstract}
The semileptonic decays of $\Lambda_c$ and $\Lambda_b$ are treated in
the framework of a constituent quark model. Both nonrelativistic and
semirelativistic Hamiltonians are used to obtain the baryon wave
functions from a fit to the spectra, and the wave functions are
expanded in both the harmonic oscillator and Sturmian bases. The
latter basis leads to form factors in which the kinematic dependence
on $q^2$ is in the form of multipoles, and the resulting form factors
fall faster as a function of $q^2$ in the available kinematic
ranges. As a result, decay rates obtained in the two models using the
Sturmian basis are significantly smaller than those obtained using the
harmonic oscillator basis. In the case of the $\Lambda_c$, decay rates
calculated using the Sturmian basis are closer to the experimentally
reported rates. However, we find a semileptonic branching fraction for
the $\Lambda_c$ to decay to excited $\Lambda^*$ states of 11\% to
19\%, in contradiction with what is assumed in available experimental
analyses. Our prediction for the $\Lambda_b$ semileptonic decays is
that decays to the ground state $\Lambda_c$ provide a little less than
70\% of the total semileptonic decay rate. For the decays
$\Lambda_b\to\Lambda_c$, the analytic form factors we obtain
satisfy the relations expected from heavy-quark effective
theory at the non-recoil point, at leading and next-to-leading orders in
the heavy-quark expansion. In addition, some features of the
heavy-quark limit are shown to naturally persist as the mass of the
heavy quark in the daughter baryon is decreased.

\flushright{JLAB-THY-05-304}
\end{abstract}
\maketitle
\setcounter{page}{1}

\section{Introduction and Motivation}

Many of the parameters of the Standard Model (SM), including the
Cabbibo-Kobayashi-Maskawa (CKM)~\cite{CKM} matrix elements, are not
yet determined with `satisfactory' precision. Very precise knowledge
of these matrix elements is important as they play a crucial role in
the search for answers to some fundamental questions, such as the
nature of $CP$ violation and the unitarity of the CKM
matrix. Semileptonic decays of hadrons have been, and will continue to
be, the main source of information on the CKM matrix elements. The
precision with which the CKM matrix elements are extracted from these
semileptonic decays is strongly dependent on how well the form factors
that describe the matrix elements of the hadronic currents are
known. The vast literature on these form factors is a testament to the
importance of these parameters.

The semileptonic decays of heavy mesons have been studied extensively
in the last two decades. Wirbel, Stech and Bauer~\cite{WSB,BW} assumed
monopole type form factors for the decays of heavy mesons. In
Ref.~\cite{ISGW,ISGW1}, a non-relativistic quark model (NRQM) was used
to treat the semileptonic decays of $B$ and $D$ mesons, and relatively
simple forms for the form factors were presented.  The first of those articles, 
along with the work of Shifman and Voloshin~\cite{voloshin}, ultimately lead
to the development of the heavy quark effective theory (HQET). In
addition, Ivanov and Santorelli~\cite{ISA} used a relativistic quark
model to find the form factors. These are just a very few of the very
large number of articles that treat semileptonic decays of mesons in
some kind of model.

Weak decays of hadrons involving one or more heavy quarks
$(m_Q>>\Lambda_{QCD})$ have an additional symmetry in the effective Lagrangian
which was first pointed out by Isgur and Wise~\cite{IW}. There, they used the
additional heavy quark symmetry to obtain normalized, model-independent
predictions for all the form factors for the decays of heavy hadrons to
daughter hadrons that are also heavy. This led to many subsequent calculations
by many authors.

For the hadronic matrix elements of the electroweak currents between
two heavy mesons, the application of HQET provides a number of
features that simplify the extraction of CKM matrix elements from such
decays. First, the number of form factors is reduced, so that the six
form factors that describe the decays of heavy pseudoscalar mesons to
heavy pseudoscalar and vector mesons are replaced by a single form
factor, at leading order in HQET.  This form factor has become known
as the Isgur-Wise function. Second, the absolute normalization of this
form factor at the so-called non-recoil point is known. Third,
corrections to this normalization do not arise at order $1/m_Q$ in the
heavy quark expansion, but at order $1/m_Q^2$. This is known as Luke's
Theorem~\cite{luke}, and is an analog of the Ademollo-Gatto
theorem~\cite{gatto}. This means that some predictions made at leading order are
more robust than might be expected. Finally, the corrections that do
arise can be estimated systematically in the heavy quark expansion. As
a result, HQET has become the tool of choice in the extraction of
$|V_{cb}|$~\cite{pdg}.

For the semileptonic decays of a heavy meson to a light meson, the predictions
of HQET are not quite as powerful: there is no reduction in the number of
form factors needed to describe the decay, nor are the normalizations of any
of the form factors known. However, the heavy quark symmetry, along with
SU(2) or SU(3) flavor symmetry for the light mesons, can be used to relate
the form factors for $D\to\pi$ and $D\to\rho$ decays, for instance, to those
for $B\to\pi$ and $B\to\rho$ decays, respectively. Thus, even though it is
not as predictive in the decays of heavy to light mesons, there is
still a great deal of reliance on HQET for extracting $V_{ub}$ from meson
decays.

For the semileptonic decays of a heavy baryon to another heavy baryon,
HQET makes predictions that are completely analogous to those made for
heavy-to-heavy meson decays: (i) the six form factors that describe the
decays to the ground-state heavy baryons are replaced by a single form
factor, the Isgur-Wise function; (ii) the normalization of the Isgur-Wise
function is known at the non-recoil point; (iii) corrections to this
normalization first appear at order $1/m_Q^2$; (iv) corrections can be
systematically estimated in a $1/m_Q$ expansion.

In the case of a heavy baryon decaying to a light baryon, HQET makes
predictions that are not as powerful as in the heavy-to-heavy case,
but which are significantly more powerful than for the heavy-to-light
transitions of mesons.  Among the baryons, the leading-order
prediction is that the number of independent form factors decreases
from six to two. In addition, as with mesons, the heavy quark symmetry
can be used to relate the form factors for the decay $\Lambda_c^+\to
n$ to those for $\Lambda_b\to p$, for instance. This, in principle,
could facilitate the extraction of $V_{ub}$ from semileptonic decays
of the $\Lambda_b$, and since the number of unknown form factors is
reduced from six to two, the theoretical uncertainty in the extraction
from these decays should be significantly smaller than extractions
from meson decays.

While HQET has been tremendously successful and useful in treating
semileptonic decays of heavy hadrons, it is not without its limitations.
It is a limit of QCD that applies only to hadrons containing heavy quarks.
For the decays of such hadrons, it only predicts the relationships among
form factors, not their kinematic dependence; ans\"atze, models of one
kind or another, or lattice simulations, are still needed for this. In
addition, the predictions of HQET are valid only as long as the energy of
the daughter hadron is not comparable to the mass of the heavy quark. For
heavy to heavy decays, this means that the predictions are valid for all
of the available phase space, but for heavy to light decays, such as $B\to
\pi$, a large portion of the available phase space is beyond the region of
reliable applicability of HQET.  These limitations mean that the
predictions of HQET must be complemented/supplemented by information
arising from other approaches to hadron structure.

While some work has been done in modeling the form factors for the
semileptonic decays of heavy baryons, to the best of our knowledge little
has been done in treating the decays to excited baryons. Predictions for
the number of independent form factors for decays to excited states have
been made in the framework of HQET~\cite{robertsa}, and Leibovich and
Stewart~\cite{LS} have examined the form factors for decays to the $1/2^-$
and $3/2^-$ states, using large $N_c$ arguments. In the semileptonic
decays of $B$ mesons to those with charm, it is known that $B^0$ decays to
the ground state pseudoscalar and vector mesons provide only about 75\% of
the total semileptonic decay rate, while for the $B^\pm$, the
corresponding fraction is about 85\%. Any assumption that decay of a heavy
baryon to the ground state will saturate the semileptonic decay rate is
therefore subject to potentially large corrections.

In some of the work done in this area, the predictions of HQET, along with
various ans\"atze for the form factors, have been used to estimate some
decay rates. Leibovich and Stewart~\cite{LS} follow such a procedure to
estimate the rates for decays of the $\Lambda_b$ to the $J^P=1/2^-$ and
$3/2^-$ $\Lambda_c$ states. Polarization effects in semileptonic
$\Lambda_b$ and $ \Lambda_c$ decays have been studied by K\"{o}rner and
Kr\"{a}mer~\cite{KK}, using the predictions of HQET to estimate the
dominant form factors for both $b\to c$ and $c \to s$ transitions. They
have also calculated the asymmetry parameters that characterize the
angular dependence of the decay distributions.

A number of authors have constructed explicit quark models of the form
factors for the decays of $\Lambda_Q$ baryons to ground state
baryons. The decays of $\Lambda_b$ and $\Omega_b$ have been treated by
Singleton using a spectator quark model~\cite{SING}.  He also
discusses the polarization of the $W$ boson and the daughter baryon in
these processes. Albertus {\it et al.}~\cite{Albertus} use a NRQM to
evaluate the form factors for $\Lambda_b\to\Lambda_c$, explicitly
applying heavy quark symmetry to their trial wave functions. To date,
there appear to be only two lattice studies of the semileptonic decays
of heavy baryons. A first study of $\Lambda_b$ and $\Xi_b$
semileptonic decays was made by Bowler {\it et al.}~\cite{Richards},
while Gottlieb and Tamhankar~\cite{gottlieb} have examined the decay of the
$\Lambda_b$. P\'erez-Marcial and collaborators~\cite{huerta} have
studied the semileptonic decays of a number of charmed baryons, both
in a non-relativistic quark model and in the MIT bag model. There have been
light-front calculations~\cite{lightfront}, as well as ones using sum rules
\cite{sum}, Bethe-Salpeter formalisms~\cite{BSF}, bag models~\cite{bag}, and 
quark model calculations \cite{qmc}. Large $N_c$ arguments have also
been applied to these form factors~\cite{largen}, as well as perturbative QCD arguments
\cite{pqcd}. For the decays of a heavy baryon to a
light one, work has been done using QCD sum rules~\cite{sumlight}, and there's one
quark model calculation~\cite{qmclight} apart from the work of Scora~\cite{scora}, to
the best of our knowledge.

The experimental status of heavy baryon semileptonic decays is somewhat
rudimentary. The semileptonic decay rate for $\Lambda_c\to\Lambda$ has been
measured by the CLEO and Argus collaborations~\cite{argus,CLEO}, while the
Delphi collaboration has only recently published an analysis of the exclusive
semileptonic decay of the $\Lambda_b$~\cite{delphi}. Prior to this, only the
inclusive semileptonic  branching fraction $\Lambda_b\to\Lambda_c\ell\nu X$ had
been reported in the PDG~\cite{pdg}. In their analysis of the
$\Lambda_c\to\Lambda$ semileptonic decay, the CLEO collaboration have assumed
that the ground state $\Lambda$ saturates the semileptonic decays of the
$\Lambda_c$, and cite the absence of any final states of the form
$\Lambda\ell\nu$ with additional decay products from the $\Lambda_c$ to support
their assumption~\cite{CLEO}. No experiments have yet reported results for the
decay $\Lambda_c\to n$.

The major difficulty in the baryon sector is that there is no source
of heavy baryons as there is for mesons. Electron-positron colliders
have produced billions of $B$ mesons, utilizing the fact that the
$\Upsilon(4s)$ is just above the $B\overline{B}$ threshold. In
principle, a similar abundance can be duplicated among $D$ mesons, by
using the $\Psi(3s)$. With baryons, production at such machines will
be continuum production, as there are no (known) resonances to enhance
the rate of production. Hadron colliders can provide larger yields,
but they provide large yields of everything, and the heavy baryons
will then have to be separated from everything else that is
produced. However, the recent CLEO measurement suggests that some
optimism regarding the future measurement of these decays might be
warranted. In addition, there might be prospects for such studies at
Jefferson Laboratory upgraded to 12 GeV or higher, or at E907 at FNAL.
The advantage in these cases is that the target will be a baryon,
unlike the continuum production of $e^+e^-$ machines.

In this paper we study the semileptonic decay of $\Lambda_Q$ baryons, the
motivation for which is two-fold.  One of our motivations is the importance of
the CKM matrix elements $V_{ub}$ and $V_{cb}$, and that baryon semileptonic
decays can provide complementary extractions of these quantities, despite the
difficulties mentioned above.  In particular, a model such as ours, coupled
with constraints provided by HQET, may lead to a more precise extraction of
$V_{ub}$ than provided by meson decays.

Our second motivation is to examine the predictions for these decays
of a quark model developed very much in the spirit of the work by
Capstick and Isgur~\cite{CI}, which builds on the work of Isgur and
Karl~\cite{IsgurKarl-,IsgurKarl}. Such a model has been applied, with some
success, to the strong~\cite{capstickroberts} and
electromagnetic~\cite{capstickkeister, Capstick} couplings of baryons,
and the semileptonic decays of baryons is a useful complementary
extension of such a model. Indeed, a similar model, applied to the
semileptonic decays of mesons~\cite{ISGW}, gave rise to HQET. We note
that the thesis of Scora~\cite{scora} treats a number of baryon
semileptonic decays in a framework very similar to that used in the
treatment of mesons in~\cite{ISGW}. We use a similar framework, but we
extend the model to examine the decays to excited baryons, whereas
Scora~\cite{scora} examined only decays to ground state baryons. We
also use a more sophisticated treatment of baryon structure.

This manuscript is organized as follows: in Section II we discuss the
hadronic matrix elements and decay rates. Section III presents a brief
outline of heavy quark effective theory as it relates to the decays
that we discuss. In Section IV we describe the model we use to obtain
the form factors, including some description of the Hamiltonian.  Our
analytic results are discussed in Section V, our numerical results are
given in section VI, and Section VII presents our conclusions and
outlook. A number of details of the calculation, including the
explicit expressions for the form factors, are shown in a number of
Appendices.

\section{Matrix Elements and Decay Rates}

\subsection{Matrix Elements}

The transition matrix element for semileptonic decay of $\Lambda_Q$ 
$(\Lambda_Q \rightarrow \Lambda_q\ell\nu_\ell)$ is
\beq
T=\frac{G_F}{\sqrt{2}} V_{Qq} \overline{u_\ell} \gamma^\mu (1 - \gamma_5) 
u_{\nu_\ell}\<\Lambda_{q}(p', s')| J_{\mu}| \Lambda_Q(p, s)\>,
\eeq
where $G_F/\sqrt2=g^2/(8M_W^2)$ is the Fermi coupling constant, $M_W$ is the 
intermediate vector boson mass, $V_{Qq}$ is the CKM matrix element, and 
$\overline{u_{\ell}} \gamma^{\mu}(1 - \gamma_{5}) u_{\nu_\ell}$ is the lepton 
current. Since quarks are confined, the matrix element of the hadron current is
described in terms of a number of form factors. We will build a model of the 
baryons we wish to study, and obtain approximations to the form factors that
describe the hadronic matrix elements. For transitions between ground state 
$(J^P=1/2^+)$ baryons, the hadronic matrix elements of the vector and axial
currents are
\begin{eqnarray}
\langle\Lambda_{q}(p', s')| V_{\mu}| \Lambda_{Q}(p, s)\rangle &=& 
\overline{u}(p', s')\left(F_1(q^2) \gamma_{\mu} + F_2(q^2)
\frac{ p_\mu}{m_{\Lambda_Q}} +  F_3(q^2)\frac{ p'_\mu}{m_{\Lambda_q}}
\right) u(p,s),\label{vectorme}\\
\langle\Lambda_{q}(p', s')| A_{\mu}| \Lambda_{Q}(p, s)\rangle &=& 
\overline{u}(p', s')\left(G_1(q^2) \gamma_{\mu} + G_2(q^2)
\frac{ p_\mu}{m_{\Lambda_Q}} +  G_3(q^2)\frac{ p'_\mu}{m_{\Lambda_q}}\right)
\gamma_{5} u(p,s),\label{axialvme}
\end{eqnarray}
where the $F_i$ and $G_i$'s are baryon form factors which depend on the square 
of the momentum transfer $q=p-p^\prime$ between the initial and the final 
baryons. Similarly, the matrix elements for decays to a daughter 
baryon with $J^P=3/2^-$ are
\begin{eqnarray}
\langle\Lambda_{q}^{3/2}(p', s')| V_{\mu}| \Lambda_{Q}(p, s)\rangle &=&
\overline{u}^\alpha(p', s')\left[
\frac{p_\alpha}{m_{\Lambda_Q}}\left(F_1\gamma_{\mu} + F_2\frac{
p_\mu}{m_{\Lambda_Q}} +  F_3\frac{ p'_\mu}{m_{\Lambda_q^{3/2}}}\right)+F_4 g_{\alpha\mu}\right]
u(p,s),\nonumber\\
\langle\Lambda_{q}^{3/2}(p', s')| A_{\mu}| \Lambda_{Q}(p, s)\rangle &=& 
\overline{u}^\alpha(p', s')\left[
\frac{p_\alpha}{m_{\Lambda_Q}}\left(G_1 \gamma_\mu + 
G_2\frac{ p_\mu}{m_{\Lambda_Q}} +  G_3
\frac{ p'_\mu}{m_{\Lambda_q^{3/2}}}\right)+G_4 g_{\alpha\mu}\right]\gamma_{5} u(p,s).
\end{eqnarray}
The spinor $\overline{u}^\alpha(p', s')$ satisfies the conditions
\beq
p^\prime_\alpha\overline{u}^\alpha(p', s')=0, \,\,\,\,
\overline{u}^\alpha(p', s')\gamma_\alpha=0,\,\,\,\, \overline{u}^\alpha(p', s')
\slash{p}^\prime=m_{\Lambda_q^{3/2}}\overline{u}^\alpha(p', s').
\eeq
The corresponding matrix elements for decay to a baryon with $J^P=5/2^+$ are
\begin{eqnarray}
\langle\Lambda_{q}^{5/2}(p', s')| V_{\mu}| \Lambda_{Q}(p, s)\rangle &=&
\overline{u}^{\alpha\beta}(p',
s')\frac{p_\alpha}{m_{\Lambda_Q}}\left[\frac{p_\beta}{m_{\Lambda_Q}}\left(F_1\gamma_\mu+
F_2\frac{p_\mu}{m_{\Lambda_Q}} +
F_3\frac{p'_\mu}{m_{\Lambda_q^{5/2}}}\right)+F_4 g_{\beta\mu}\right]u(p,s),\nonumber\\
\langle\Lambda_{q}^{5/2}(p', s')| A_{\mu}| \Lambda_{Q}(p, s)\rangle &=& 
\overline{u}^{\alpha\beta}(p', s')\frac{p_\alpha}{m_{\Lambda_Q}}
\left[\frac{p_\beta}{m_{\Lambda_Q}}\left(G_1 \gamma_\mu + 
G_2\frac{ p_\mu}{m_{\Lambda_Q}} +  G_3\frac{ p'_\mu}{m_{\Lambda_q^{5/2}}}
\right)+G_4 g_{\beta\mu}\right]\gamma_{5}u(p,s),
\end{eqnarray}
where the spinor $\overline{u}^{\alpha\beta}(p', s')$ is symmetric in the
indices $\alpha$ and $\beta$, and satisfies
\beqy
p^\prime_\alpha\overline{u}^{\alpha\beta}(p', s')&=&
p^\prime_\beta\overline{u}^{\alpha\beta}(p', s')=0,\nonumber\\
\overline{u}^{\alpha\beta}(p', s')\gamma_\alpha&=&\overline{u}^{\alpha\beta}(p', s')
\gamma_\beta=0,\nonumber\\
\overline{u}^{\alpha\beta}(p', s')
\slash{p}^\prime&=&m_{\Lambda_q^{5/2}}\overline{u}^{\alpha\beta}(p',
s'),\nonumber\\
\overline{u}^{\alpha\beta}(p', s')g_{\alpha\beta}&=&0.
\eeqy

Here we have only presented the form factor equations involving spinors having
natural parity. The equations for unnatural parity spinors can be constructed 
in a similar manner by switching $\gamma_5$ from the equations defining the $G_i$ to
the equations defining the $F_i$.
 
\subsection{Decay Rates}

The decay rate that arises from any of these matrix elements is
\beq
d\Gamma =\frac{1}{2m_{\Lambda_Q}} \frac{ 
G_{F}^{2}}{2}|{V_{Qq}}|^{2}\left( \prod_f \frac{ d^{3}p_{f}}{(2\pi)^3 
2E_f}\right)  L^{\mu\nu} H_{\mu\nu}(2\pi)^4 \delta^{(4)}(p_A - 
\sum{p_f}),
\eeq
where $A$ refers to the initial hadron. The leptonic tensor $L^{\mu\nu}$ 
is
\beq
L^{\mu\nu}  = 8[p_\ell^{\mu} p_{\nu_\ell}^{\nu} + p_{\nu_\ell}^{\mu} 
p_\ell^{\nu} - 
g^{\mu\nu} p_\ell\cdot p_{\nu_\ell} + i\epsilon^{\mu\nu\alpha\beta} 
p_{\ell\alpha} p_{\nu_\ell\beta}].
\eeq
The hadronic tensor $H_{\mu\nu}$ is
\beq
H_{\mu\nu} = \sum_{{\rm spin}} \langle \Lambda_{Q} 
\vert{J_\nu^\dagger}
\vert \Lambda_{q}\rangle \langle \Lambda_{q} \vert 
J_{\mu}\vert 
\Lambda_{Q}\rangle,
\eeq
where $\Lambda_Q$ and $\Lambda_q$ refer to the initial and final 
baryons, 
respectively. The tensor $H_{\mu\nu}$ must have the Lorentz structure
\begin{eqnarray}
H_{\mu\nu} &=& -\alpha g_{\mu\nu} + \beta_{++}(p+p')_\mu (p+p')_\nu + 
\beta_{+-}(p+p')_\mu (p-p')_\nu\nonumber \\
&+& \beta_{-+}(p-p')_\mu (p+p')_\nu + \beta_{--}(p-p')_\mu 
(p-p')_\nu\nonumber\\
 &+&  i \gamma \epsilon_{\mu\nu\rho\sigma}(p+p')^\rho 
(p-p')^\sigma.\nonumber
\end{eqnarray}
The complete expression for the differential decay rate is
\begin{eqnarray}
\label{rate}
\frac{d^2\Gamma}{dx dy} &=& |V_{Qq}|^2
\frac{G_F^2m_{\Lambda_Q}^5}{64\pi^3} \left[\alpha C_{\alpha}+ \beta_{++}
C_{\beta++}+\beta_{-+} C_{\beta-+} +\beta_{+-} C_{\beta+-}
+\beta_{--}C_{\beta--}+\gamma C_\gamma\right], 
\end{eqnarray} 
where
\begin{eqnarray} 
C_{\alpha} &=& {2\over m_{\Lambda_Q}^2}\left(y-{m_\ell^2\over
m_{\Lambda_Q}^2}\right),\nonumber\\
C_{\beta_{++}} &=&8\left[ x(2x_m+y)- 2x^2 -y/2\right]-{m_\ell^2\over 
 m_{\Lambda_Q}^2}\left( {m_\ell^2\over m_{\Lambda_Q}^2}-{4m_{\Lambda_q}^2
 \over m_{\Lambda_Q}^2}-8x +3y\right),\nonumber\\
C_{\beta_{-+}} &=& C_{\beta_{+-}}={m_\ell^2\over m_{\Lambda_Q}^2}\left[4(x- 
x_m)-y-
{m_\ell^2\over m_{\Lambda_Q}^2}\right],\nonumber\\ 
C_{\beta_{--}} &=& {m_\ell^2\over m_{\Lambda_Q}^2}\left(y-{m_\ell^2\over 
m_{\Lambda_Q}^2}\right),\nonumber\\
C_\gamma  &=& \mp 2y\left[2 x_m -4x +y +{m_\ell^2\over m_{\Lambda_Q}^2}(2
x_m+y)\right]. 
\end{eqnarray}
In these expressions, $x=E_\ell/m_{\Lambda_Q}$, where $E_\ell$ is the lepton
energy, $x_m
=(m_{\Lambda_Q}^2-m_{\Lambda_q}^2)/(2m_{\Lambda_Q}^2)$, 
and $y=q^2/m_{\Lambda_Q}^2 = (p
-p')^2/m_{\Lambda_Q}^2$. The $\mp$ sign in $C_\gamma$ is determined by the
charge of the lepton, with the upper (negative) sign corresponding to decays 
to $\ell\nu_\ell$.
The lepton energy has the range
\begin{eqnarray} 
{-K\over 2\sqrt{y}}(y-m_\ell^2/m_{\Lambda_Q}^2)^{1/2}+{(2
x_m+y)\over 4y}\left(y+{m_\ell^2\over m_{\Lambda_Q}^2}\right)
\le x \le 
{K\over 2\sqrt{y}}(y-m_\ell^2/m_{\Lambda_Q}^2)^{1/2}+{(2
x_m+y)\over 4y}\left(y+{m_\ell^2\over 
m_{\Lambda_Q}^2}\right)\nonumber
\end{eqnarray}
with $K={1\over 2}[(2x_m-y)^2-4m_{\Lambda_q}^2/m_{\Lambda_Q}^2]^{1/2}$, and $y$  has
the kinematic range  $m_\ell^2/m_{\Lambda_Q}^2 \le y \le
(m_{\Lambda_Q}-m_{\Lambda_q})^2/m_{\Lambda_Q}^2$. If the lepton mass is neglected, the terms in $\beta_{+-}$,
$\beta_{-+}$ and $\beta_{--}$ vanish, and the differential decay rate  becomes 
\begin{eqnarray} 
\frac{d^2\Gamma}{dx dy} &=& |V_{Qq}|^2
\frac{G_F^2m_{\Lambda_Q}^5}{32\pi^3} \left[\frac{\alpha   y}{m_{\Lambda_Q}^2} +
2\beta_{++}\left[2x(2 x_m +y)- 4x^2  - y\right]\mp \gamma y\left(2 x_m -4x
+y\right)\right], 
\end{eqnarray}  %
where the lepton energy is now constrained by $-K/2+(2 x_m+y)/4\le x \le K/2+(2 
x_m+y)/4$, and the lower limit on y is zero. In this case the differential rate 
depends only on $\alpha$, $\beta_{++}$ and $\gamma$. The explicit expressions
for  $\alpha$, $\beta_{++}$ and $\gamma$ in  terms of form factors for
different final baryon spins are given in Appendix \ref{hadrontensor}.

\section{Heavy Quark Effective Theory} 

Heavy quark effective theory (HQET)~\cite{HQET} has been a very useful tool
in the study of electroweak decays of heavy hadrons. This effective theory has
been applied to a number of processes, both inclusive and exclusive, to higher
and higher order in the $1/m_Q$ expansion, where $m_Q$ is the mass of the heavy
quark. In most applications, the aim has been to constrain the hadronic
uncertainties in the extraction of CKM matrix elements such as $V_{ub}$ and
$V_{cb}$. In this section, we take a different tack; we examine the predictions
of HQET for decays of a heavy $\ll$ into any of the allowed excited daughter
baryons, whether this daughter baryon is heavy or light, with the aim of 
comparing these predictions with the form factors that we obtain in our model.

\subsection{Heavy to Heavy}

In a heavy excited baryon, the light quark system has some total angular
momentum $j$, so that the total angular momentum of the baryon can be 
$J=j\pm 1/2$. These two states are degenerate because of the heavy quark spin
symmetry. It is useful to show explicitly the representation we use for these
two degenerate baryons. In the notation of Falk~\cite{Falk}, we write
$u^{\mu_1\dots\mu_j}_{j+1/2}(\vp)=\gursp-u^{\mu_1\dots\mu_j}_{j-1/2}(\vp)$,
with 
\beq \gursp=A^{\mu_1\dots\mu_j}(\vp)u_Q(\vp). 
\eeq 
Here, $u_Q(v)$ is the spinor of the heavy quark, and $A^{\mu_1\dots\mu_j}(\vp)$
is a tensor that describes the spin-$j$ light quark system. This tensor is
symmetric in all of its Lorentz indices, meaning that the $\gursp$ is also
symmetric in all its Lorentz indices. Both
$u^{\mu_1\dots\mu_j}_{j\pm 1/2}(\vp)$ satisfy the conditions 
\beqy 
\slash v^\prime\gursp&=&\gursp,\nonumber\\ 
\vp_{\mu_i}u^{\mu_1\dots\mu_i\dots\mu_j}&=&0,\,\,
g_{\mu_k\mu_l}{u}^{\mu_1\dots\mu_j}(v)=0, 
\eeqy 
where $\mu_k$ and $\mu_l$ indicate any pair of the indices $\mu_1\dots\mu_j$.
The state with $J=j+1/2$ also satisfies 
\beq \gamma_{\mu_i}u^{\mu_1\dots\mu_i\dots\mu_j}_{j+1/2}=0. 
\eeq 
Further details of the structure and properties of these tensors are given in
Falk's article~\cite{Falk}.

At this point, it is useful to discuss the parity of the states, which is
determined by the parity of the light component. A spin-$j$ light quark
component with parity $(-1)^j$ is said to have `natural' parity, unnatural
parity otherwise. The natural-parity light quark systems therefore have
$j^P=(2n)^+$ or $j^P=(2n+1)^-$, with $n$ a positive integer or zero. The
natural-parity light quark systems are represented by tensors, while those with
unnatural parity are represented by pseudo-tensors. Since the parity of the
baryon is that of the light quark system, we may refer to the baryons as being
tensors or pseudo-tensors, with the understanding that this really refers to
the light-quark component of the baryon. It is thus convenient to divide the
decays we discuss into two classes, those in which the daughter baryons are
tensors, and those in which they are pseudo-tensors. We begin with the
discussion of the tensor decays.

In general, we are interested in the matrix element
\beq
{\cal A} =<\llcs(\vp,j)|\bar{c}\Gamma b|\lb(v)>,
\eeq
where $c$ and $b$ are the heavy quark fields, and $\Gamma$ is an arbitrary
combination of Dirac matrices.
With the use of HQET, we may write this as
\beq
<\llcs(\vp,j)|\bar{c}\Gamma b|\lb(v)>=\gbursp\Gamma\uv M_{\mdm},
\eeq
to leading order. In writing this form, we are omitting multiplicative QCD
corrections of order unity that arise from matching of the effective theory to
full QCD at different mass scales. Here, $M_{\mdm}$ is the most general tensor that
we can construct, given the kinematic variables at our disposal. Clearly,
$M_{\mdm}$ may not contain any factors of $\vp_{\mu_i}$ or $g_{\mu_i\mu_j}$, and
therefore takes the form
\beq
M_{\mdm}=\eta^{(j)}(\vvp)v_{\mu_1}\dots v_{\mu_j}. 
\eeq 
Thus, a single form factor, $\eta^{(j)}(\vvp)$ is needed to this order, regardless of the spin of
the final baryon. In addition, spin symmetry allows us to relate the form
factors for $\Gamma=\gamma_\mu$ to those for $\Gamma=\gamma_\mu\g5$.

The case of $\jp=1/2^+$, $j=0$ requires a special comment. These states may be
thought of as radial excitations of the ground state $\llc$. Because of the
heavy quark symmetry, and the orthogonality of these states with respect to the
ground state, we must have
\beq
<\llcs(\vp,j^P=0^+)_{(n)}|\bar{c}\Gamma b|\lb(v)>= (\vvp-1)\eta^{(0)}_{(n)}(\vvp)
\bar u(\vp)\Gamma\uv,
\eeq
where the subscripts $(n)$ denote the $n$th radial excitation. That is, these
amplitudes must vanish as $\vp\to v$. This result has been pointed out by
Isgur, Wise and Youssefmir~\cite{IWY}. Note, too, that all of the other
amplitudes ($j\ne 0$) vanish trivially at the non-recoil point.

For the pseudo-tensor decays, we write exactly the same form, but $M_{\mdm}$
must now be a pseudo-tensor object, and must therefore be constructed by using
the $\ep$ tensor. Inspection shows that no such pseudo-tensor can be
constructed, given that we have only two kinematic variables at our disposal,
namely $v$ and $\vp$, and that the spinor-tensor used to describe the daughter
baryon is symmetric in its indices. Thus, decay amplitudes for transitions to
pseudo-tensor daughter baryons vanish at leading order in HQET.

Applying these results to the specific case of $j^P=1^-$, we find,
for $J^P=1/2^-$, 
\begin{equation}
F_1=\frac{w-1}{\sqrt{3}}\eta^{(1)}(w),\,\,F_2=G_2=-\frac{2}{\sqrt{3}}\eta^{(1)}(w),\,\,
G_3=F_3=0,\,\,G_1=\frac{w+1}{\sqrt{3}}\eta^{(1)}(w). 
\end{equation}
For $3/2^-$, 
\begin{equation}
F_2=F_3=G_2=G_3=F_4=G_4=0,\,\,F_1=G_1=\eta^{(1)}(w). 
\end{equation}
In these two sets of equations $\eta^{(1)}$ is a universal function of the
Isgur-Wise type, and $w=v\cdot v^\prime$.

For $j^P=2^+$, we find for $J^P=3/2^+$,
\begin{equation}\label{spinthp}
F_3=G_3=F_4=G_4=0,\,\,F_1=\frac{2(w-1)}{\sqrt{10}}\eta^{(2)}(w),\,\,F_2=G_2=
-\frac{4}{\sqrt{10}}\eta^{(2)}(w),\,\,G_1=\frac{2(w+1)}{\sqrt{10}}\eta^{(2)}(w),
\end{equation}
and for $5/2^+$
\begin{equation}\label{spinfhp}
F_2=F_3=F_4=G_4=G_2=G_3=0,\,\,F_1=G_1=\eta^{(2)}(w). 
\end{equation}
As with the previous example, the function $\eta^{(2)}$ is an Isgur-Wise form factor
common to both decays.

For the elastic decays, as well as for decays to the
$1/2^-,\,\,3/2^-$ doublet, the matrix elements have been evaluated at order
$1/m_c$ and $1/m_b$ in the heavy quark expansion~\cite{LS}. When we present our results
for the form factors, we will compare our expressions with the predictions of
HQET.

\subsection{Heavy to Light}

For the heavy to light transitions, we may no longer describe the daughter
baryons in terms of the spin structure of the light quark system that helps to
make up the baryon. Instead, we are forced to use the total angular momentum of
the baryon concerned, as well as its parity. As before, we may represent one of
these baryons, denoted $\Lambda^*$, by a generalized Rarita-Schwinger field $\gpurs$, where the
auxiliary conditions now are
\beqy
\slash p \gpurs&=&m_{\lls}\gpurs, \,\, \gamma_{\mu_1} \gpurs=0,\nonumber\\
p_{\mu_1}\gpurs&=&0,\,\, {u_\mu}^{\mu\dots\mu_n}(p)=0,
\eeqy
and a baryon with angular momentum and parity $J^P$ is represented by a spinor-tensor with
$n=J-1/2$ indices. As was the case with the heavy to heavy transitions, we
need to divide the possible transitions into two classes, which we call tensor
and pseudo-tensor, with the obvious meaning.

As before, we begin with the transitions to tensor states. Here, we say a 
state of total angular momentum $J$ is a tensor if its parity is
$(-1)^{(J-1/2)}$, and is a pseudo-tensor otherwise. The matrix element of interest is
\beq \label{lite1}
<\lls(p)_{\jp}|\bar s\Gamma c|\llc(v)>=\gpburs M_{\mdn}\Gamma\uv,
\eeq
where $M_{\mdn}$ is the most general tensor that one can construct, and $n=J-1/2$. 
As with the heavy to heavy transitions, we may not use any factors of 
$\gamma_{\mu_i}$, $p_{\mu_i}$ or $g_{\mu_i\mu_j}$ in constructing $M_{\mdn}$, which 
must therefore have the form 
\beq
M_{\mdn}=v_{\mu_1}\dots v_{\mu_n}{\cal A}_n. 
\eeq 
Here, ${\cal A}_n$ is the most general Lorentz scalar that we can build. On inspection, 
we find that 
\beq
{\cal A}_n=\xi_1^{(n)}+\slash v \xi_2^{(n)}, 
\eeq 
so that each of these transitions is described by two form factors, at leading order 
in HQET.

For the transitions into pseudo-tensor daughter baryons, we write
exactly the same form as in Eq.~(\ref{lite1}), but now $M_{\mdn}$ must
be a pseudo-tensor. This may involve the use of the $\ep$ tensor, but
since $\gpurs$ is symmetric in its indices, at most one of these
indices may be contracted with the indices of the $\ep$ tensor. With
some patience, and the use of a few well chosen identities, one can
show that any pseudo-tensor term constructed with the $\ep$ tensor may
always be reduced to an ordinary tensor multiplying a $\g5$ matrix. We
will therefore leave out much of the tedium, and simply write for
these transitions
\beq
M_{\mdn}=v_{\mu_1}\dots v_{\mu_n}\left(\zeta_1^{(n)}+\slash v \zeta_2^{(n)}\right)\g5,
\eeq
where the $\xi_i$  and $\zeta_i$ are functions of the kinematic variable
 $v\cdot p^\prime$. Thus, any of the heavy to light transitions is described by a 
pair of form factors, to this order in HQET. Note that for both sets of heavy to light
transitions, we may use the spin symmetry of HQET to relate the two form
factors necessary for $\Gamma=\gamma_\mu$ to those for $\Gamma=\gamma_\mu\g5$.

For $1/2^+$, we find
\begin{equation}\label{lighthp}
F_3=G_3=0,\,\,F_2=G_2=2\xi_2^{(0)},\,\,F_1=\xi_1^{(0)}-\xi_2^{(0)}
,\,\,G_1=\xi_1^{(0)}+\xi_2^{(0)},
\end{equation}
while for $1/2^-$, the form factors are
\begin{equation}
F_3=G_3=0,\,\,F_2=G_2=-2\zeta_2^{(0)},\,\,F_1=-\left(\zeta_1^{(0)}+\zeta_2^{(0)}\right)
,\,\,G_1=-\left(\zeta_1^{(0)}-\zeta_2^{(0)}\right).
\end{equation}
For $3/2^-$,
\begin{equation}
F_3=G_3=F_4=G_4=0,\,\,F_2=G_2=2\xi_2^{(1)},\,\,F_1=\xi_1^{(1)}-\xi_2^{(1)},\,\,G_1=\xi_1^{(1)}+
\xi_2^{(1)}.
\end{equation}
For $3/2^+$,
\begin{equation}
F_3=G_3=F_4=G_4=0,\,\,F_2=G_2=-2\zeta_2^{(1)},\,\,F_1=-\left(\zeta_1^{(1)}+\zeta_2^{(1)}\right),
\,\,G_1=-\left(\zeta_1^{(1)}-\zeta_2^{(1)}\right).
\end{equation}
For $5/2^+$,
\begin{equation}
F_3=G_3=F_4=G_4=0,\,\,F_2=G_2=2\xi_2^{(2)},\,\,F_1=\xi_1^{(2)}-\xi_2^{(2)},\,\,G_1=\xi_1^{(2)}+
\xi_2^{(1)}.
\end{equation}
Note that, in principle, the form factors for the decays to $1/2^-$ have no
relationship with those for decays to $3/2^-$, in this limit.

\section{The Model} 

\subsection{Wave Function Components}

\label{wfcomponents}

Our calculation follows the spirit of the work by ISGW~\cite{ISGW}. In our model, a baryon state has the form
\begin{eqnarray} |A_Q({\bf p},s)\rangle &=&
3^{3/4} \int d^3p_\rho d^3p_\lambda C^A \Psi_{A_Q}^S |q_1({\bf
p}_1,s_1)q_2({\bf p}_2,s_2)q_3({\bf p}_3,s_3)\rangle,\nonumber 
\end{eqnarray} 
where ${\bf p}_\rho= \frac{1}{\sqrt2}({\bf p}_1 - {\bf p}_2)$, ${\bf 
p}_\lambda =
\frac{1}{\sqrt6}({\bf p}_1 + {\bf p}_2 - 2{\bf p}_3)$ are the Jacobi 
momenta, 
$C^A$ is the antisymmetric color wave function and $\Psi_{A_Q}^S = 
\phi_{A_Q} 
\psi_{A_Q} \chi_{A_Q}$ is a symmetric combination of flavor, momentum 
and spin 
wave functions.

For $\Lambda_Q$ the flavor wave function we use is
\begin{eqnarray}
\phi_{\Lambda_Q} =  \frac{1}{\sqrt{2}}(ud -du)Q,\nonumber
\end{eqnarray}
which is antisymmetric in quarks $1$ and $2$. The momentum-spin
portion of the wave function must therefore be antisymmetric in quarks
$1$ and $2$. For states like the neutron and proton, we use the
`$uds$' basis used in Refs.~\cite{IsgurKarl-,CI}. In that basis, the
wave function of the proton is simply $uud$, while that for the
neutron is $ddu$. This flavor wave function provides some
simplification in dealing with matrix elements of the
Hamiltonian. However, the treatment of current matrix elements, such
as those that describe semileptonic decays, will require some extra
care, as will be explained later.

The total spin of the three spin-$1/2$ quarks can be either $3/2$ or $1/2$.  The
spin wave functions for the maximally stretched state in each case  are 

\begin{eqnarray}
\chi_{3/2}^S(+3/2)  &=& |\uparrow\uparrow\uparrow\rangle, \nonumber\\
\chi_{1/2}^\rho(+1/2)  &=& 
\frac{1}{\sqrt{2}}(|\uparrow\downarrow\uparrow\rangle - 
|\downarrow\uparrow\uparrow\rangle), \nonumber\\
\chi_{1/2}^\lambda(+1/2)  &=&- 
\frac{1}{\sqrt{6}}(|\uparrow\downarrow\uparrow
\rangle  + |\downarrow\uparrow\uparrow\rangle -2|
\uparrow\uparrow\downarrow\rangle), \nonumber
\end{eqnarray}
where $S$ labels the state as totally symmetric, while $\lambda(\rho)$ denotes the
mixed symmetric states that are symmetric (anti-symmetric) under the exchange
of quarks $1$ and $2$. The  momentum wave function for total
$L=\ell_\rho+\ell_\lambda$ is constructed from a Clebsch-Gordan sum of the wave
functions of the two Jacobi coordinates ${\bf p}_\rho$ and ${\bf p}_\lambda$, and takes
the  form 
\begin{eqnarray}
\psi_{LMn_{\rho}\ell_{\rho}n_{\lambda}\ell_\lambda}({\bf p}_\rho, {\bf 
p}_\lambda) = 
\sum_m\langle LM|\ell_{\rho}m,\ell_\lambda M-m\rangle\psi_{n_\rho \ell_\rho m}
({\bf p}_\rho) \psi_{n_\lambda \ell_\lambda M-m}({\bf p}_\lambda).\nonumber
\end{eqnarray}
The momentum and spin wave functions are then coupled to
give symmetric wave functions corresponding to total spin $J$ and parity
$(-1)^{(l_\rho+l_\lambda)}$,
\begin{eqnarray}
\Psi_{JM}= \sum_{M_L}\< JM|LM_L, SM-M_L\>\psi_{LM_Ln_{\rho}
\ell_{\rho}n_{\lambda}\ell_\lambda}\chi_{S}(M-M_L).\nonumber
\end{eqnarray}
The full wave function for a state $A$ is built from a linear 
superposition of such
components as 
\begin{equation}
\Psi_{A,J^PM}=\phi_A\sum_i \eta_i^A \Psi_{JM}^i.
\end{equation}
Here $\phi_A$ is the flavor wave function of the state $A$, and the  $\eta_i^A$
are coefficients that are determined by diagonalizing a Hamiltonian in  the
basis of the $\Psi_{JM}$. For this calculation, we limit the expansion  in the
last equation to components that satisfy $N\le 2$, where
$N=2(n_\rho+n_\lambda)+\ell_\rho+\ell_\lambda.$  Consistent with this is the 
fact that the states we discuss all correspond to $N\le 2$. With this
limitation,  the wave function for a $\Lambda_Q$ with $J^P=1/2^+$ takes the
form

\begin{eqnarray}
\Psi_{\Lambda_Q,1/2^+M}&=&\phi_{\Lambda_Q}\left(\vphantom{\sum_i}
\left[\eta_1^{\Lambda_Q}\psi_{000000}({\bf p}_\rho, {\bf 
p}_\lambda)
+\eta_2^{\Lambda_Q}\psi_{001000}({\bf p}_\rho, {\bf 
p}_\lambda)
+\eta_3^{\Lambda_Q}\psi_{000010}({\bf p}_\rho, 
{\bf p}_\lambda)\right]\chi_{1/2}^\rho(M)\right.\nonumber \\
&+&\eta_4^{\Lambda_Q}\psi_{000101}({\bf p}_\rho, 
{\bf p}_\lambda)\chi_{1/2}^\lambda(M)
+\eta_5^{\Lambda_Q}\left[\psi_{1M_L0101}({\bf p}_\rho, {\bf 
p}_\lambda)\chi_{3/2}^S(M-M_L)\right]_{1/2,M} \\
&+&\left.\eta_6^{\Lambda_Q}\left[\psi_{1M_L0101}({\bf p}_\rho, {\bf 
p}_\lambda)\chi_{1/2}^\lambda(M-M_L)\right]_{1/2,M}
+\eta_7^{\Lambda_Q}\left[\psi_{2M_L0101}({\bf p}_\rho, {\bf p}_\lambda)
\chi_{3/2}^S(M-M_L)\right]_{1/2,M}\right),\nonumber 
\end{eqnarray}
where 
$\left[\psi_{LM_Ln_\rho\ell_\rho n_\lambda\ell_\lambda}({\bf p}_\rho, {\bf 
p}_\lambda)\chi_S(M-M_L)\right]_{J,M}$ is a shorthand notation that denotes
the Clebsch-Gordan sum $\sum_{M_L}\< JM|LM_L, SM-M_L\>
\psi_{LM_Ln_\rho\ell_\rho n_\lambda\ell_\lambda}({\bf p}_\rho, {\bf 
p}_\lambda)\chi_S(M-M_L)$. When we diagonalize the Hamiltonian, this 
expansion will provide the wave functions for
seven states with $J^P=1/2^+$, the lowest of which will be taken to be 
the ground state of the system. 

A simplified version of the model would truncate this expansion after 
the first
component, giving
\begin{eqnarray}
\Psi_{\Lambda_Q,1/2^+M} &=& \phi_\Lambda \psi_{000000}({\bf p}_\rho, 
{\bf p}_\lambda)\chi_{1/2}^\rho(M),\nonumber
\end{eqnarray}
while the first radial excitation of interest in this model would be 
\begin{eqnarray}
\Psi_{\Lambda_Q,1/2^+_1M} &=& \phi_\Lambda \psi_{000010}({\bf p}_\rho, 
{\bf p}_\lambda)\chi_{1/2}^\rho(M).\nonumber
\end{eqnarray}
There exists a second radial excitation which, in the truncated basis 
would be
\begin{eqnarray}
\Psi_{\Lambda_Q,1/2^+_2M} &=& \phi_\Lambda \psi_{001000}({\bf p}_\rho, 
{\bf p}_\lambda)\chi_{1/2}^\rho(M).\nonumber
\end{eqnarray}

The latter state has its radial excitation in the $\rho$ coordinate, which 
means that it has a very small overlap with the ground state in the spectator
model that we  use. For some states, this truncation provides a very good
approximation to  the wave function, but there are important configuration
mixing effects for a  number of states. In the spectator assumption that we
use, not  all of these states have an overlap with the initial ground-state
$\Lambda_Q$.  The possible states  which can be connected to the ground state
are the states with $J^P= 1/2^+, 1/2^+_1, 1/2^-, 3/2^-, 3/2^+, 5/2^+$,
where $1/2^+$ and $1/2^+_1$ denote the ground state and the first (radially)
excited state. 

It is useful for us to list the single-component representations of these
states. The states with $J^P=1/2^+$ have already been given. For the remaining
states, we have
\begin{eqnarray}
\Psi_{\Lambda_Q,1/2^-M} &=& \phi_\Lambda \left[\psi_{1M_L0001}({\bf p}_\rho, 
{\bf p}_\lambda)\chi_{1/2}^\rho(M-M_L)\right]_{1/2,M},\nonumber\\
\Psi_{\Lambda_Q,3/2^-M} &=& \phi_\Lambda \left[\psi_{1M_L0001}({\bf p}_\rho, 
{\bf p}_\lambda)\chi_{1/2}^\rho(M-M_L)\right]_{3/2,M},\nonumber\\
\Psi_{\Lambda_Q,3/2^+M} &=& \phi_\Lambda \left[\psi_{2M_L0002}({\bf p}_\rho, 
{\bf p}_\lambda)\chi_{1/2}^\rho(M-M_L)\right]_{3/2,M},\nonumber\\
\Psi_{\Lambda_Q,5/2^+M} &=& \phi_\Lambda \left[\psi_{2M_L0002}({\bf p}_\rho, 
{\bf p}_\lambda)\chi_{1/2}^\rho(M-M_L)\right]_{5/2,M}.
\end{eqnarray}
From these representations, the multiplet structure expected in the heavy quark
limit is easily identified, with the $1/2^-$ and $3/2^-$ states forming a
multiplet, and the $3/2^+$ and $5/2^+$ states forming another. Both of the
$1/2^+$ states we consider are singlets.

\subsubsection{Expansion Bases}

A common choice for constructing baryon wave function is the harmonic
oscillator basis. One advantage of using this basis is that it
facilitates calculation of the required matrix elements. However, it
leads to form factors that fall off too rapidly at large values of
momentum transfer. We therefore also use the so-called Sturmian
basis~\cite{KP}. In this basis, form factors have multipole dependence
on $q^2$, which is what is expected experimentally. The full wave
functions in momentum space are

\begin{eqnarray}\label{hoa}
\psi^{\rm h.o.}_{nLm} ({\bf p})&=& \left[\frac{2\,n!}{\left(n + L +\half\right)!}
\right]^{\half} (i)^L(-1)^n \frac{1}{\alpha^{L+\thalf}} 
e^{-\frac{p^2}{(2\alpha^2)}}
L_n^{L+\half}(p^2/\alpha^2){\cal Y}_{Lm}({\bf p})
\end{eqnarray}
in the harmonic oscillator basis, and 
\begin{eqnarray} \label{sta}
\psi^{\rm St}_{nLm} ({\bf p}) &=& {2\left[ n! (n + 2L + 2)!\right]^{\half}
\over \left(n + L + \half\right)!}  (i)^L\frac{1}{\beta^{L+\thalf}}
\frac{1}{\left(\frac{p^2}{\beta^2} + 1\right)^{L+2}}
P_n^{\left(L +\thalf, L + \half\right)} \left(\frac{p^2 
-\beta^2}{p^2 
+\beta^2}\right) {\cal Y}_{Lm}({\bf p})
\end{eqnarray}
in the Sturmian basis. The $L_n^\nu(x)$ are generalized Laguerre
polynomials and the $P_n^{(\mu,\nu)}(y)$ are Jacobi polynomials, with
$p=\left|{\bf p}\right|$.  The corresponding wave functions in
coordinate space are

\begin{eqnarray}
\psi^{h.o.}_{nLm} ({\bf r})&=& \left[\frac{2\,n!}{\left(n + L +\half\right)!}
\right]^{\half} \alpha^{L+\thalf} e^{-\frac{\alpha^2r^2}{2}}
L_n^{L+\half}(\alpha^2r^2){\cal Y}_{Lm}({\bf r})\nonumber
\end{eqnarray}
in the harmonic oscillator basis, and 
\begin{eqnarray} 
\psi^{St}_{nLm} ({\bf r}) &=& \left[\frac{n!}{(n + 2L + 2)!}\right]^{\half}
 (2\beta)^{L+\thalf}e^{-\beta r}
L_n^{2L +2} \left(2\beta r\right) {\cal Y}_{Lm}({\bf r})\nonumber
\end{eqnarray}
in the Sturmian basis.

\subsubsection{Hamiltonian}

\label{Hamiltonian}

We use a non-relativistic quark model similar to that of Isgur and
Karl~\cite{IsgurKarl-,IsgurKarl}, with some of the modifications suggested by
Capstick and Isgur~\cite{CI,simon}. The Isgur-Karl model evolved from the
pioneering work of others; an extensive list of references to the origins of
the model can be found in Ref.~\cite{CI}.

The phenomenological Hamiltonian we use takes the form
\beq
H=\sum_i K_i + \sum_{i<j} 
\left( V^{ij}_{\rm conf}+H^{ij}_{\rm hyp}\right),
\eeq
where $\sum_iK_i$ is the kinetic part of the Hamiltonian. For this, we use two
forms, the usual non-relativistic form given by 
\beq
K_i= \left( m_i+\frac{p_i^2}{2m_i} \right),
\eeq
and a semirelativistic form given by
\beq
K_i= \sqrt{p_i^2+m_i^2}.
\eeq
The spin independent confining potential is a simplified version of
that used by Capstick and Isgur~\cite{CI}, with
\beq
V^{ij}_{\rm conf}=C_{qqq}+{br_{ij}\over 2}-{2\alpha_{\rm Coul}\over3r_{ij}},
\eeq
with $r_{ij}=\vert\rmb{r}_i-\rmb{r}_j \vert$. Here $H^{ij}_{\rm hyp}$ is the 
hyperfine interaction, assumed to have the form
\beq
H^{ij}_{\rm hyp}={2\alpha_{\rm hyp}\over 3 m_i m_j} \left\{
{8\pi\over 3} \rmb{S}_i\cdot\rmb{S}_j\delta^3(\rmb{r}_{ij})
+{1\over r_{ij}^3}
\left[ {3(\rmb{S}_i\cdot\rmb{r}_{ij})(\rmb{S}_j\cdot\rmb{r}_{ij})
         \over r_{ij}^2} - \rmb{S}_i\cdot\rmb{S}_j \right]
\right\}
\eeq
The first term is a contact term, while the second is a tensor term.
The spin-orbit interaction is neglected. We note here that 
$\alpha_{\rm Coul}$, $\alpha_{\rm hyp}$, $b$, $C_{qqq}$, and $m_i$ are not fundamental, but
are phenomenological parameters obtained from a fit to the spectrum of baryon states.
%
\subsection{Obtaining the Form Factors}

\subsubsection{$\Lambda_Q \to \Lambda_q$}

\label{obtaining}

Here, we illustrate the procedure we follow to obtain the form
factors, using the decay of the $\Lambda_Q$ to the ground state
$\Lambda_q$ as an example. We begin with the vector current matrix
element from Eq.~(\ref{vectorme}), with the assumption that the parent
$\Lambda_Q$ is at rest and the daughter $\Lambda_q$ has three momentum
${\bf p}$.
%
%
The left-hand side of Eq.~(\ref{vectorme}) is evaluated using the
quark model, after the operator $V_\mu=\bar{q} \gamma_\mu Q$ has been
reduced to its Pauli (non-relativistic) form. Specific values for the
index $\mu$ are chosen, as well as specific values of $s$ and
$s^\prime$. By making three sets of such choices, three equations for
the $F_i$ in terms of the quark-model matrix elements of three
operators are obtained. This system of equations is then solved to
obtain the expressions for the form factors. In the specific case at
hand, choosing $s=s^\prime=+1/2$ and $\mu=0$, for instance, leads to
\begin{eqnarray}
\langle \Lambda_q({\bf p},+)|\bar{q}\gamma_0 Q|\Lambda_Q(0,+)\rangle  
&=&  \int d^3p'_\rho d^3p'_\lambda d^3p_\rho
d^3p_\lambda  C^{A*} C^A  \Psi_{\Lambda_q}^{*S}(+)\nonumber\\
&\times&   \langle q'_1q'_2q|q^\dagger \gamma_0 Q|q_1q_2Q\rangle
\Psi_{\Lambda_Q}^S(+)\nonumber\\
&=&F_1+F_2+F_3,
\end{eqnarray}
where 
\beq
\langle q'_1q'_2q|q^\dagger \gamma_0 Q|q_1q_2Q\rangle =  
\langle q'_1q'_2|q_1q_2\rangle \langle q|q^\dagger \gamma_0 Q|Q\rangle.\nonumber
\eeq
The matrix element $\langle q'_1q'_2|q_1q_2\rangle$ gives $\delta$-functions
in spin, momentum and flavor in the spectator approximation, while the operator
$\bar{q}\gamma_0Q=1+{\cal O}\left(\frac{1}{m_qm_Q}\right)$. Using the
$\delta$-functions, the integral is simplified to
\beq
\left< \Lambda_q({\bf p},+)\left|\left[1+{\cal
O}\left(\frac{1}{m_qm_Q}\right)\right]\right|
\Lambda_Q({\bf 0},+)\right> =  \int d^3p_\rho d^3p_\lambda   
\psi_{\Lambda_q}^*({\bf p}'_\rho, {\bf p}'_\lambda)\left[1+{\cal O}
\left(\frac{1}{m_qm_Q}\right)\right]
\psi_{\Lambda_Q}({\bf p}_\rho, {\bf p}_\lambda),
\label{integral}
\eeq
with ${\bf p}'_\rho = {\bf p}_\rho$, ${\bf p}'_\lambda = {\bf
p}_\lambda - 2\sqrt{3/2}\,m_\sigma {\bf p}/m_{\Lambda_q}$, 
where $m_\sigma$ is the mass of the light quark. This leaves
the momentum integration, which is performed by using  both bases for the
momentum wave function shown earlier. The analytic results  for the form
factors for $\Lambda_Q$ decaying into various $\Lambda_q$ final states are
given in Appendix \ref{formfactors}. For decays to excited states, the
calculation of  the form factors is a little more involved, but the basic idea
is as outlined here.

\subsubsection{$\Lambda_Q \to N$}

For decays in which the daughter baryon is a nucleon, the procedure is much the
same as outlined in the previous subsection, with one modification. To
illustrate, let us take the specific example of $\Lambda_b\to p$. The flavor wave
functions of these two states have been chosen to be
\beq
\phi_{\Lambda_b}=\frac{1}{\sqrt{2}}(ud-du)b,\,\,\,\, \phi_p=uud.
\eeq
For the transition to occur, the third quark in the parent baryon, the $b$
quark, undergoes the transition $b\to u$, leaving a final state that is 
$\frac{1}{\sqrt{2}}(ud-du)u$. This has no overlap with the flavor wave function
that we use for the proton. We must now permute the third quark with the first and second
quarks, giving
\beq
\{13\} \frac{1}{\sqrt{2}}(ud-du)u=\frac{1}{\sqrt{2}}(udu-uud),\,\,\,\
\{23\} \frac{1}{\sqrt{2}}(ud-du)u=\frac{1}{\sqrt{2}}(uud-duu),
\eeq
both of which now have some overlap with the proton flavor wave function we use.
This requires that the sum of matrix elements
\begin{eqnarray}
\langle N({\bf p},+)|\{13\}O_i|\Lambda_Q({\bf 0},+)\rangle +\langle N({\bf
p},+)|\{23\}O_i|\Lambda_Q({\bf 0},+)\rangle\nonumber
\end{eqnarray}
be evaluated, where we apply the permutation to the wave function of the
daughter nucleon. The permutation operators also transform the spin and 
momentum wave function of the nucleon. The transformed spin wave functions are
\beq
\{13\} \chi^\lambda(s)=-\frac{\sqrt{3}}{2}\chi^{\rho}(s)-\frac{1}{2}
\chi^{\lambda}(s),\,\,\,\,
\{23\} \chi^\lambda(s)=\frac{\sqrt{3}}{2}\chi^{\rho}(s)-\frac{1}{2}
\chi^{\lambda}(s).
\eeq

After carrying out the transformation on the nucleon wave function,
and using the fact that the ground state momentum space wave function is
totally symmetric, we find
\beq
\langle p({\bf p},s)|O_i|\Lambda_Q({\bf 0},s^\prime)\rangle 
= (-\sqrt{3/4}) \int d^3p_\rho d^3p_\lambda   
\psi_{p}^*({\bf p}'_\rho, {\bf p}'_\lambda) A^{ss^\prime}(O_i)  
\psi_{\Lambda_Q}({\bf p}_\rho, {\bf p}_\lambda),
\eeq
where $A^{ss^\prime}(O_i)$ is the Pauli reduction of the operator
$O_i$. The integrations required for the $\Lambda_b$ to proton form
factors are the same as those in Eq.~(\ref{integral}) in the previous
subsection, and so the form factors are the same up to a
multiplicative factor. For excited states, however, the procedure is
slightly more involved, and is easily illustrated by examining the
decays to the radially excited nucleon.

Assuming single components, the wave function of the radially excited state is
\begin{equation}
\Psi_{N,1/2^+_1M} = \phi_N\psi_{000010}({\bf p}_\rho, 
{\bf p}_\lambda)\chi_{1/2}^\lambda(M).
\end{equation}
The $\{13\}$ transformation, acting on the spin-space part of this wave 
function, produces
\begin{eqnarray}
\{13\} \Psi_{000010}({\bf p}_\rho, {\bf
p}_\lambda)\chi_{1/2}^\lambda(M)&=& \Psi_{000010}({\bf p}_\rho^\prime, {\bf
p}_\lambda^\prime)
\left[-{\sqrt{3}\over 2}\chi_{1/2}^\rho(M)
-{1\over 2}\chi_{1/2}^\lambda(M)\right]\nonumber\\
&=&
-\frac{1}{8}\left[\sqrt{27}\,\psi_{001000}({\bf p}_\rho, {\bf p}_\lambda)
\chi_{1/2}^{\rho}(M)+
3\,\psi_{001000}({\bf p}_{\rho}, {\bf
p}_{\lambda})\chi_{1/2}^{\lambda}(M)\right.\nonumber\\
&+&\sqrt{3}\,\psi_{000010}({\bf p}_{\rho}, {\bf p}_{\lambda})\chi_{1/2}^{\rho}(M)
+\psi_{000010}({\bf p}_{\rho}, {\bf
p}_{\lambda})\chi_{1/2}^{\lambda}(M)\nonumber\\
&+&\left.\sqrt{18}\,\psi_{000101}({\bf p}_{\rho^\prime}, {\bf p}_{\lambda})\chi_{1/2}^{\rho}(M)
+\sqrt{6}\,\psi_{000101}({\bf p}_{\rho}, 
{\bf p}_{\lambda})\chi_{1/2}^{\lambda}(M)\right],
\end{eqnarray}
with a similar expression for the $\{23\}$ transformation. Here
${\bf p}_{\rho^\prime}= \frac{1}{\sqrt2}({\bf p}_3 - {\bf p}_2)$, 
${\bf p}_{\lambda^\prime} = \frac{1}{\sqrt6}({\bf p}_3 
+ {\bf p}_2 - 2{\bf p}_1)$ 
are the Jacobi coordinates in the transformed basis. Of these
components, only the first, third and fifth have spin wave functions
that overlap with the decaying $\Lambda_Q$, while only the first and
third have non-zero spatial overlaps. The integrals that arise from
the first component are simply a numerical factor ($\sqrt{27}/8$)
times those that arise in the $\Lambda_Q\to\Lambda_q$ matrix
elements, for the radially excited $\Lambda_q$. The integrals that
arise from the third term are also a numerical factor ($\sqrt{3}/8$)
times the $\Lambda_Q\to\Lambda_q$ ground-state integrals, multiplied
by a factor that arises from the spectator overlap. In this case, this
overlap is expected to be small, since the spectators are in a
radially excited state in the daughter baryon, but in their ground
state in the parent.

The above procedure is relatively straightforward to
implement in the harmonic oscillator basis, largely due to the fact
that the Moshinsky rotations have been treated by a number of authors,
and are also fairly simple to calculate. In particular, the fact that
the `permuted' wave function can be written in terms of a finite set
of transformed wave function components is another feature that makes
the harmonic oscillator basis attractive for calculations like
these. In the Sturmian basis, however, the permutation of particles
requires an infinite sum of transformed wave functions. This sum could
be truncated at some point in a calculation such as this. However, at
this point we do not examine decays to daughter nucleons in the
Sturmian basis.

\section{Analytic Results and Comparison with HQET}

The analytic expressions that we obtain for the form factors are shown in
Appendix \ref{formfactors}, for both the Sturmian and harmonic oscillator bases.
The results shown there are valid when the wave function for a
particular state is written as a single component, in either expansion basis.

As mentioned earlier, one of the advantages of the Sturmian basis is that it
leads to form factors that behave like multipoles in the kinematic variable, and
this is seen in the forms that we display. At this point, it is instructive to
compare, as far as possible, these analytic forms with the predictions of HQET.
While HQET does not give the explicit forms of the form factors, a number of
relationships among the form factors are expected, and any model should
reproduce these relationships. In what follows, we restrict our comparison to
the predictions that are valid at the non-recoil point, as we have ignored any
kinematic dependence beyond the Gaussian or multipole factors shown in Appendix
\ref{formfactors}. In addition, we focus mainly on the predictions for heavy to
heavy transitions. 

\subsection{Natural Parity Daughter Baryons}

We begin by discussing the form factors for decays to daughter baryons of
natural parity. In this work, this means daughter baryons with $J^P=1/2^+$ (both
ground state and first excited state), $J^P=1/2^-$ and $3/2^-$ (which constitute
a degenerate doublet when the daughter baryons are also treated as heavy) and
$J^P=3/2^+$ and $5/2^+$ (also a doublet). In our discussion of these results, we
implicitly assume that the wave functions for the states are dominated by a
single component of the wave function expansions that we use. These
single-component wave functions have been described in section 
\ref{wfcomponents}.

For elastic decays, predictions have been made at least to order
$1/m_q^2$ and $1/m_Q^2$. However, we will restrict our discussion to the predictions
valid to order $1/m_q$ and $1/m_Q$. To this order, using the results of Falk and
Neubert~\cite{falkneubert}, the relationships among form factors are
\begin{eqnarray}
F_2&=&G_2=\frac{m_Q}{m_q}F_3=-\frac{m_Q}{m_q}G_3,\nonumber\\
F_1&=&G_1-F_2\left(1+\frac{m_q}{m_Q}\right).
\end{eqnarray}
Our expressions for the form factors satisfy these relationships, in both bases,
to the appropriate order. In fact, the analytic forms obtained exactly match the
structure predicted by HQET~\cite{falkneubert}.

For the $\left(1/2^-,3/2^-\right)$ doublet, there are 14 form factors in
general, which Leibovitch and Stewart~\cite{LS} write in terms of a number of universal
functions and constants, valid at order $1/m_q$ and $1/m_Q$. Using their
expressions, and writing form factors for the $1/2^-$ state as primed
quantities, the relationships expected are
\begin{eqnarray}
F_1^\prime&=&\frac{1}{2\sqrt{3}m_q}\left(3m_Q-m_q\right)F_4,\nonumber\\
F_3^\prime&=&3G_3^\prime+\frac{2}{\sqrt{3}}\left(G_3-2F_4\right),\nonumber\\
F_3&=&-G_3,\,\,\,\, G_2=F_2,\,\,\,\, F_1-G_2=G_3-F_2,\nonumber\\
G_4&=&-3F_4+2\sqrt{3}G_3^\prime,\,\,\,\, 
F_2^\prime-G_2^\prime=-\frac{2}{\sqrt{3}}G_3,\nonumber\\
F_2^\prime+G_2^\prime+2G_1^\prime&=&\sqrt{3}F_4\left(1+\frac{m_Q}{m_q}\right)
-2G_3^\prime-\frac{2}{\sqrt{3}}G_3,
\end{eqnarray}
where terms that vanish at the non-recoil point have been ignored. Our results
for these states also satisfy all eight of the relationships shown above, in
both bases. Thus, there is a very good correspondence between the predictions of
HQET and those of the quark model that we use, and this correspondence is
independent of the wave function basis chosen.

For the $\left(3/2^+,5/2^+\right)$ doublet, the available predictions
are at leading order, shown in Eqs.~(\ref{spinthp}) and
(\ref{spinfhp}). These are also satisfied by our analytic expressions
for the form factors, in both bases.

For the excited state with $J^P=1/2^+_1$, the predictions of HQET are that the
form factors should vanish at the non-recoil point, by reason of the
orthogonality of the wave functions. In the treatments in the literature, this
is achieved by assuming that the form factors have an explicit factor that
vanishes as $w\to 1$. In the expressions that we have obtained for the leading
order form factors, this orthogonality arises explicitly from the size
parameters of the wave functions. 

It is instructive to examine the expression for $F_1$ for this decay, in the
limit when the Hamiltonian is that of a harmonic oscillator. The expression for
$F_1$ is 
\beq
F_1 = I_H
\frac{1}{2\alpha^2_{\lambda\lambda'}}\left[(\alpha^2_\lambda-\alpha^2_{\lambda'})-\frac{m_\sigma}{3\alpha^2_{\lambda\lambda'}}\left(\frac{\alpha^2_\lambda}{m_Q}(7\alpha^2_{\lambda'}-3\alpha^2_\lambda)
-\frac{\alpha^2_{\lambda'}}{m_q}(7\alpha^2_{\lambda}-3\alpha^2_{\lambda'})\right)\right],
\eeq
where
\beq
I_H =\sqrt{\frac{3}{2}}\left(\frac{\alpha_\lambda^{3/2}\alpha_{\lambda'}^{3/2}}
{\alpha_{\lambda\lambda'}^{3}}\right)
\exp\left( -\frac{3 m^2_\sigma}{2m^2_{\Lambda_q}}\frac{p^2}
{\alpha_{\lambda\lambda'}^2}\right).
\eeq
In the above expressions, $\alpha_{\lambda}(\alpha_{\lambda'})$ is the size parameter of the
initial (final) wave function associated with the Jacobi coordinate $\lambda$,
and $\alpha_{\lambda\lambda'}^{2}=(\alpha^2_{\lambda}+\alpha^2_{\lambda'})/2$. If
the Hamiltonian is taken to be a harmonic oscillator of the form
\beq
V=\frac{K}{2}\left(\left|{\bf r}_1-{\bf r}_2\right|^2+
\left|{\bf r}_1-{\bf r}_3\right|^2+\left|{\bf r}_2-{\bf r}_3\right|^2\right)
=3K\left(\rho^2+\lambda^2\right)
\eeq
where ${\bf r}_i$ is the position of the $i$-th quark and
$\lpmb{\rho}=({\bf r}_1-{\bf r}_2)/\sqrt{2}$ and 
$\lpmb{\lambda}=({\bf r}_1+{\bf r}_2-2{\bf r}_3)/\sqrt{6}$ are the
Jacobi coordinates, then
\beq
\alpha_\lambda=\left(\frac{3K m_\sigma m_Q}{m_Q+2m_\sigma}\right)^{1/4},\,\,\,\,
\alpha_{\lambda^\prime}=\left(\frac{3K m_\sigma
m_q}{m_q+2m_\sigma}\right)^{1/4}.
\eeq

With these forms, the term in $F_1$ proportional to $m_\sigma$ vanishes
identically, while the term in $(\alpha^2_\lambda-\alpha^2_{\lambda'})$
becomes proportional to $1/m_q-1/m_Q$, and so vanishes in the heavy quark
limit. The terms in $p^2$, which we do not include here, will be those
that contribute, despite the orthogonality of the wave functions, as
expected. Note that even though the $p^2$ terms will appear with explicit
factors of $1/m_q^2$, $p$ will range from small values (of order
$\Lambda_{\rm QCD}$), to a maximum of $(m_{\Lambda_Q}^2-
m_{\Lambda_q}^2)/(2m_{\Lambda_Q})$. Such terms are therefore not necessarily
negligible. However, in the non-relativistic model that we use for the form
factors, we have neglected such terms.

\subsection{Unnatural Parity Daughter Baryons}

For the decays to baryons with unnatural parity, HQET predicts that the form
factors should vanish at leading order. In the present model, we first have to
identify such states, which we do in the heavy quark limit, using the
single-component wave functions. The wave functions of interest are
\begin{eqnarray}
\Psi_{\Lambda_Q,1/2^+M} &=& \phi_\Lambda \left[\psi_{000101}({\bf p}_\rho, 
{\bf p}_\lambda)\chi_{1/2}^\lambda(M-M_L)\right]_{1/2,M},\nonumber\\
\Psi_{\Lambda_Q,3/2^+M} &=& \phi_\Lambda \left[\psi_{000101}({\bf p}_\rho, 
{\bf p}_\lambda)\chi_{3/2}^S(M-M_L)\right]_{3/2,M},\nonumber\\
\Psi_{\Lambda_Q,3/2^-M} &=& \phi_\Lambda \left[\psi_{1M_L0100}({\bf p}_\rho, 
{\bf p}_\lambda)\chi_{3/2}^S(M-M_L)\right]_{3/2,M},\nonumber\\
\Psi_{\Lambda_Q,5/2^-M} &=& \phi_\Lambda \left[\psi_{1M_L0100}({\bf p}_\rho, 
{\bf p}_\lambda)\chi_{3/2}^S(M-M_L)\right]_{5/2,M}.
\end{eqnarray}
In the spectator assumption that we use, none of these states have any
overlap with the ground state parent $\Lambda_Q$. In fact, there is a `two-fold'
orthogonality at play. The spin wave function of the two spectator
quarks is orthogonal to the corresponding wave function in the parent baryon.
The spatial wave functions of these two quarks are also orthogonal in parent and
daughter. Thus, decays to these states will only occur through configuration
mixing in the wave function, induced by various terms in the Hamiltonian. 

In the model that we use, configuration mixing in the spin wave functions
arises from hyperfine terms involving the heavy quark, which means that such
mixing will be small. Thus we expect that decays to such states should be
significantly suppressed. Interestingly, the suppression of the decays to
these unnatural parity doublets persists as the mass of the heavy quark in the
daughter baryon is decreased, as such configuration mixing remains small. In
this case, even though the definition of unnatural parity is different for
light states, there are still a number of decays (in $\Lambda_c\to\Lambda$, for
instance) that are predicted to be significantly suppressed. We will comment on this
later, when we examine the numerical results of our model.

\section{Numerical Results}

\subsection{Model Parameters, Mass Spectra and Wave Functions}

In Section~\ref{Hamiltonian}, we introduced the two Hamiltonians we diagonalize to
obtain the baryon spectrum. The two Hamiltonians differed only in the form chosen for
the kinetic portion, one of which was nonrelativistic (NR), while the other was
semirelativistic (SR). In addition, we use two different expansion bases to obtain the
wave functions: the harmonic oscillator (HO) basis, and the  Sturmian (ST) basis. In
the following, the four spectra we obtain will be denoted HONR, HOSR,  STNR and STSR,
in what should be an obvious notation.

There are eight free parameters to be obtained for each spectrum: four
quark masses ($m_u=m_d$, $m_s$, $m_c$ and $m_b$), and 4 parameters of
the potential ($\alpha_{\rm hyp}$, $\alpha_{\rm Coul}$, $b$ and
$C_{qqq}$). We have investigated the effects of a tensor interaction
in the two harmonic oscillator models, and found the effects to be
small. In the results we present, the tensor interaction has therefore
been ignored. The eight parameters are determined from a `variational
diagonalization' of the Hamiltonian. The variational parameters are
the size parameters $\alpha_\rho$ and $\alpha_\lambda$ of
Eq.~(\ref{hoa}), or $\beta_\rho$ and $\beta_\lambda$ of
Eq.~(\ref{sta}). This variational diagonalization is accompanied by a
fit to the known spectrum. In this fit, the eight parameters mentioned
before are varied. The values we obtain for the Hamiltonian parameters
are shown in Table~\ref{parameter1}, while some of the wave function
size parameters are shown in Table~\ref{parameter2}.
\begin{center}
\begin{table}[h]
\caption{Hamiltonian parameters obtained from the four different fits. In the
first column, HO refers to the harmonic oscillator basis, while ST
refers to the Sturmian basis. In the same column, NR indicates a
non-relativistic Hamiltonian, while SR indicates a semirelativistic
one. The form of the Hamiltonian is described in
Section~\ref{Hamiltonian}.
\label{parameter1}}
\begin{tabular}{|l|cccccccc|}
\hline
 model & $m_\sigma$ (GeV)& $m_s$ (GeV) & $m_c$ (GeV) & $m_b$ (GeV) & $b$ (GeV$^2$) 
 & $\alpha_{\rm Coul}$ \,\,\,\,\,\,&
$\alpha_{\rm hyp}$ & $C_{qqq}$ (GeV) \\\hline
 HONR & 0.40& 0.65& 1.89 & 5.28 & 0.14 &0.45 & 0.81 & -1.20 \\
 HOSR & 0.38& 0.59& 1.83 & 5.17 & 0.17 &0.09 & 0.26 & -1.45 \\
 STNR & 0.40& 0.64& 1.87 & 5.28 & 0.13 &0.35 & 0.31 & -1.22 \\
 STSR & 0.34& 0.57& 1.78 & 5.22 & 0.15 &0.19 & 0.11 & -1.23  \\ \hline
\end{tabular}
\end{table}
\end{center}
We note that the value of $b$, the slope of the linear potential, tends to be
smaller than in most published studies of the baryon spectrum. The same is true
for the strength of the hyperfine interaction, $\alpha_{\rm hyp}$. In the case
of the latter, the small strength arises because the hyperfine interaction is
treated as a contact interaction, and this can lead to very strong attractive
forces between the quarks. One result of this is that, for sufficiently large
values of $\alpha_{\rm hyp}$, the masses of the lightest baryon states can become negative.
The small value of this parameter that results from our fits is therefore
driven largely by the need for positive baryon masses. One direct consequence
is that hyperfine splittings are not well reproduced in all but the HONR model, with the
$\Delta-N$ mass splitting being about one third of its experimental value.
\begin{center}
\begin{table}[h]
\caption{Wave function size parameters, $\alpha_\rho$ and $\alpha_\lambda$, for
states of different $J^P$, in different models. All values are in GeV. For the
Sturmian basis, the size parameters have been denoted $\beta$ in the text.
\label{parameter2}}
\begin{tabular}{|l|l|cccc|}
\hline
$J^P$ &  model & $\Lambda_b$ & $\Lambda_c$ & $\Lambda$ & $N$  \\ 
&  & $\left(\alpha_\lambda,\,\,\,\,\alpha_\rho\right)$ &
$\left(\alpha_\lambda,\,\,\,\,\alpha_\rho\right)$ &
$\left(\alpha_\lambda,\,\,\,\,\alpha_\rho\right)$ &
$\left(\alpha_\lambda,\,\,\,\,\alpha_\rho\right)$  \\ \hline
$1/2^+$ &HONR &  (0.59, 0.61)& (0.55, 0.58) & (0.49, 0.53) & 0.48\\
$1/2^+$ &HOSR & (0.68, 0.68)& (0.60, 0.61) & (0.52, 0.57)& 0.54\\
$1/2^+$ & STNR &  (0.44, 0.66)& (0.41, 0.69) & (0.35, 0.75)&- \\
$1/2^+$ & STSR &  (0.46, 0.64)& (0.43, 0.67) & (0.38, 0.72) &-\\ \hline
$1/2^-$ & HONR &  -& (0.47, 0.49) & (0.40, 0.47) &0.37\\
$1/2^-$ & HOSR &  -& (0.55, 0.59) & (0.48, 0.54) &0.46 \\
$1/2^-$ &STNR & -& (0.60, 0.50) & (0.55, 0.54) &- \\
$1/2^-$ &STSR & -& (0.61, 0.49) & (0.58, 0.51) &-\\ \hline
$3/2^+$ &HONR & -& - & - & 0.35\\
$3/2^+$ &HOSR & -& -& -& 0.44\\\hline
$5/2^+$ &HONR & -& - & - & 0.35\\
$5/2^+$ &HOSR & -& -& -& 0.46\\\hline
\end{tabular}
\end{table}
\end{center}
In general, we allow the values of $\alpha_\rho$ to be different from
$\alpha_\lambda$. The exceptions occur in cases when the three quarks
are identical, as they are in the nucleon. In that case, the
variational diagonalization automatically selects
$\alpha_\rho=\alpha_\lambda$. In Table~\ref{parameter2}, we show only
some values of the size parameters. The other size parameters, for the
states that are significant for this work, are related to those
presented. For instance, for the $1/2^+_1$ states, the size parameters
are the same as for the $1/2^+$ states. Furthermore, since we do not
include a spin-orbit interaction in our Hamiltonian, the size
parameters for the $1/2^-$ and $3/2^-$ states are identical. We do not
show the size parameters for the $\Lambda_Q$ states with $Q=b$, $c$,
or $s$ and $J^P=$ $3/2^+$ or $5/2^+$, mainly because we find that
semileptonic decays to these states are very small.

\subsubsection{Mass Spectra}

Portions of the four mass spectra we obtain are shown in
Table~\ref{baryonspec}. In this table, the first two columns identify
the state and its experimental mass, while the next four columns show
the masses that result from the models that we use. The small
hyperfine interaction that we alluded to in the previous subsection
has resulted in ground state nucleons that are too heavy, in all
models. In addition, the ground state $\Delta$ (not shown in the
table) is too light in all models. Similar patterns emerge when the
various $\Lambda_Q$ and $\Sigma_Q$ (not shown) states are compared.
The size of this interaction also results in `radial' excitations that
are too heavy, even heavier than usually result in models like these.

We note, too, that the different models give very similar results for many of the
states such as the $N(1/2^+)$, $N(1/2^-)$, $\Lambda(1/2^+)$ and $\Lambda_b(1/2^+)$,
for instance, but for some states such as $N(1/2^+_1)(1440)$, there are striking differences in the masses
obtained.
\begin{center}
\begin{table}[h]
\caption{Baryon masses in GeV fitted in different quark models. The first two columns 
identify the state and its experimental mass, while the next four columns
show the masses that result from the models that we use.
\label{baryonspec}}
\begin{tabular}{|l|c|llll|}
\hline
State& Experimental Mass& HONR& HOSR& STNR& STSR\\ \hline
$N(1/2^+)$&0.94&1.00&1.08&1.08&1.08\\
$N(1/2^+_1)$&1.44&1.76&1.60&1.81&1.70\\
$N(1/2^-)$&1.54&1.45&1.44&1.50&1.47\\
$N(3/2^-)$&1.52&1.45&1.44&1.50&1.47\\
$N(3/2^+)$&1.72&1.72&1.69&1.78&1.77\\ 
$N(5/2^+)$&1.68&1.72&1.69&1.78&1.77\\ \hline
$\Lambda(1/2^+)$&1.12&1.23&1.23&1.12&1.10\\
$\Lambda(1/2^+_1)$&1.60&1.73&1.81&1.61&1.55\\
$\Lambda(1/2^-)$&1.41&1.54&1.62&1.50&1.56\\
$\Lambda(3/2^-)$&1.52&1.54&1.62&1.50&1.56\\
$\Lambda(3/2^+)$&1.89&1.81&1.81&1.77&1.87\\ 
$\Lambda(5/2^+)$&1.82&1.82&1.81&1.77&1.87\\ \hline
$\Lambda_c(1/2^+)$&2.28&2.35&2.32&2.26&2.22\\
$\Lambda_c(1/2^-)$&2.59&2.61&2.70&2.61&2.68\\
$\Lambda_c(3/2^-)$&2.63&2.61&2.70&2.61&2.68\\\hline
$\Lambda_b(1/2^+)$&5.62&5.62&5.62&5.62&5.62\\ \hline
\end{tabular}
\end{table}
\end{center}

\subsubsection{Wave Functions}

For many of the states that we treat, the wave functions that result are, to a
very good approximation, the single component wave functions shown in Section
\ref{wfcomponents}. This turns out to be a particularly good approximation for
the orbitally excited states such as the $1/2^-$ and $3/2^-$ states,
for all but the nucleon states. For the $\Lambda(1/2^-)$ and
$\Lambda(3/2^-)$, for instance, the dominant component has a
coefficient [the $\eta_i$ of Eq.~(\ref{lthm})] of at least 0.985 in
all of the models. We treat such states as being single component
states, and this will introduce errors of about a few percent
(typically less than three percent for the particular states
mentioned, usually much less for the states containing a $c$ or $b$
quark).

\begin{center}
\begin{table}[h]
\caption{Mixing coefficients ($\eta_i$) of the two lowest lying
$1/2^+$ states in different flavor sectors. The $\eta_i$ are defined 
in Eq.~(\ref{lhfpl}) of Appendix~\ref{wavefunctions}.
\label{mixingwf}}
\begin{tabular}{|l|ccc|ccc|ccc|ccc|}
\hline
Baryon states&& HONR &&& HOSR&  && STNR&& &STSR &\\
& $\eta_1$& $\eta_2$& $\eta_3$& $\eta_1$& $\eta_2$& $\eta_3$& $\eta_1$& $\eta_2$
& $\eta_3$& $\eta_1$& $\eta_2$& $\eta_3$\\ \hline
$N(1/2^+)$&0.979&-0.150&0.034&0.989& -0.110&0.028&-&-&-&-&-&-\\
$N(1/2^+_1)$& 0.022 &  0.522&  0.825 & -0.026& 0.579& 0.800&-&-&-&-&-&-\\
$\Lambda(1/2^+)$&0.994&0.005&-0.069&0.998&0.003&-0.035& 0.900&0.208& 0.382
&0.875&0.313&0.368\\
$\Lambda(1/2^+_1)$& 0.047&  0.149& 0.962&0.018& 0.650& 0.750&-0.177&
0.977&-0.115&
-0.279&0.950& -0.152\\
$\Lambda_c(1/2^+)$&0.999&0.001&-0.020& 0.999&$<$0.001&-0.012&0.917&0.137&0.374
&0.877&0.289&0.382\\
$\Lambda_c(1/2^+_1)$&0.017& 0.100& 0.993&0.010& 0.361&0.931&-0.138& 0.989&
-0.059&
-0.257&0.957& -0.132\\
$\Lambda_b(1/2^+)$&0.999&$<$0.000&-0.003& 0.999&$<$0.001&-0.004&0.915&0.141&0.378
 &0.876&0.286&0.390\\ \hline
\end{tabular}
\end{table}
\end{center}
Significant mixing occurs only in the $1/2^+$ sector, for all flavors,
particularly in the Sturmian models. Table~\ref{mixingwf} shows the
wave function coefficients for the two lowest $1/2^+$ states, in each
flavor sector, for all four models (in the case of the nucleon, we
show only the results from the HO models). The mixing shown in this
table complicates the extraction of the form factors. However, in all
results that we show for the form factors and the decay rates, this
mixing is properly accounted for. Note that in each of these wave
functions, there is also some contribution from the term in
$\eta_4$. However, this component of the wave function has negligible
overlap with the wave function of the parent baryon, and so is
neglected here.

\subsection{Form Factors and Decay Rates}

In our calculation of the form factors, we have assumed that we can use 
non-relativistic approximations for the operators. This means that we have
ignored terms in the various quark model operators that appear at order
$1/m_q^2$, $1/m_Q^2$, and above. Such terms have also been ignored in writing
the hadronic matrix elements. However, in extracting the form factors, we have
kept, and shown, terms that are of order $1/(m_q m_Q)$. To examine the validity
of this treatment, we write each form factor as
\beqy
\label{ffcomps}
F_i&=& F_i^{(0)}+\frac{1}{m_q}F_i^{(q)}+\frac{1}{m_Q}F_i^{(Q)}
+\frac{1}{m_qm_Q}F_i^{(qQ)},\nonumber\\
&\equiv&{\cal F}_i^{(0)}+{\cal F}_i^{(q)}+{\cal F}_i^{(Q)}
+{\cal F}_i^{(qQ)}
\eeqy
and show the values for ${\cal F}_i^{(0)}$, ${\cal F}_i^{(q)}$, etc., in 
Table~\ref{formfactors2}. In this table, we show only the results for the
HONR and STNR models.

\begin{center}
\begin{table}[h]
\caption{Form factor components ${\cal F}_i$ and ${\cal G}_i$ as defined in
Eq.~(\protect{\ref{ffcomps}}), evaluated at the non-recoil point. The
components are shown for the HONR (HO) and STNR (St.) models. The
columns labeled `$\Lambda_c$' are for the $\Lambda_c\to\Lambda^{(*)}$
form factors, while those labeled `$\Lambda_b$' are for the
$\Lambda_b\to\Lambda_c^{(*)}$ form factors.
\label{formfactors2}}
\begin{tabular}{|l|llll|llll|llll|}
\hline
& \multicolumn{4}{c|}{$J^P=1/2^+$} & \multicolumn{4}{c|}{$J^P=1/2^-$} &
\multicolumn{4}{c|}{$J^P=3/2^-$} \\
form& \multicolumn{2}{c}{$\Lambda_c$} &\multicolumn{2}{c|}{$\Lambda_b$} &
\multicolumn{2}{c}{$\Lambda_c$} &\multicolumn{2}{c|}{$\Lambda_b$}
& \multicolumn{2}{c}{$\Lambda_c$} &\multicolumn{2}{c|}{$\Lambda_b$}\\
factor & H.O.\,\,\,\,\,\, & St. \,\,\,\,\,\, & H.O.\,\,\, & St. \,\,\,\,\,\, 
& H.O.\,\,\, & St. & H.O.\,\,\, & St.\,\,\,\,\,\,  & H.O.\,\,\, & St.\,\,\,\,\,\,  & H.O.\,\,\, &
St.\\ \hline
${\cal F}_1^{(0)}$&0.98&0.97& 0.99 &0.99& 0 & 0 & 0 & 0 &-1.08 &-1.48&-1.16
&-1.38 \\
${\cal F}_1^{(q)}$& 0.54&0.78 & 0.20 &0.28 & 0.36 &0.32&0.16 &0.12 &
-0.46&-0.76&
-0.23&-0.25  \\
${\cal F}_1^{(Q)}$ &0.23& 0.15&0.08 &0.05 & -0.04 &-0.04&-0.04 &-0.01&-0.10
&-0.37&
-0.03&-0.12  \\
${\cal F}_1^{(qQ)}$&0&0& 0 &0 &0  &0  & 0& 0 &0  &0  & 0& 0\\ \hline
${\cal F}_2^{(0)}$&0 &0&  0 &0&-1.24&-1.71  &-1.34 &-1.60 &0 & 0 &0 & 0\\
${\cal F}_2^{(q)}$&-0.54&-0.72&0.20&-0.26   &0.36  &0.32&0.16&0.12
&0.46&0.76&0.23&0.25  \\
${\cal F}_2^{(Q)}$&0&0& 0 &0 &-0.34&-0.43 &-0.11 &-0.14 &$<$0.01 & $<$0.01
&$<$0.01 & $<$0.01\\
${\cal F}_2^{(qQ)}$&0.05&-0.03& 0.01 &$<$0.01 &0.06&0.05  &0.01&0.02
&0.08&0.07&0.02&0.01   \\ \hline
${\cal F}_3^{(0)}$&0&0&0 &0  &0  &0  &0 &0 &0  &0  &0 &0  \\
${\cal F}_3^{(q)}$&0&0& 0 &0 &0  &0  &0 & 0  &0  &0  &0 & 0\\
${\cal F}_3^{(Q)}$&-0.21&-0.11&-0.07 &-0.04  & 0.34 &0.43& 0.08&0.14&
0.37&0.43&0.13&0.15\\
${\cal F}_3^{(qQ)}$&0&0&0 &0  &0  &0   & 0 &0 &0  &0  &0 & 0\\ \hline
${\cal F}_4^{(Q)}$&-&-&-&-&- &- & -&- &-0.14&-0.13&-0.06 &-0.05  \\\hline
${\cal G}_1^{(0)}$&0.98&0.97&0.99 &0.99  &1.24 &1.71 &1.34&1.60  &
-1.08&-1.48&-1.16&-1.38  \\
${\cal G}_1^{(q)}$&0&0&  0&0 & 0  & 0 &0 &0 & 0  & 0 &0 &0  \\
${\cal G}_1^{(Q)}$&0&0& 0 &0 & 0.04&0.02 & 0.02&0.01  & 0.07&0.06&0.03 & 0.02
\\
${\cal G}_1^{(qQ)}$&0.02&-0.01& $<$0.01 &$<$0.01 & 0.06 &0.02 & 0.01 &$<$0.01
& 0.05&0.07& 0.01 & 0.01 \\ \hline
${\cal G}_2^{(0)}$&0&0&0 &0  &-1.24 &-1.71 &-1.34&-1.60  & 0&0  & 0&0\\
${\cal G}_2^{(q)}$&-0.54&-0.72& -0.20  &-0.26&0.36 &0.32&0.16&0.12  &
0.46&0.76&0.23&0.25  \\
${\cal G}_2^{(Q)}$&0&0& 0 &0 &0.08  &0.06&0.04 &0.03 & 0& 0& 0& 0\\
${\cal G}_2^{(qQ)}$&-0.14&0.06&  -0.02&-0.01 & 0 & $<$0.01 & $<$0.01& $<$0.01&
0.14& 0.20 &0.02& 0.02 \\ \hline
${\cal G}_3^{(0)}$&0&0& 0 &0 &0  &0  &0 &0 &0  &0  &0 &0  \\
${\cal G}_3^{(q)}$&0&0& 0 &0 & 0 &0  & 0&0  &0  &0  &0 &0\\
${\cal G}_3^{(Q)}$&0.23&0.11&0.08 &0.04  &0.08&0.07  &0.04&0.03  &-0.37
&-0.43&-0.13&-0.15\\
${\cal G}_3^{(qQ)}$&0.13&0.09& 0.02&0.01  &  -0.06&-0.02&-0.01 &0.01 &
-0.17&-0.1& -0.03&-0.02 \\ \hline
${\cal G}_4^{(Q)}$&-&-&-   &-&-  &-  &-  &-&0.14&0.18&0.06 &0.06  \\
${\cal G}_4^{(qQ)}$&-&- &- &- &-  &- &-  &- &0.12&0.13&0.02&0.02   \\ \hline
\end{tabular}
\end{table}
\end{center}

For the elastic decays, the form factors $F_1$ and
$G_1$ are dominant, while all other form factors are sub dominant. For $1/2^-$
final states, $F_2$, $G_1$ and $G_2$ are dominant, while for $3/2^-$, $F_1$ and
$G_1$ are the dominant form factors. In each case, we see that the ${\cal F}^{(0)}$
or ${\cal G}^{(0)}$ term is significantly larger than the `higher order' terms, as
expected. The numbers in this table suggest that the convergence in $1/m_q$ is
rapid, modulo the model dependence.

\subsubsection{$\Lambda_c \to \Lambda^{(*)}$}

In Table~\ref{formfactors1} we show the values of the form factors at
the non-recoil point, for the decays $\Lambda_c\to\Lambda$, for both
elastic and inelastic channels. In this table, the results from all
four models are presented. The results we obtain for the elastic
channel are consistent with the predictions of HQET as estimated by
Scora~\cite{scora}.

\begin{center}
\begin{table}[h]
\caption{The form factors for $\Lambda_c\to\Lambda^{(*)}$ transitions,
calculated at the non-recoil point, in the four models used here.
\label{formfactors1}}
\begin{tabular}{|l|l|lllclllc|}
\hline
spin & model & $F_1$\,\,\, & $F_2$\,\,\,  & $F_3$ \,\,\, & $F_4$ \,\,\, & 
$G_1$ \,\,\,  & $G_2$ \,\,\, & $G_3$ \,\,\, & $G_4$\,\,\,  \\ \hline
$1/2^+$ &HONR & 1.75 & -0.54 & -0.23 & -&0.98 & -0.54 & 0.23  & - \\
$1/2^+$ &HOSR & 1.76 & -0.55 & -0.24 & - & 0.98 & -0.55 & 0.24 & - \\
$1/2^+$ & STNR & 1.90 & -0.72 & -0.11 & - & 0.97 & -0.72 & 0.11 & - \\
$1/2^+$ & STSR &1.78 & -0.66 & -0.09 & - & 0.92 & -0.66 & 0.09 & -   \\ \hline
$1/2^-$ &HONR & 0.32 & -1.22 & 0.34 & - & 1.20 & -0.80 & 0.08 & -\\
$1/2^-$ &HOSR & 0.42 & -1.02 & 0.30 & - & 1.14 & -0.61 & 0.10 & - \\
$1/2^-$ & STNR & 0.28 & -1.82 & 0.43&- &  1.73 & -1.42 & 0.07 & - \\
$1/2^-$ & STSR  & 0.36 & -1.30 & 0.31 & - & 1.38 & -1.04 & 0.08 & -\\ \hline
$3/2^-$ & HONR & -1.83 & 0.46 & 0.37 & -0.14 &-1.00 & 0.46 & -0.37 & 0.14 \\
$3/2^-$ &HOSR &-1.81 & 0.52 & 0.35 & -0.18 &-0.94 & 0.52 & -0.35 & 0.16\\
$3/2^-$ & STNR & -2.61 & 0.76 & 0.43 & -0.13 & -1.42 & 0.76 & -0.47 & 0.13 \\
$3/2^-$ &STSR & -2.03 & 0.58 & 0.34 & -0.13 &-1.11 & 0.57 & -0.38 & 0.13 \\
\hline
\end{tabular}
\end{table}
\end{center}
In their treatment of the process $\Lambda_c\to\Lambda e^+\nu$, the CLEO
Collaboration have used the leading order predictions of HQET to analyze the
decay rate in terms of two form factors, $\xi_1$ and $\xi_2$. In terms of the
form factors that we have been using, these HQET form factors are 
\beqy \label{xiforms}
\xi_1&=&F_1+F_2/2,\,\,\,\, \xi_2=F_2/2,\nonumber\\
\xi_1&=&G_1-G_2/2,\,\,\,\, 
\xi_2=G_2/2
\eeqy
The two sets of equations above arise from inverting
Eqs.~(\ref{lighthp}) either in terms of the $F_i$ or the $G_i$. In
Table~\ref{ffratio}, we show the values we obtain for the ratio
$\xi_2/\xi_1$, evaluated at the non-recoil point. We also show the
value obtained by the CLEO Collaboration in their analysis. We note
that CLEO present a single value for the ratio of form factors, while
we have two sets of values, arising from the two equations above.
These two expressions give values for this ratio that are different,
but not disturbingly so. The vector ratio (involving the $F_i$) tends
to be smaller than the axial-vector ratio (involving the $G_i$), and
both are smaller than the ratio extracted by the CLEO
collaboration. The differences among the numbers we obtain using the
two methods can be traced back to the $1/m_Q$ terms in $F_1$; if those
terms are ignored, both methods give the same value for the ratio.

\begin{center}
\begin{table}[h]
\caption{The ratio $\xi_2/\xi_1$ for $\Lambda_c \to \Lambda(1/2^+)$. 
The first row is obtained using the vector relation defined in the
text, while the second row is obtained using the axial-vector
relation.
\label{ffratio}}
\begin{tabular}{|l|lllll|}
\hline
$\xi_2/\xi_1$ & HONR & HOSR & STNR & STSR & CLEO \\ \hline
Vector & -0.18 & -0.18 & -0.23 & -0.23 & -0.31 \\ \hline
Axial Vector & -0.21 & -0.22 & -0.27 & -0.26 & -0.31 \\ \hline
\end{tabular}
\end{table}
\end{center}
\begin{figure}[h]
\centerline{\epsfig{file=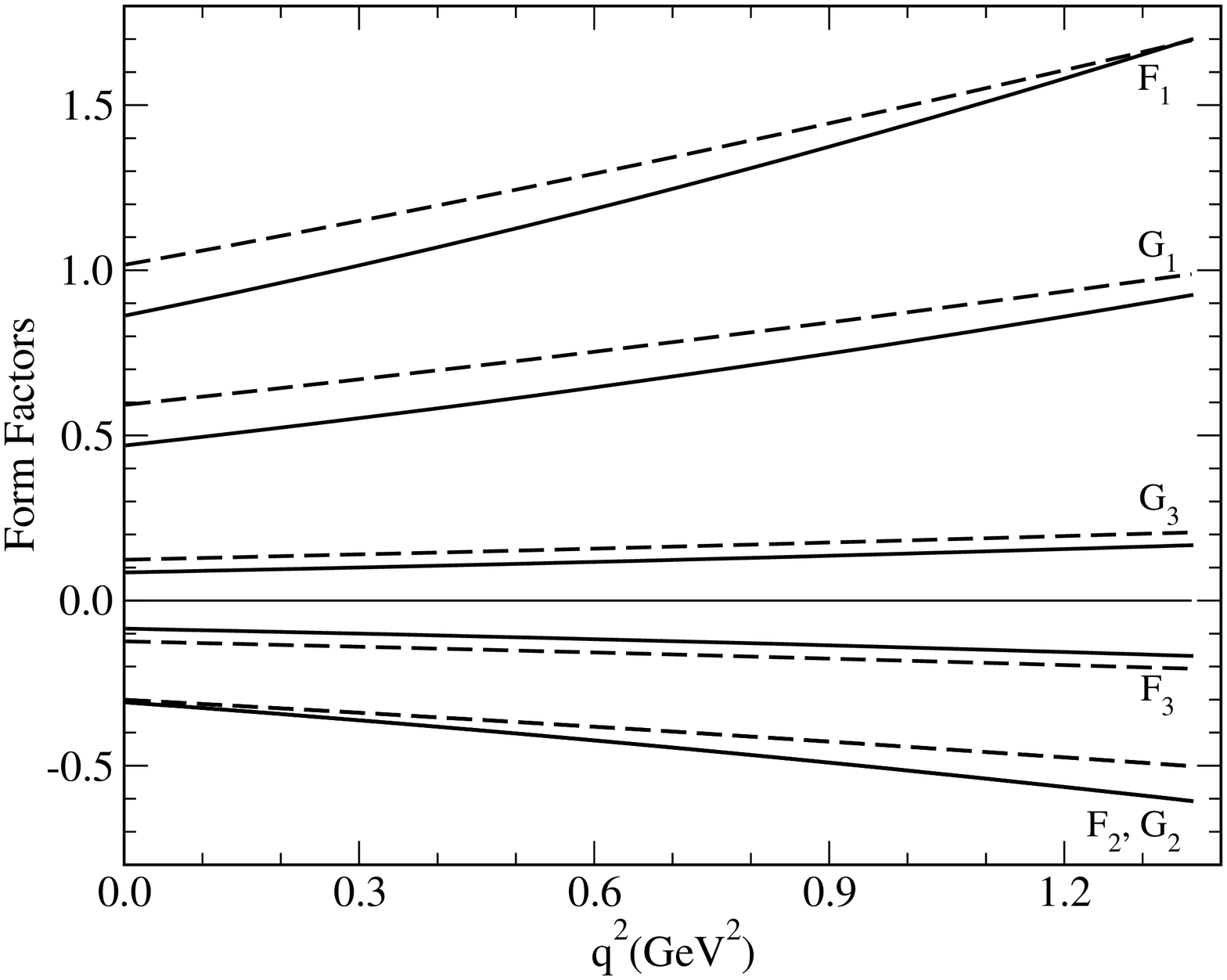,width=10cm}
\hspace*{-0.5in}\epsfig{file=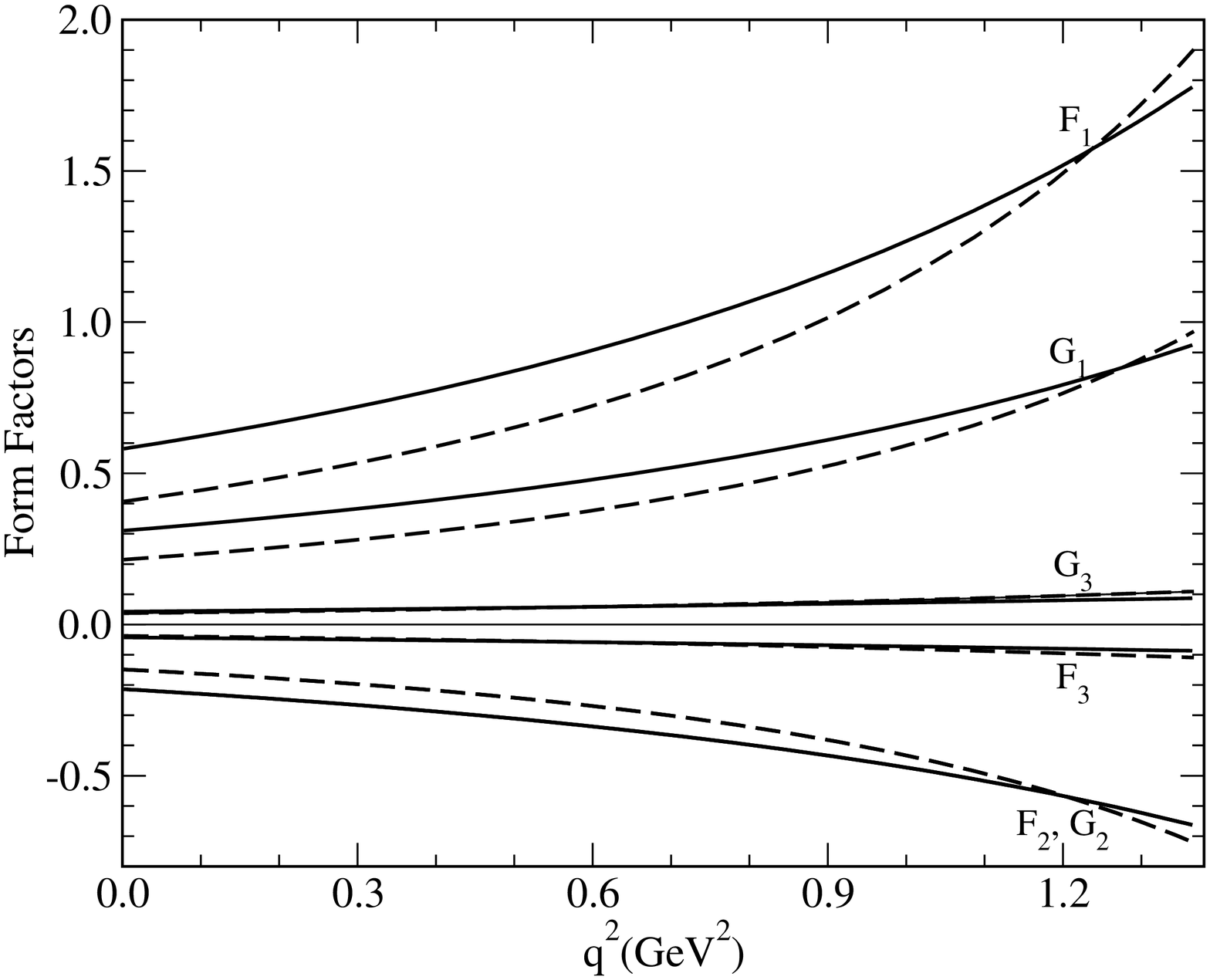,width=10cm}}
\caption{Form factors for $\Lambda_c\to\Lambda(1/2^+)$ obtained using harmonic
oscillator wave functions (left panel, HOSR and HONR models) and Sturmian wave 
functions (right panel, STSR and STNR models). In each panel, the solid curves 
arise from the semirelativistic version of the model, while the dashed curves 
arise from the nonrelativistic version. Note that $F_2$ is indistinguishable 
from $G_2$ in all cases.
\label{formfactor1}}
\end{figure}
Figure~\ref{formfactor1} shows the $q^2$ dependence of the form
factors for the elastic transition $\Lambda_c\to\Lambda(1/2^+)$, calculated 
in the HONR and
HOSR models on the left, and in the STSR and STNR models on the
right. In each panel, the solid curves arise from the SR version of
the model, while the dashed curves are from the NR version. If we
compare the form factors shown in Figure~\ref{formfactor1}, we see
that those calculated using the Sturmian wave functions have
larger slopes near the non-recoil point (maximum $q^2$) than those
calculated using the harmonic oscillator wave functions. The form
factors calculated in the different models all have similar values
near the non-recoil point (as seen in Table~\ref{formfactors1}). The
larger slopes in the case of the Sturmian model form factors means
that we can expect smaller integrated rates from the STSR and STNR
models.

\begin{figure}[h]
\centerline{\epsfig{file=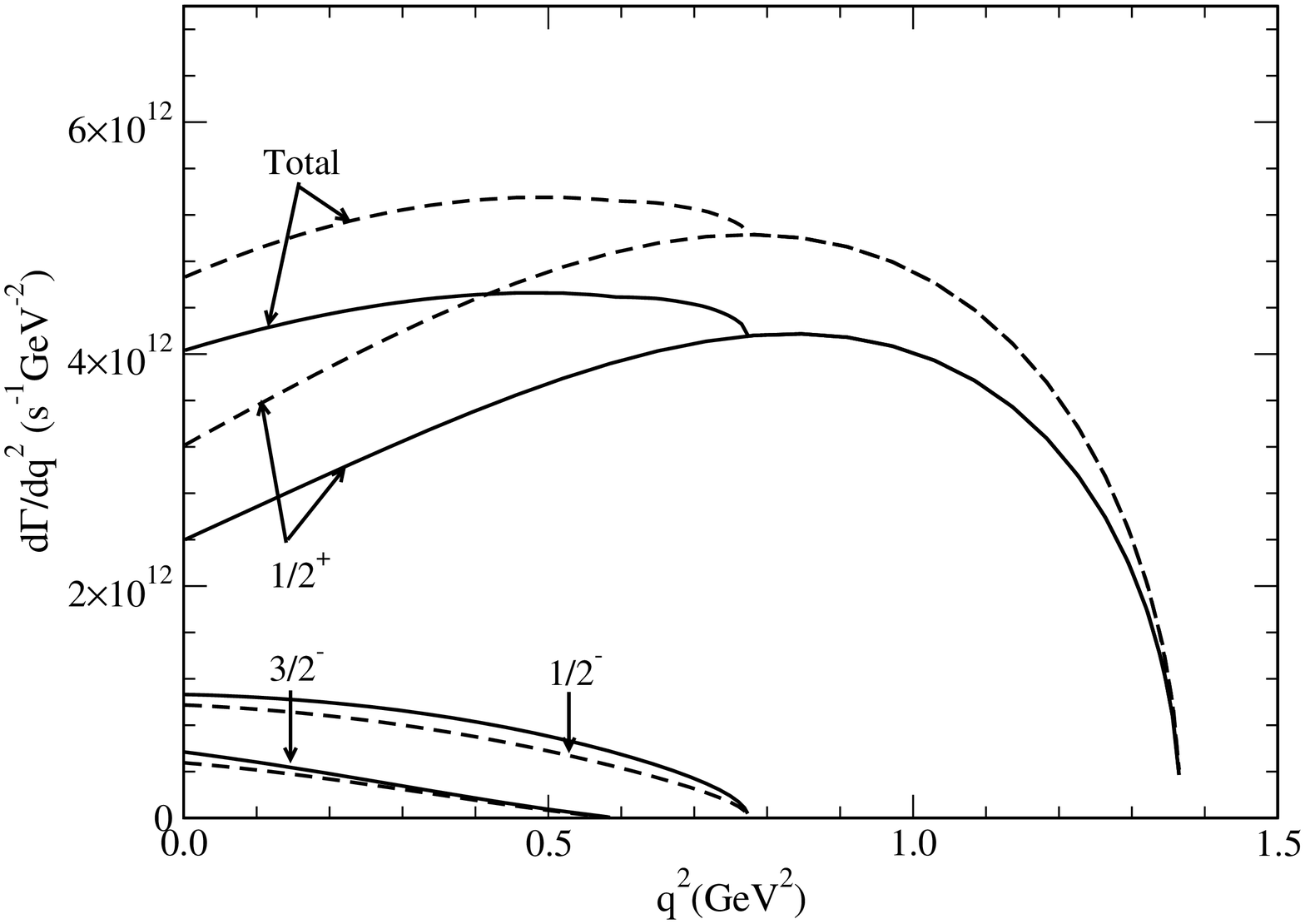,width=10cm}\hspace*{-0.25in}
\epsfig{file=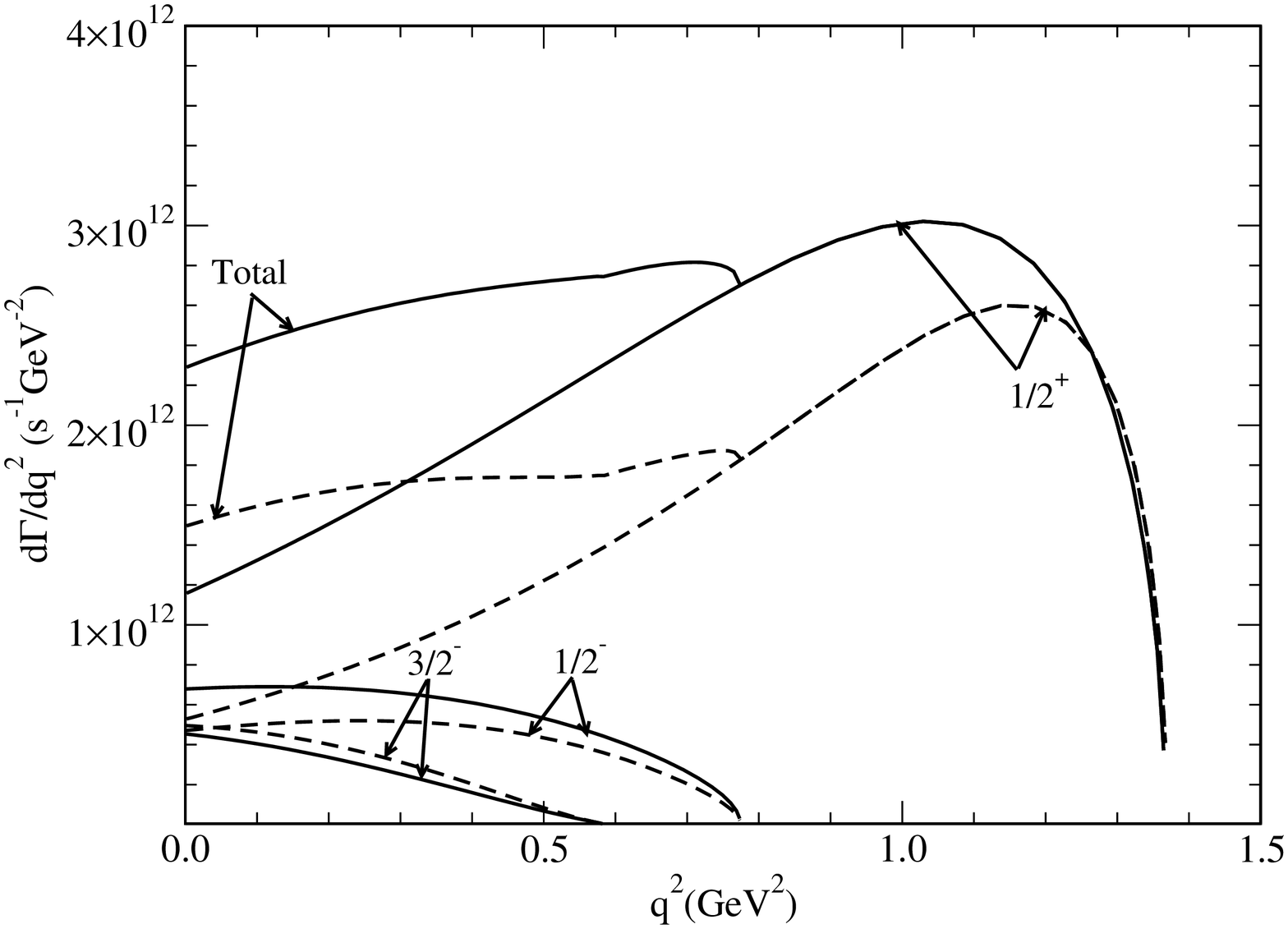,width=10cm}}
\caption{The differential decay rates for different $\Lambda_c\to\Lambda^{(*)}$
transitions, in the different models that we use. The curves on the
left arise from the two versions of the harmonic oscillator model,
while those on the right are from the Sturmian models. The curves are
for exclusive final states with $J^P=1/2^+$, $1/2^-$ and $3/2^-$. Also
shown are the differential decay rates obtained by adding the
exclusive modes described (labeled as `total'). In each panel, the
solid curves arise from the semirelativistic versions of the models,
while the dashed curves arise from the nonrelativistic
versions.\label{decayrate1}}
\end{figure}

The differential decay rates, $d\Gamma/dq^2$, that we obtain in the four models
are shown in Figure~\ref{decayrate1}. For these rates, we use $|V_{cs}|=0.974$.
In these figures, we show the differential
rates for decays to the elastic channel, as well as  for two orbital
excitations, the states with $J^P=1/2^-$ and $3/2^-$. We have also examined the
differential decay rates to the $3/2^+$ and $5/2^+$ orbitally excited states,
as well as to the $1/2^+$ radially excited state. With the exception of the
latter, we find these rates to be significantly smaller than those shown in this
figure.

As expected from the plots for the form factors, the differential decay rates
that arise from the Sturmian wave functions for the ground state show a larger
variation over the allowed $q^2$ range. We also point out that the most 
noticeable difference between the NR and SR versions of a particular model is
seen in the differential rate for the elastic decay. 

The integrated decay rate for the different final states in the
different models are shown in Table~\ref{ratelc}. As anticipated
above, the total semileptonic decay rates that we obtain in the
harmonic oscillator models are significantly larger than those
obtained in the Sturmian models. This effect is largest in the elastic
decays, where the HO models predict decay rates that are more than
twice as large as the ST models. We note that the elastic rates
predicted by the ST models are much closer to the experimentally
reported rate~\cite{CLEO} than those predicted by the HO models.
\begin{center}
\begin{table}
\caption{Integrated decay rates for $\Lambda_c\to\Lambda^{(*)}$ in
units of $10^{11}s^{-1}$, for different $\Lambda$ states in the four
models we consider. The last row shows the `elastic fraction' obtained in our
model, where the decays shown in
the table are assumed to saturate the semileptonic decays.
\label{ratelc}}
\begin{tabular}{|l|ll|ll|l|}
\hline
Spin & $\Gamma$(HONR) & $\Gamma$(HOSR) & $\Gamma$(STNR) & $\Gamma$(STSR) &
Expt.~\protect{\cite{CLEO}}\\ \hline
$1/2^+$ & $2.10$ &  $2.36$ & 0.79  & 1.11 & 1.05$\pm$ 0.35\\ \hline
$1/2^-$ & $0.19$ &  $0.29$ & 0.12 & 0.15 & -\\ \hline
$3/2^-$ & $0.05$ &  $0.06$ &  0.06& 0.05 & -\\ \hline
$1/2^+_1$ & $0.02$ &  $0.02$ &  $<$0.01 & $<$0.01 & -\\ \hline
total & 2.36& 2.73 & 0.97 & 1.31 & - \\ \hline
$\Gamma_\Lambda/\Gamma_{\rm total}$ & 0.89 & 0.86 & 0.81 & 0.85 & 1.0
(assumed) \\ \hline
\end{tabular}
\end{table}
\end{center}
From Table~\ref{ratelc}, it is clear that, while the elastic channel dominates 
the decay rate of the $\Lambda_c$,
it does not saturate the decay. In each model, we find that the decay rate to
the $1/2^-$ state is roughly one tenth of the elastic decay rate, while
the decay rate to the $3/2^-$ state is about five percent of the elastic. 
Decays to these two excited states account for about
15\% of the total decays of the $\Lambda_c$, assuming that decays to other
excited states are negligible. It is also interesting to note that the ratio 
$\Gamma_\Lambda/\Gamma_{\rm total}$ is almost independent of the model that we 
use, even
though the absolute rates are very different in the different models.

The assumption that the channels we explore saturate the resonant
decays of the $\Lambda_c$ is certainly consistent with the results we
have obtained with the other states that we consider. First we point
out that phase space limits how many excited $\Lambda$ states can be
considered, and the higher the excitation, the more limited the phase
space available for producing such a state. For some final states for
which there might be sufficient phase space to allow the decay, the
spin-space structure of the state allows little overlap with the
initial baryon, and configuration mixing that could involve components
with larger overlap with the initial baryon is very small. In
addition, angular momentum factors (in orbitally excited states) lead
to suppression of the decay rate.

We can compare our predictions for decays to the excited $\Lambda$
states with the assumption made by the CLEO Collaboration~\cite{CLEO},
that the elastic channel saturates the semileptonic decays of the
$\Lambda_c$. In our models, we find that between 11\% and 19\% of the
$\Lambda_c$ semileptonic decays are to excited states. In addition,
our branching fraction (of 81\% to 89\%) to the ground state $\Lambda$
must represent an upper limit, as we have not included any
non-resonant production of multi-particle final states. It appears
difficult to understand the lack of evidence for any decays to excited
states in Ref.~\cite{CLEO}. This article reports no signal for decays
of the kind $\Lambda_c\to\Lambda Xe^+\nu$, and this is taken as
evidence of saturation. However, the excited $\Lambda$ states that we
consider do not decay to $\Lambda\pi$, the most obvious decay mode to
search for, as this decay is isospin violating. They will
predominantly decay to $\Sigma\pi$ final states.  In fact, the $1/2^-$
state, the $\Lambda(1405)$, has a 100\% branching fraction to
$\Sigma\pi$, while the $\Lambda(1520)$, the $3/2^-$ state, has roughly
equal dominant branching ratios to $\Sigma\pi$ and $NK$, with only
about ten percent going into $\Lambda\pi\pi$. Thus, our suggestion is
that CLEO should investigate final states like $\Sigma\pi\ell\nu$ and
$NK\ell\nu$, and not states like $\Lambda\pi\pi\ell\nu$.

The results discussed above are obtained using the assumption that the lightest
of the $J^P=1/2^-$ $\Lambda$ states, identified with the $S_{01}$ state $\Lambda
(1405)$ found in analyses of scattering data, is a three-quark state. There are
a number of other descriptions of this state in the literature, such as a
dynamically generated bound state~\cite{oset}, and a multi-quark state
\cite{choe}. If the CLEO Collaboration (or other groups) search for decays of
the $\Lambda_c$ to excited $\Lambda$ states, especially the $\Lambda(1405)$,
and find no such decays, this would be a strong hint that this state is not a 
simple three quark state, as we have assumed.

Our estimate of the fraction of $\Lambda_c$ decays to excited states
has important consequences for the absolute normalization of the
branching fractions to the more than sixty observed final states in
$\Lambda_c^+$ decay. Most of these branching fractions are measured
relative to the decay mode $\Lambda_c^+\to pK^-\pi^+$, and the
absolute branching fraction of this mode cannot be extracted from data
without introducing model dependence. One of the two important
techniques for this extraction is based on
measurements~\cite{Argus96,CLEO91} of the cross section for
$\Lambda_c^+X$ production in $e^+e^-$ annihilation, with the
subsequent semileptonic decay $\Lambda_c^+\to \Lambda
\ell^+\nu_\ell$. The extraction relies on the assumption that the
fraction $f$ of decays $\Lambda_c^+\to X_s \ell^+\nu_\ell$ that have
$X_s$ as the ground state $\Lambda$ is unity (the elastic channel
saturates the semileptonic decays), with a significant
uncertainty. Our calculated value $f=0.85$, with an error of $0.04$
estimated by evaluating $f$ in four different models, changes the
central value of this parameter and may allow a reduction in the
assumed error from model dependence in the extracted absolute
branching fractions.

\subsubsection{$\Lambda_b \to \Lambda_c^{(*)}$}

In Table~\ref{formfactors3} we show the values of the form factors at
the non-recoil point, for the decays $\Lambda_b\to\Lambda_c^{(*)}$,
where this notation means that the $\Lambda_c$ may be in an excited
state. The results from all four models are shown, along with the
results from a lattice study~\cite{Richards}. The lattice results are
actually given as multiples of $\xi(w)$, evaluated at the non-recoil
point, and Ref.~\cite{Richards} reports a number of different values
for $\xi(w)$. In the `physical' limit, values
$\xi^{(A)}(1)=1.03^{+0.18}_{-0.19}$ and $\xi^{(V)}(1)=0.87\pm 0.22$
are quoted, where the two extractions are from the axial and vector
currents, respectively.  The results we obtain for the elastic decays
are consistent with the predictions of HQET as estimated by Scora
\cite{scora}, as well as with these lattice simulations.
\begin{center}
\begin{table}
\caption{Form factors of $\Lambda_b \to \Lambda_c^{(*)}$, calculated
at the non-recoil point, in the four models we use. Also shown are the lattice
estimates for the elastic form factors, taken from~\cite{Richards}. The lattice
numbers are in fact multiples of their estimate of $\xi(w)$, for which they
explore a number of scenarios. 
\label{formfactors3}}
\begin{tabular}{|l|l|lllclllc|}
\hline
$J^P$ & model & $F_1$\,\,\, & $F_2$\,\,\,  & $F_3$ \,\,\, & $F_4$ \,\,\, & 
$G_1$ \,\,\,  & $G_2$ \,\,\, & $G_3$ \,\,\, & $G_4$\,\,\,  \\ \hline
$1/2^+$ &HONR & 1.27 & -0.20 & -0.08 & - & 0.99 & -0.20 & 0.08 & - \\
$1/2^+$ &HOSR & 1.24 & -0.18 & -0.08 & - & 0.97 & -0.18 & 0.08 & -  \\
$1/2^+$ & STNR &  1.28 & -0.26 & -0.04 & - & 0.98 & -0.26 & 0.04 & - \\
$1/2^+$ & STSR &   1.20 & -0.22 & -0.03 & - & 0.92 & -0.22 & 0.03 & - \\\hline
$1/2^+$ & Lattice & 1.28$\pm$0.06 & -0.19$\pm$0.04 & -0.06$^{+0.02}_{-0.01}$ 
&- &  0.99 & -0.24$^{+0.05}_{-0.04}$ & 0.09$\pm 0.02$ & -\\ \hline
$1/2^-$ &HONR & 0.12 & -1.20 & 0.11 & - & 1.21 & -1.05 & 0.03 & -\\
$1/2^-$ &HOSR & 0.15 & -0.95 & 0.09 & - & 1.01 & -0.82 & 0.04 & - \\
$1/2^-$ & STNR & 0.10 & -1.63 & 0.14 &- &  1.61 & -1.50 & 0.03 & - \\
$1/2^-$ & STSR  & 0.11 & -1.21 & 0.10 &- &  1.24 & -1.12 & 0.03 & - \\ \hline
$3/2^-$ & HONR & -1.33 & 0.17 & 0.13 & -0.06&-1.03 & 0.17 & -0.13 & 0.06 \\
$3/2^-$ &HOSR &-1.13 & 0.15 & 0.12 & -0.05&-0.87 & 0.15 & -0.12 & 0.05\\
$3/2^-$ & STNR & -1.75 & 0.25 & 0.15 & -0.05 & -1.36 & 0.25 & -0.22 & 0.05 \\
$3/2^-$ &STSR & -1.31 & 0.16 & 0.11 & -0.05 & -1.04 & 0.16 & -0.18 & 0.05\\
\hline
\end{tabular}
\end{table}
\end{center}

\begin{figure}[h]
\centerline{\epsfig{file=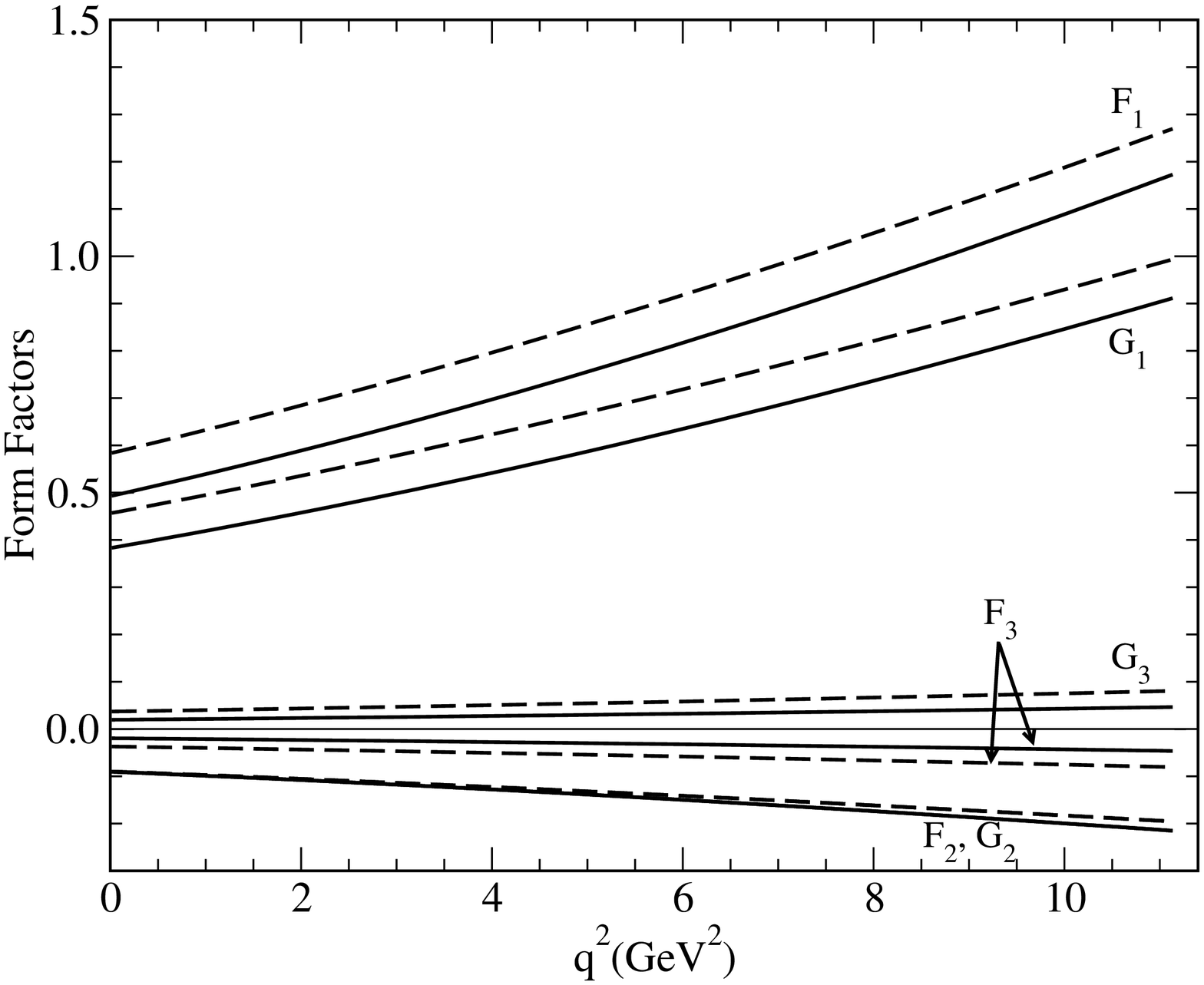,width=10cm}
\hspace*{-0.5in}\epsfig{file=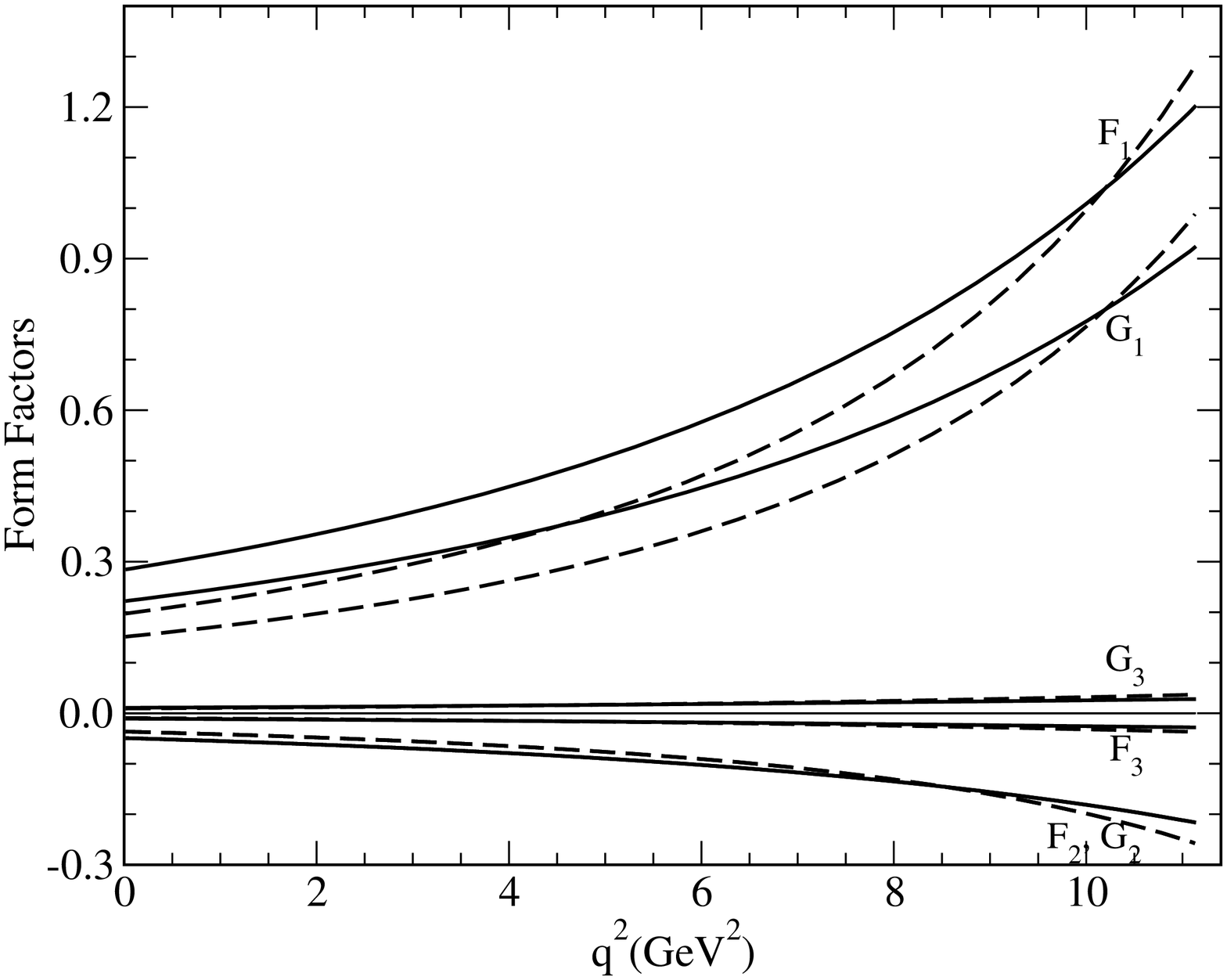,width=10cm}}
\caption{Form factors for $\Lambda_b\to\Lambda_c(1/2^+)$ obtained using harmonic
oscillator wave functions (left panel, HOSR and HONR models) and Sturmian wave 
functions (right panel, STSR and STSR models). In each panel, the solid curves 
arise from the semirelativistic version of the model, while the dashed curves 
arise from the nonrelativistic version. Note that
$F_2$ is indistinguishable from $G_2$ in all cases.
\label{formfactor2}}
\end{figure}
Figure~\ref{formfactor2} shows the $q^2$ dependence of the form
factors for the elastic decay of the $\Lambda_b$, calculated in the HONR and
HOSR models on the left, and in the STSR and STNR models on the
right. In each panel, the solid curves arise from the SR version of
the model, while the dashed curves are from the NR version. As we
noted in the case of the $\Lambda_c\to\Lambda(1/2^+)$, the form
factors obtained in the Sturmian basis have significantly larger
slopes than the corresponding form factors calculated in the harmonic
oscillator basis, at the non-recoil point.

In terms of the Isgur-Wise function $\xi(w)$ for the elastic decay of the
$\Lambda_b$, the form factor $F_1$ is
\beq
F_1=\left[1+\bar\Lambda\left(\frac{1}{2m_c}+\frac{1}{2m_b}\right)\right]\xi(w),
\eeq
where $\bar\Lambda=m_{\Lambda_b}-m_b=m_{\Lambda_c}-m_c$ at leading order in the
heavy quark expansion. From the forms given in Appendix \ref{formfactors}, and with the
identification $\bar\Lambda\approx 2m_\sigma$, we can extract
\beq
\xi(w)\approx\exp\left( -\frac{3 m^2_\sigma}{2m^2_{\Lambda_c}}\frac{p^2}{\alpha^2}
\right)
\eeq
in the harmonic oscillator basis, or 
\beq
\xi(w)\approx\frac{1}{\left[1 + \frac{3m^2_\sigma}{2m^2_{\Lambda_c}}
\frac{p^2}{\beta^2}\right]^2}
\eeq
in the Sturmian basis (assuming single-component wave functions), and we have assumed
that $\alpha_\lambda=\alpha_{\lambda'}\equiv\alpha$, 
$\beta_\lambda=\beta_{\lambda'}\equiv\beta$ in the heavy quark limit.
Writing
\beq
p^2=m_{\Lambda_c}^2(w^2-1)\approx 2m_{\Lambda_c}^2(w-1),
\eeq
the above expressions become
\beq
\xi(w)\approx\exp\left( -\frac{3 m^2_\sigma}{\alpha^2}(w-1)
\right)
\eeq
in the harmonic oscillator basis, or 
\beq
\xi(w)\approx\frac{1}{\left[1 + \frac{3m^2_\sigma}{\beta^2}
(w-1)\right]^2}
\eeq
in the Sturmian basis. 

\begin{figure}[h]
\centerline{\epsfig{file=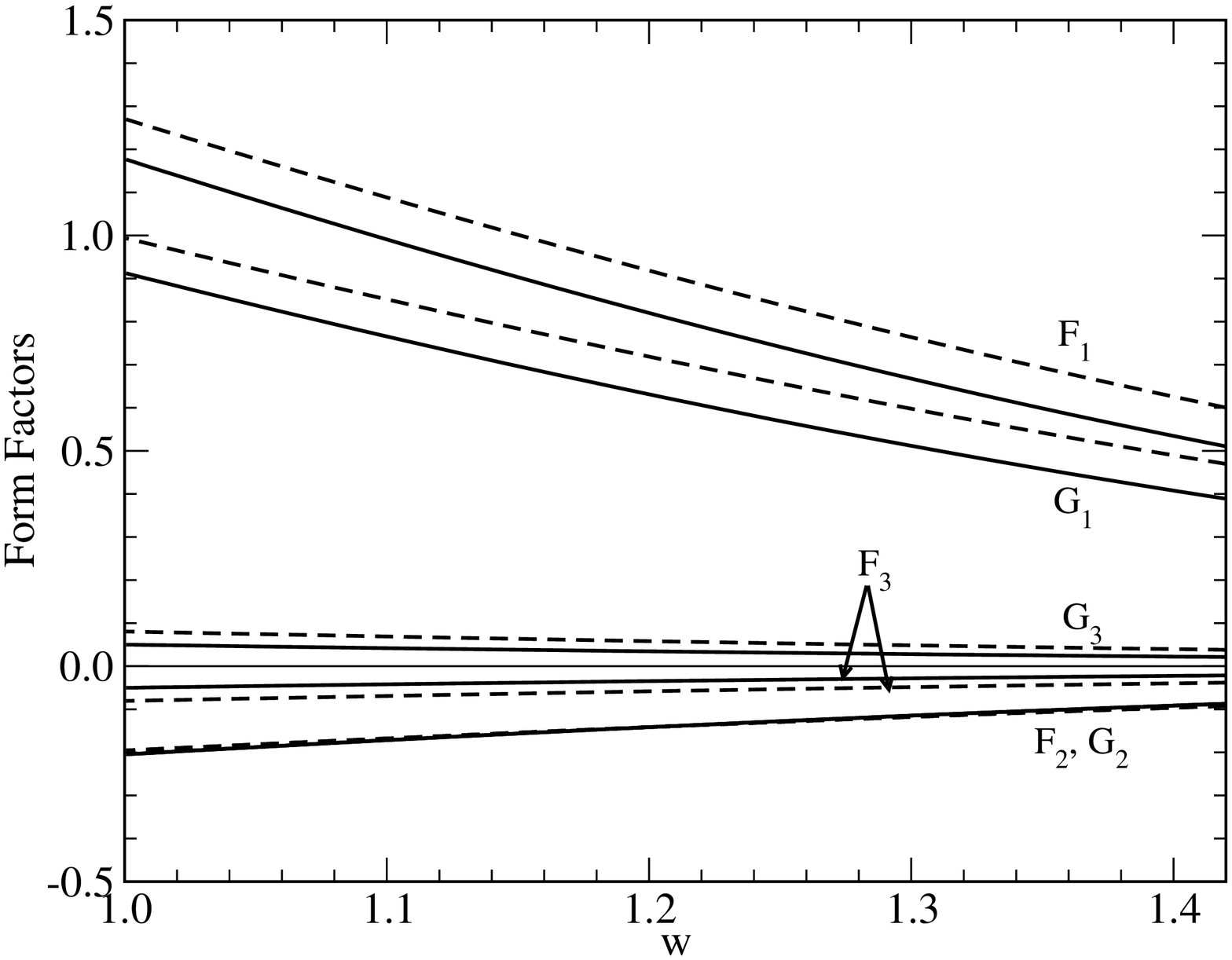,width=10cm}
\hspace*{-0.5in}\epsfig{file=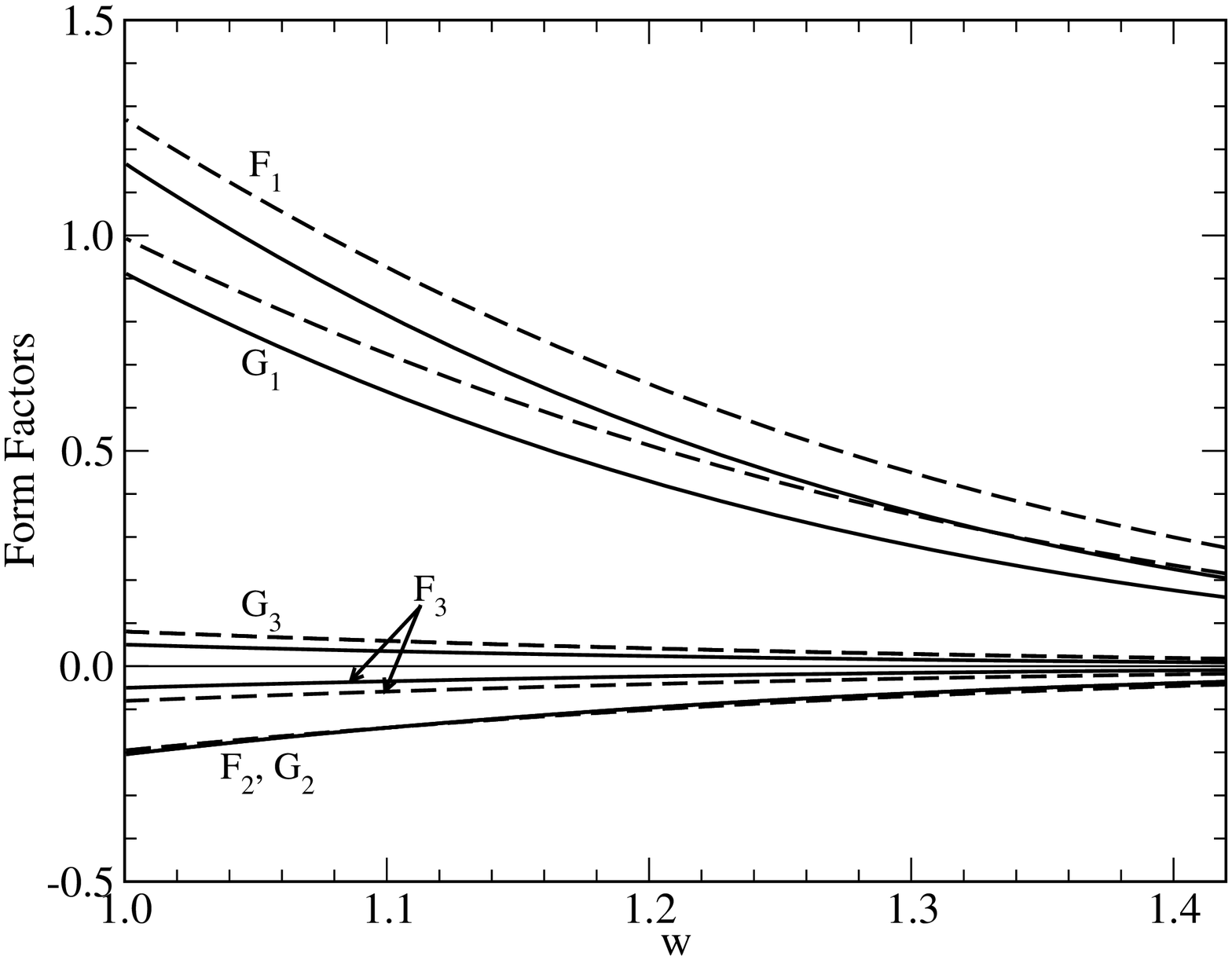,width=10cm}}
\centerline{\epsfig{file=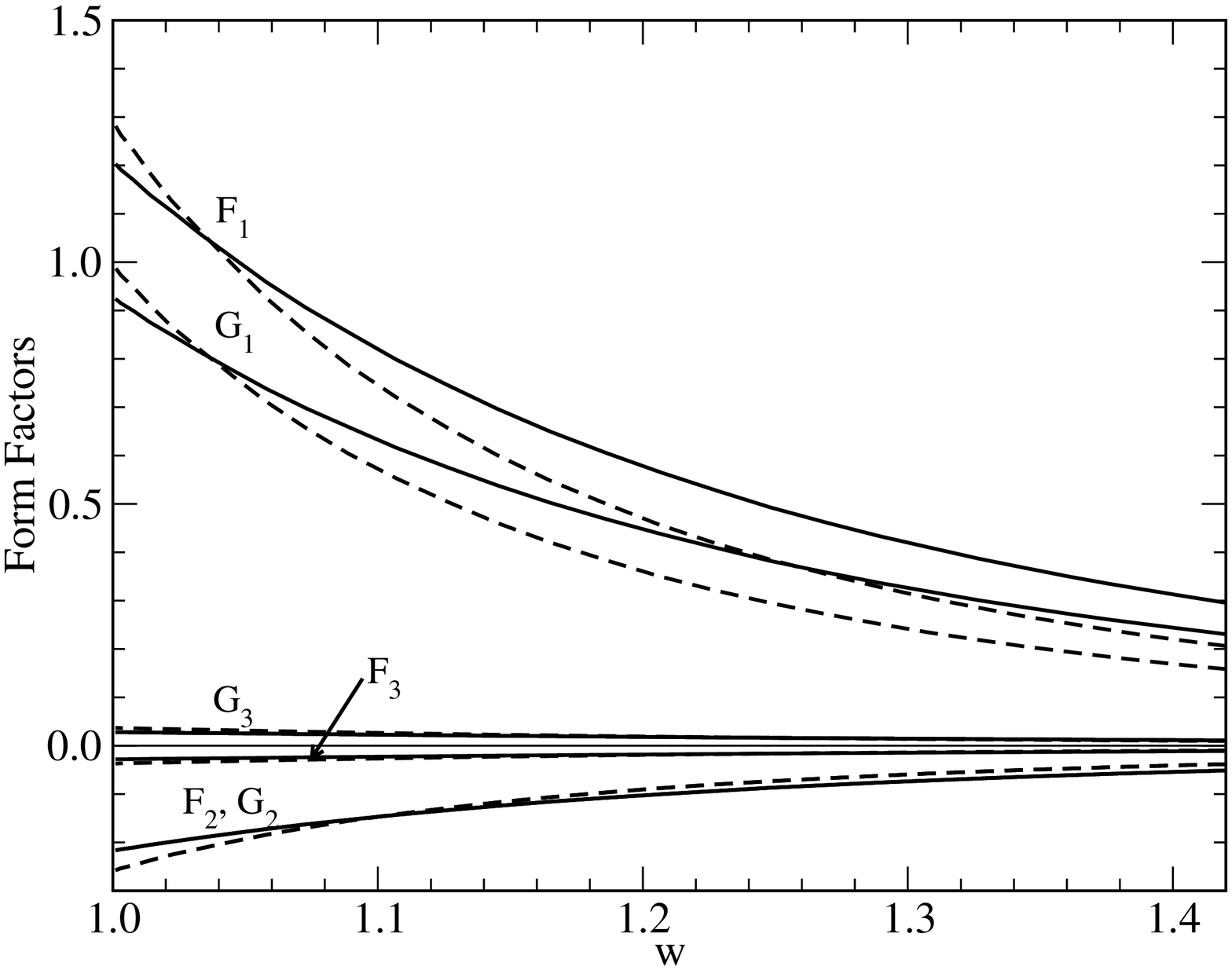,width=10cm}}
\caption{The elastic form factors for the decay of the $\Lambda_b$ as functions
of $w$. The upper two graphs arise from the harmonic oscillator model,
while the lower graph is from the Sturmian version of our model.
Among the upper graphs, the panel on the left shows the form factors
obtained in this work, while those on the right incorporate a
`relativistic' factor in the exponential (see text). In each panel,
the solid curves arise from the semirelativistic version of the model.
\label{kappaeffect}}
\end{figure}

The Isgur-Wise function may be expanded as
\beq
\xi(w)=1-\rho^2(w-1)+\frac{\sigma^2}{2}(w-1)^2+\dots,
\eeq
where the slope of the form factor at the non-recoil point has been
denoted $\rho^2$, and the curvature is denoted $\sigma^2$. Rigorous
bounds have been placed on the values of both the slope and curvature
parameters for meson decays, and some models have difficulty in
satisfying those bounds. In particular, in the model of
ISGW~\cite{ISGW}, a factor $\kappa$ was introduced by hand (see the
discussion between Eqs.~(B2) and~(B3) of Ref.~\cite{ISGW}) to modify
the $q^2$ dependence of the form factors. In our model, the equivalent
procedure would be to change $I_H$ in Eq. (\ref{elasticff}) from
\begin{eqnarray}
I_H =\left(\frac{\alpha_\lambda^{3/2}\alpha_{\lambda'}^{3/2}}{\alpha_{\lambda\lambda'}^{3}}
\right)\exp\left( -\frac{3 m^2_\sigma}{2m^2_{\Lambda_q}}\frac{p^2}{\alpha_{\lambda\lambda'}^2}
\right)\nonumber
\end{eqnarray}
as calculated to
\begin{eqnarray}
I_H =\left(\frac{\alpha_\lambda^{3/2}\alpha_{\lambda'}^{3/2}}{\alpha_{\lambda\lambda'}^{3}}
\right)\exp\left( -\frac{3 m^2_\sigma}{2m^2_{\Lambda_q}}\frac{p^2}{\kappa^2\alpha_{\lambda\lambda'}^2}
\right).\nonumber
\end{eqnarray}
The argument used by ISGW was that this factor of $\kappa$ would take
into account `relativistic effects'. The effect of this change is
shown in Figure~\ref{kappaeffect}, where the form factors for
$\Lambda_b\to\Lambda_c$ are plotted as functions of $w=v\cdot
v^\prime$, for the two harmonic oscillator models (upper graphs). For
comparison, the lower graph shows form factors obtained in the
Sturmian basis, also as functions of $w$. The graph on the upper left
shows our calculated form factors, while that on the upper right shows
form factors including the factor of $\kappa$.

Table~\ref{slopetable} lists the slope of the Isgur-Wise
function that we have extracted, at the non-recoil point, in both the
harmonic oscillator and Sturmian models, as well as in the
`relativistically modified' harmonic oscillator model (using the
factor $\kappa$). The slopes of the form factor near the non-recoil
point are larger in the Sturmian models than in the harmonic
oscillator models. This is easily understood by noting that the value
of $\rho^2$ is
\beq
\rho^2=3\frac{m_\sigma^2}{\alpha^2}
\eeq
in the HO models, and
\beq
\rho^2=6\frac{m_\sigma^2}{\beta^2}
\eeq
in the ST models. The extra factor of two in the latter case arises
because the form factors in the ST models have a dipole dependence on
$w$. A corresponding monopole form would give the same slope as the HO
models. Since the values of $m_\sigma$ are similar in the two sets of
models, and the values of $\alpha$ are not very different from the
values of $\beta$, the ST models will give slopes that are roughly
twice as large as the HO models. In the same way, it is easily shown
that the ST models lead to curvatures that are about six times as
large as those obtained in the HO models.
\begin{center}
\begin{table}[h]
\caption{Slope of the Isgur-Wise function, evaluated at the
non-recoil point, for the elastic decay of the $\Lambda_b$.
\label{slopetable}}
\begin{tabular}{|l|llllll|}
\hline
model & HONR & HOSR & HONR$\kappa$ & HOSR$\kappa$ & STNR & STSR \\\hline
$d\xi(w)/dw$ & -1.38 &-1.33 &-2.82 &-2.71 &-5.71 & -3.27\\ \hline
\end{tabular}
\end{table}
\end{center}

The $\kappa$-modified harmonic oscillator model leads to slopes that are
similar to those obtained in the Sturmian models, since the value chosen for
$\kappa$ was 0.7 (so that $1/\kappa^2\approx 2$). Relativistic effects do not
need to be invoked to obtain the large slopes obtained in the Sturmian models.
The differences in the slopes are simply artifacts of the expansion bases used
for the wave functions.

In the follow-up article to Ref.~\cite{ISGW}, Scora and
Isgur~\cite{isgw2} rewrite the quark model form factors, explicitly
replacing the exponential factor that arises with the harmonic
oscillator wave functions. The change they make is
\beq \label{isgwfactor}
\exp\left\{-\frac{1}{6}r^2_{\rm wf}[(m_B-m_D)^2-q^2]\right\}\longrightarrow
\frac{1}{\left\{1+\frac{1}{6N}r^2[(m_B-m_D)^2-q^2]\right\}^N},
\eeq
where $r^2_{\rm wf}$ is the value obtained from the harmonic oscillator wave
functions, and
\beq
r^2=\frac{3}{4m_Qm_q}+ r^2_{\rm wf}+r^2_{\rm QCD},
\eeq
where the last term arises from matching of currents in HQET with
full QCD. In Eq.~(\ref{isgwfactor}), the integer $N=2+n+n'$, where $n$
and $n'$ are the harmonic oscillator principal quantum numbers for the
initial and final wave functions. The final forms that they used are
therefore very similar to the forms that we have obtained in the
Sturmian models.

The values we have obtained for the slope of the Isgur-Wise function
in our Sturmian models are significantly larger than the value
obtained recently by Huang {\it et al.}~\cite{huang} using a HQET
approach based on QCD sum rules: their value for $\rho^2$ is less than
1.5, similar to the values we obtain in the HO models. In a recent
analysis of the $\Lambda_b$ form factor measured in hadronic Z decays,
the DELPHI Collaboration~\cite{delphi} found
$\rho^2=2.03\pm 0.46$, where the error shown is statistical. They also
reported two sets of systematic errors, each comparable to the
statistical error.  This result means that for the Sturmian models, we
will obtain integrated decay rates that are significantly smaller than
the DELPHI rate. In the lattice study by Bowler {\it et
al.}~\cite{Richards}, the reported slopes is $1.1\pm 1.0$. A more
recent lattice study with ${\cal O}\left(a^2,\alpha_sa^2\right)$
improved lattices~\cite{gottlieb} does not quote values for the
slope. However, a conservative estimate from the
graphs they present gives values for $\rho^2$ that appear to be
consistent with the large values we obtain in the ST models.

Also of some interest is the curvature of the Isgur-Wise function,
denoted $\sigma^2$. In the HO models with no modifications, the
prediction is that $\sigma^2_{\rm HO}=(\rho^2_{\rm HO})^2$, while the
ST models give $\sigma^2_{\rm ST}=3(\rho^2_{\rm ST})^2/2$. Bounds on
the curvature of the Isgur-Wise function for meson decays have been
derived by Le Yaouanc, Oliver and Raynal~\cite{curvature}. To the best
of our knowledge, no such bounds have been derived for baryon
decays. However, the values of the curvature we obtain using both the
HO and ST models easily satisfy the known bounds for meson
decays. Note that the large slope {\it and} large curvature we obtain
suggest that the common procedure of parameterizing the Isgur-Wise
function only in terms of its slope parameter, can potentially lead to
significant errors in the extraction of CKM matrix elements.

 \begin{figure}[h]
\centerline{\epsfig{file=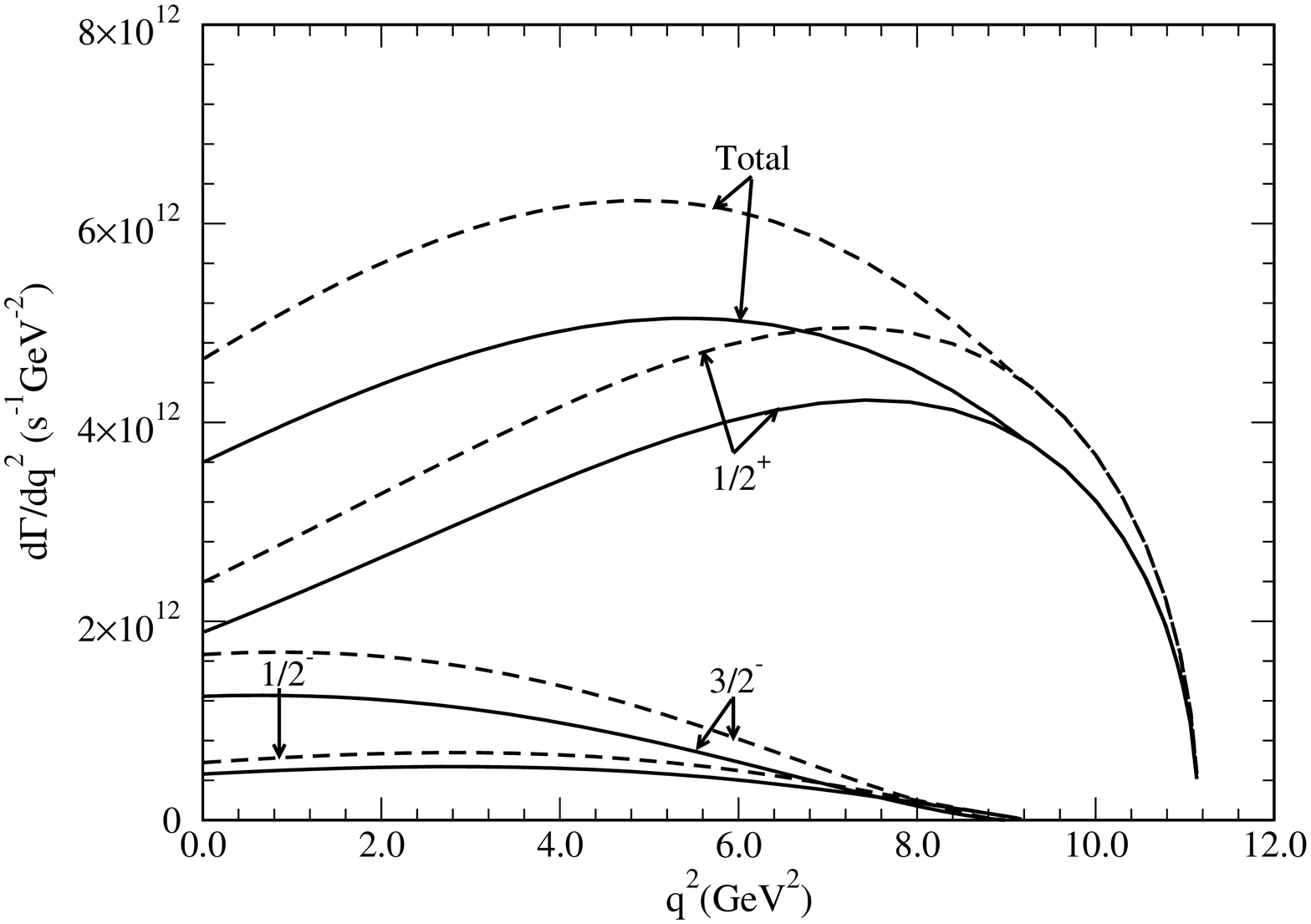,width=10cm}\hspace{-0.3in}
\epsfig{file=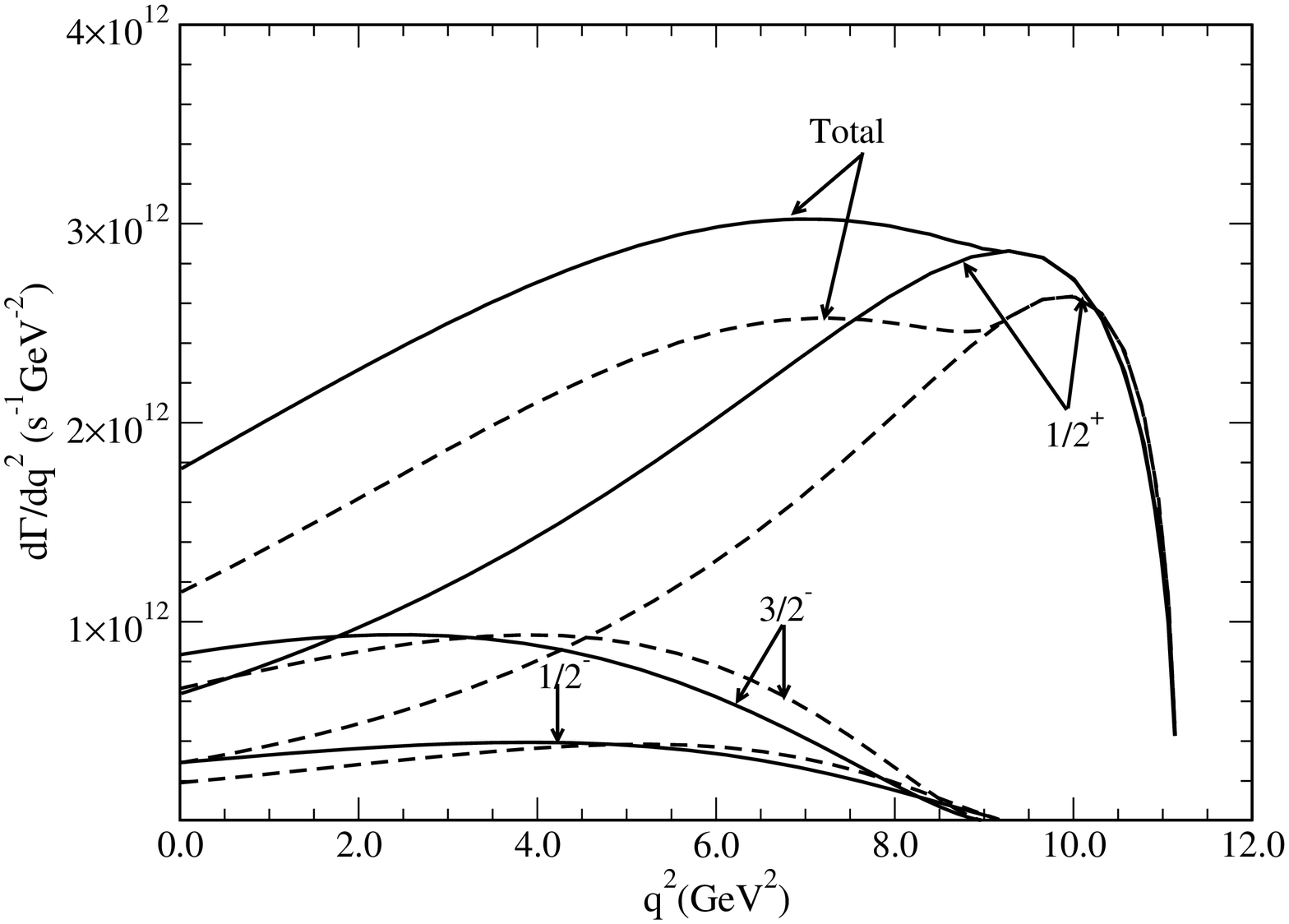,width=10cm}}
\vspace*{-0.22in}
\centerline{\epsfig{file=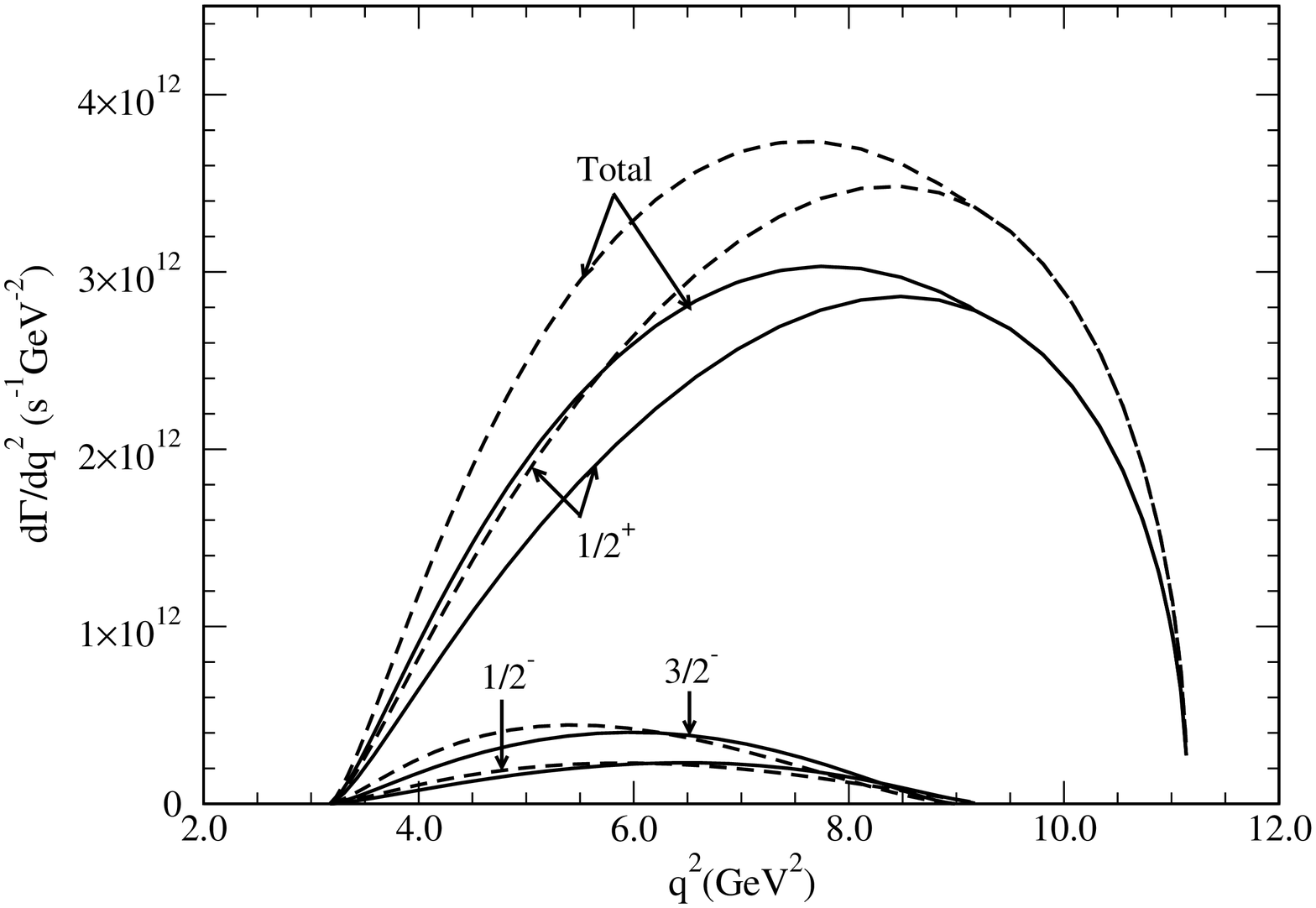,width=10cm}\hspace{-0.3in}
\epsfig{file=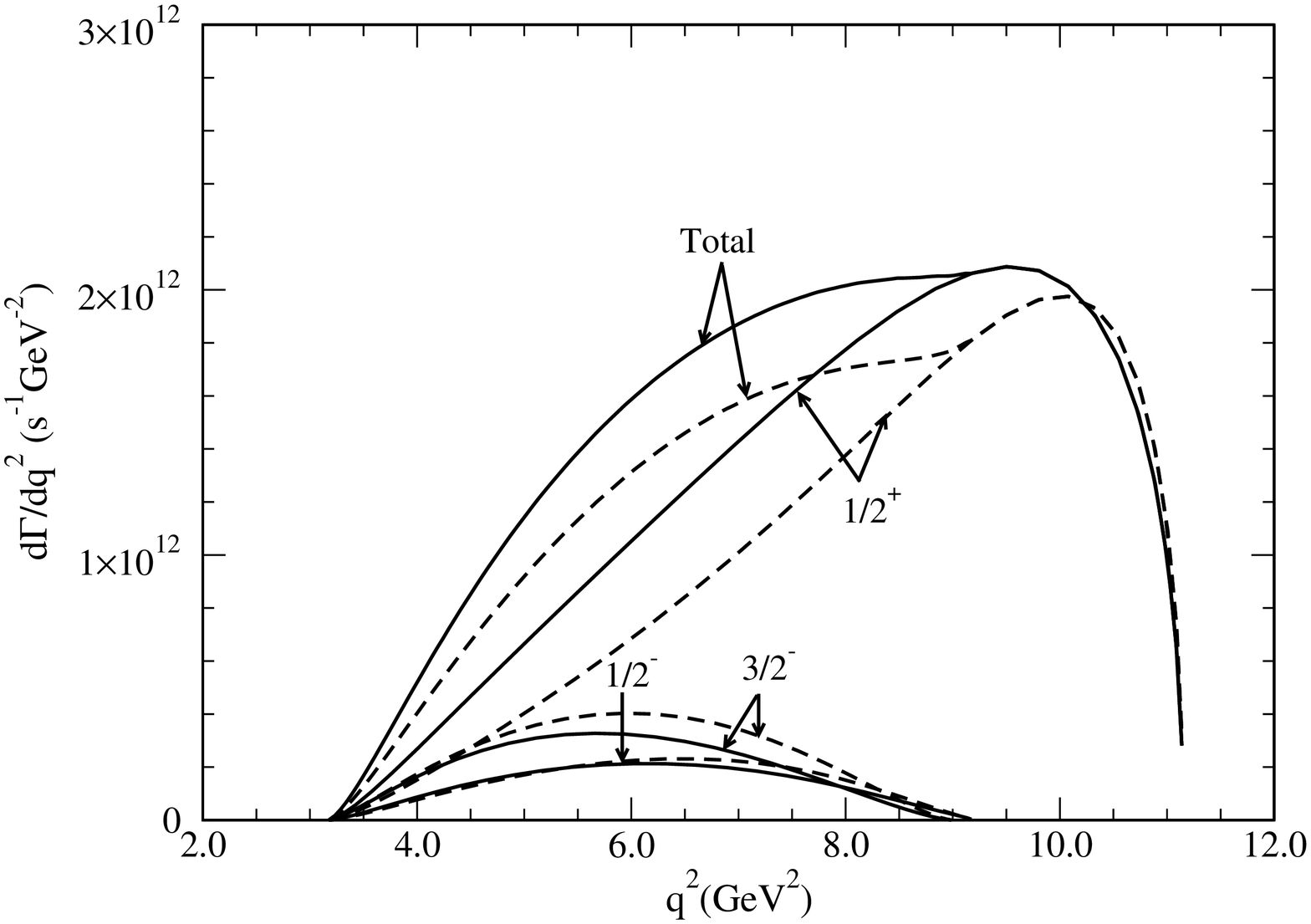,width=10cm}}
\caption{The differential decay rates for different 
$\Lambda_b\to\Lambda_c^{(*)}$ transitions, in the various models
that we use. The curves on the left arise from the two versions of the
harmonic oscillator model, while those on the right are from the
Sturmian models. The upper panels are for
$\Lambda_b\to\Lambda_c^{(*)}\ell\bar\nu_\ell$, where $\ell$ is $e^-$ or
$\mu^-$. The lower panels are for
$\Lambda_b\to\Lambda_c^{(*)}\tau\bar\nu_\tau$. The curves are for final
states with $J^P=1/2^+$, $1/2^-$ and $3/2^-$.\label{decayrate3}}
\end{figure}
The differential decay rates $d\Gamma/dq^2$ that we obtain in the four
models are shown in Figure~\ref{decayrate3} (assuming $|V_{cb}|=.041$). 
In these plots, we show
the differential rates for the elastic channel, for the radially excited $1/2^+_1$ 
state, as well as for decays to 
two orbital excitations, the states
with $J^P=1/2^-$ and $3/2^-$. We have also examined the
decay rates to the $3/2^+$, $5/2^+$ states, and found them to be smaller
than those shown in this figure, contributing of the order of one or two percent to
the total rate.

The integrated decay rate for the different final states in the different models
are shown in Table~\ref{ratelb}. As anticipated above, the total
semileptonic decay rates
that we obtain in the harmonic oscillator models are significantly larger than
those obtained in the Sturmian models. This effect is largest in the elastic 
decays, where HO models predict decay rates that are more
than twice as large as the ST models. Note that, in all models, the decay rate to the 
$3/2^-$ state is roughly twice the decay rate to the $1/2^-$ state. In the heavy quark
limit, this ratio of decay rates is expected to be two, and results from arguments that
are similar to spin-counting arguments.

\begin{center}
\begin{table}[h]
\caption{Rates for $\Lambda_b\to \Lambda_c^{(*)}$ decays in units of 
$10^{10}s^{-1}$. The first five rows are for decays with a muon or
electron in the final state, while the last four rows are for decays
with a $\tau$ in the final state. The rows labeled `total' are
obtained by adding the exclusive decay rates shown in the table, while
the row with the branching fractions assumes that the exclusive
channels shown saturate the semileptonic decays of the $\Lambda_b$. The elastic
fraction reported by the DELPHI collaboration (fifth row, sixth column) is actually
$\frac{\Gamma(\Lambda_b\to\Lambda_c\ell\bar\nu_\ell)}
{\Gamma(\Lambda_b\to\Lambda_c\ell\bar\nu_\ell)+
\Gamma(\Lambda_b\to\Lambda_c\pi\pi\ell\bar\nu_\ell)}$. The errors on
both DELPHI results are statistical and systematic, respectively.
\label{ratelb}}
\begin{tabular}{|l|cccc|c|}
\hline
$J^P$ & $\Gamma$(HONR) & $\Gamma$(HOSR) & $\Gamma$(STNR) & $\Gamma$(STSR)&
$\Gamma_{\rm DELPHI}$  \\
\hline
$1/2^+$ & $4.60$ &  $5.39$  & 1.47  & 2.00 & $4.07^{+0.90+1.30}_{-0.65-0.98}$\\
 \hline
$1/2^-$ & $0.45$ &  $0.52$ & 0.26 &0.27 & - \\ \hline
$3/2^-$ & $0.95$ &  $0.91$  & 0.63 &0.61 & - \\ \hline
Total ($\Lambda_c^{(*)}\ell^-\bar\nu_\ell$) & 5.95 & 6.82 & 2.36 & 2.88 &-
\\ \hline
$\Gamma_{\Lambda_c}/\Gamma_{\rm total}$ & 0.76 & 0.79 & 0.62 & 0.69&
$0.47^{+0.10+0.07}_{-0.08-0.06}$ \\ \hline\hline
$1/2^+$ & $1.90$ &  $2.09$  & 0.82  & 1.00 &-\\ \hline
$1/2^-$ & $0.10$ &  $0.11$ & 0.08&0.07 &-\\ \hline
$3/2^-$ & $0.15$ &  $0.13$  & 0.14 &0.12  & -\\ \hline
Total ($\Lambda_c^{(*)}\tau^-\bar\nu_\tau$)& 2.15 & 2.33 & 1.04 & 1.19 &-\\
\hline
\end{tabular}
\end{table}
\end{center}
Table~\ref{ratelb} also shows that a significant fraction of the
semileptonic decay of the $\Lambda_b$ is inelastic. This is analogous
to what has been seen in $B$ semileptonic decays, where the elastic
channels account for no more than about 80\% of the total semileptonic
decay rate. For the $\Lambda_b$, our predicted ratios are similar,
ranging from 62\% to 77\% of the total semileptonic decay rate. We have estimated
the total semileptonic decay rate by assuming that the three exclusive
modes shown in Table~\ref{ratelb} saturate the semileptonic decays
(rates to other states that we have examined are significantly smaller
than those shown in the table). Using these numbers, we obtain
predictions for the total semileptonic decay rate of the $\Lambda_b$,
also shown in Table~\ref{ratelb}.

For comparison, the PDG~\cite{pdg} gives a rate of $7.486\pm
2.105\times 10^{10}s^{-1}$ for the inclusive semileptonic decay
$\Lambda_b\to\Lambda_c\ell\bar\nu +$ anything. This is significantly
larger than any of the total semileptonic widths we obtain, but the
authors of the PDG emphasize that this value results from assumptions
about the fragmentation of $b$ quarks into baryons, and `cannot be
thought of as measurements'~\cite{pdg}. The DELPHI value for the
elastic semileptonic decay rate is also shown in
Table~\ref{ratelb}. As anticipated, the rates we obtain in the
Sturmian models are significantly smaller than the DELPHI rate, while
those obtained in the harmonic oscillator models are consistent with
the DELPHI measurement.

The above examination of the decays of the $\Lambda_c$ found that the
Sturmian models provided rates that were consistent with the CLEO
measurements, while the harmonic oscillator models gave rates that
were twice as large. This suggested that the Sturmian models might be
more reliable. For the $\Lambda_b$ decays, we see that the harmonic
oscillator models provide rates that are more consistent with the
single measurement available to date. For the Sturmian models, the
predicted rates are about 2$\sigma$ away from the reported value, if
the systematic and statistical errors are treated in quadrature.

The DELPHI Collaboration also reported on the elastic fraction of the
semileptonic decays of the $\Lambda_b$. For the ratio
$\frac{\Gamma(\Lambda_b\to\Lambda_c\ell\bar\nu_\ell)}
{\Gamma(\Lambda_b\to\Lambda_c\ell\bar\nu_\ell)+
\Gamma(\Lambda_b\to\Lambda_c\pi\pi\ell\bar\nu_\ell)}$, they find a value of 
$0.47^{+0.10+0.07}_{-0.08-0.06}$, with no evidence for resonant
decays. This ratio is smaller than we predict, in all models. However,
our predictions must be thought of as upper limits for the elastic
fraction, as we do not include any non-resonant semileptonic
decays. We note that our predicted ratios are already somewhat smaller
than those reported in the decays of $B$ mesons, while the DELPHI
ratio is smaller still, suggesting that there are significant
differences between the semileptonic decays of the heavy baryons and
those of the heavy mesons. If the DELPHI results for both for the
elastic rate and the elastic fraction are not modified by future
experiments, this aspect of the physics of heavy hadrons will require
further scrutiny.

\subsubsection{$\Lambda_Q \to N^{(*)}$ Decay}

The decays of the $\Lambda_Q$ to final states consisting solely of
light quarks are interesting as they provide an alternate means of
extracting CKM matrix elements like $V_{ub}$. The expectation from
HQET is, modulo $1/m_Q$ effects, that the form factors that describe
the $\Lambda_c\to n$ semileptonic decays will be the same as those
describing the $\Lambda_b\to p$ semileptonic decays. To explore this,
we now examine the form factors for these two decays.

\begin{figure}[h]
\centerline{\epsfig{file=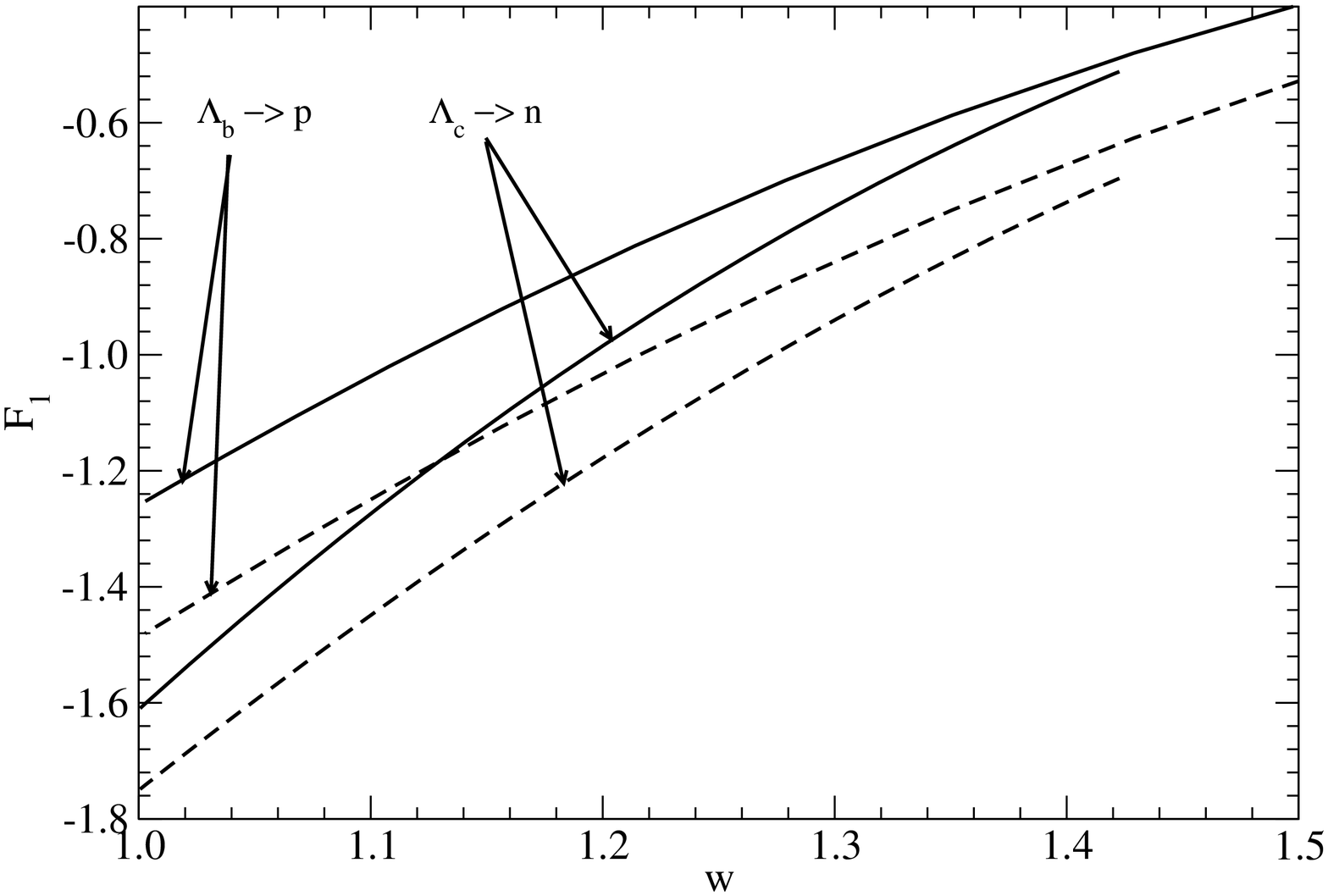,width=10cm}\hspace*{-0.4in}
\epsfig{file=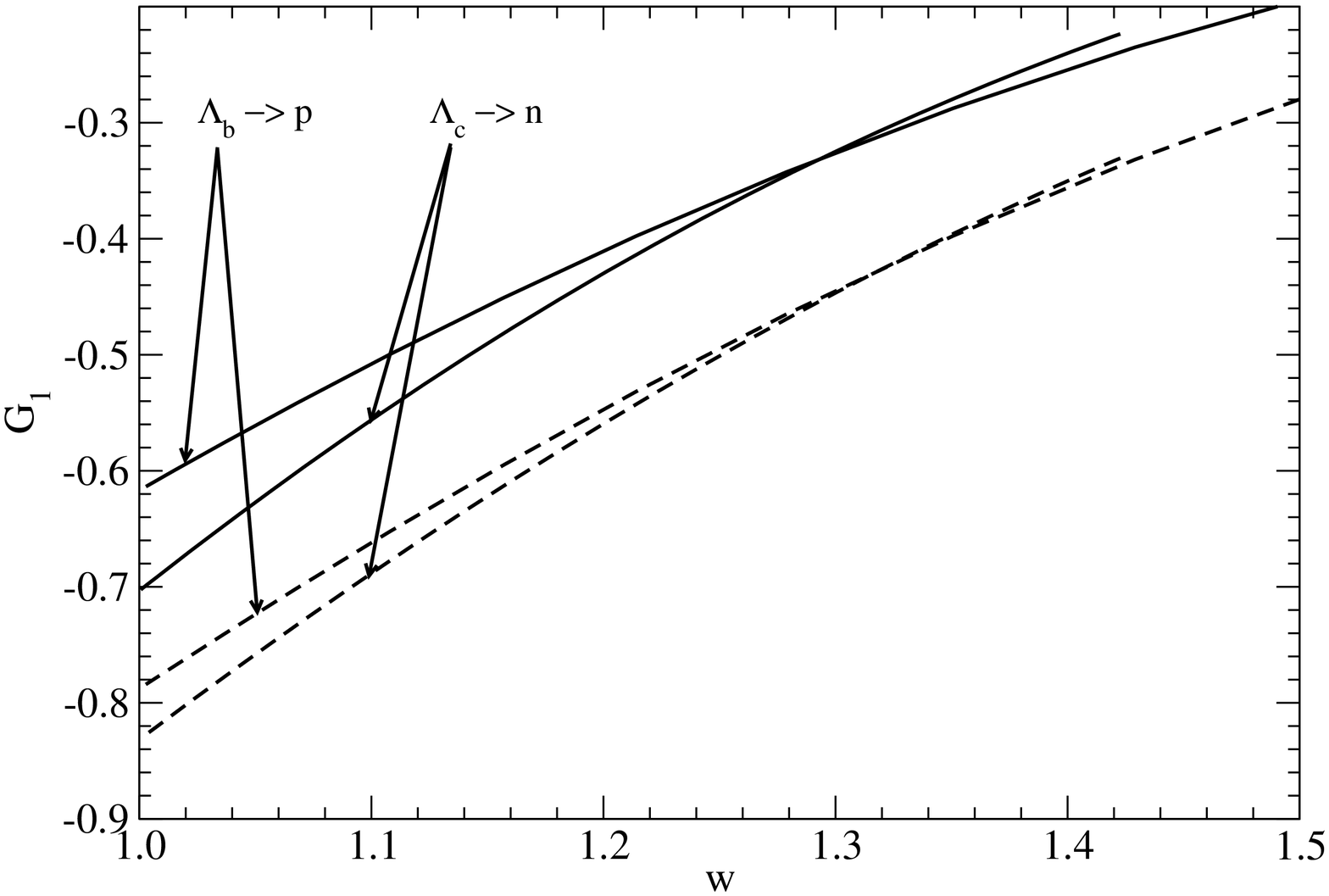,width=10cm}}
\centerline{\epsfig{file=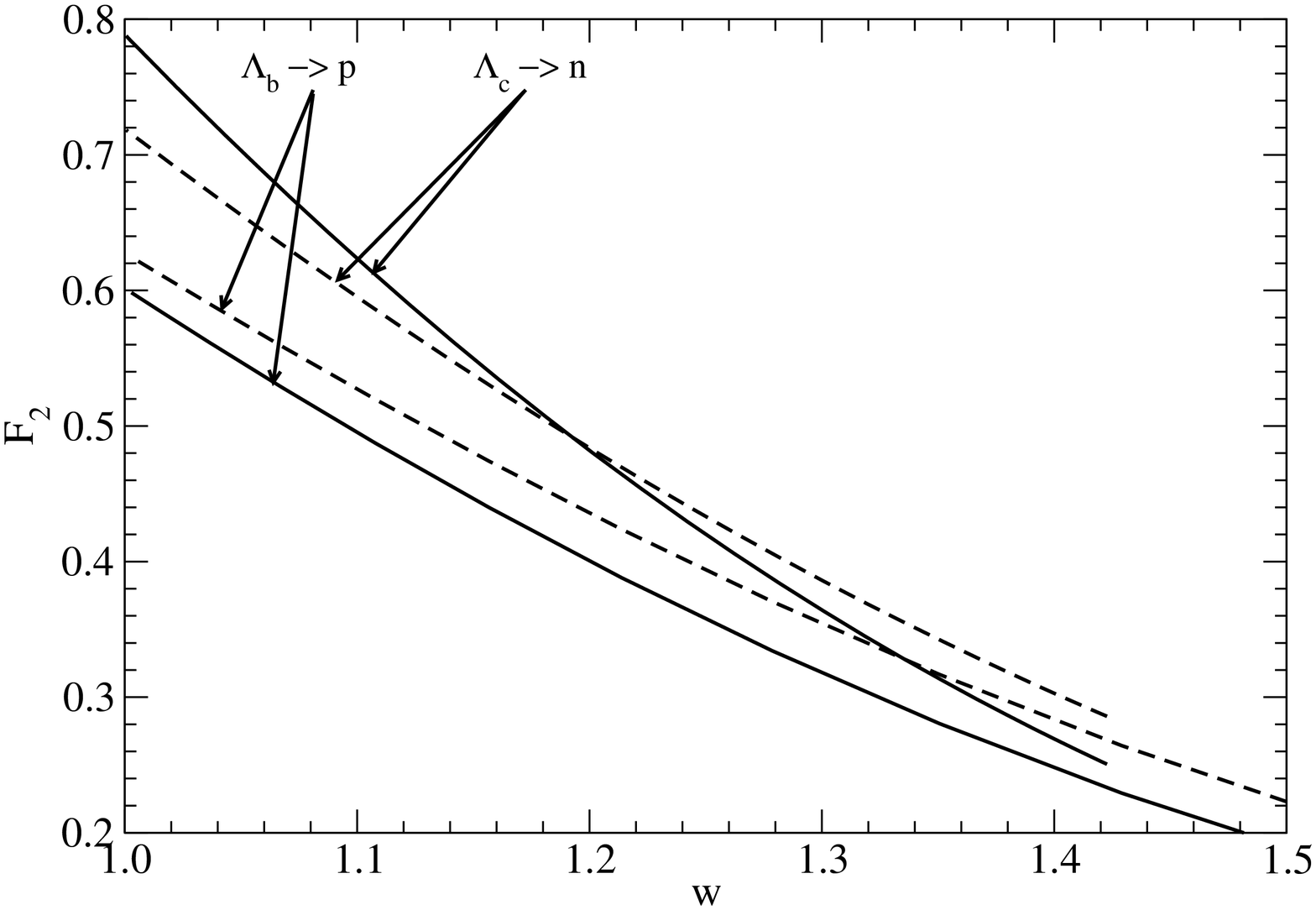,width=10cm}}
\caption{Form factors $\xi_1^{(V)}$ (top left), $\xi_1^{(A)}$ (top right) and
$\xi_2$ (bottom) for the transitions $\Lambda_c\to n$ and
$\Lambda_b\to p$. All curves are found using the harmonic oscillator
models, with the solid curves corresponding to HOSR, and the dashed
curves to HONR. The two plots for $\xi_1$ arise from the two ways of evaluating
this form factor, shown in Eq.~(\ref{xiforms}).\label{hqet}}
\end{figure}

In Figure~\ref{hqet} we show the form factors $\xi_1^{(V)}$,
$\xi_1^{(A)}$ and $\xi_2$ for the transitions $\Lambda_c\to n$ and
$\Lambda_b\to p$, obtained in the two harmonic oscillator models. The
two forms $\xi_1^{(V,A)}$ are found using the two sets of equations in
Eq.~(\ref{xiforms}). The value of $\xi_2$ is independent of which of
the two sets of equations we use, up to the order to which we
calculate the form factors. In both the nonrelativistic and
semirelativistic versions of the model, the two curves for
$\xi_1^{(A)}$ (top right plot in Fig.~\ref{hqet}) are very similar,
indicating that the HQET prediction, that this form factor should be
the same for both transitions, indeed holds up to small
corrections. For the semirelativistic version, the two curves are
closer than in the nonrelativistic case. The differences seen in the
curves for $\xi_2$, which are consistent with those in the curves for
$\xi_1^{(A)}$, arise mainly from the differences in the size
parameters ($\alpha_\rho$ and $\alpha_\lambda$) between the
$\Lambda_c$ and $\Lambda_b$ states in the models (see
Table~\ref{parameter2}).  The curves for $\xi_1^{(V)}$ (top left plot
in Fig.~\ref{hqet}) show the biggest differences in going from
$\Lambda_b\to p$ to $\Lambda_c\to n$, in both models. Here, the
differences get some contribution from the $1/m_Q$ term that is
present in $F_1$.

\begin{figure}[h]
\centerline{\epsfig{file=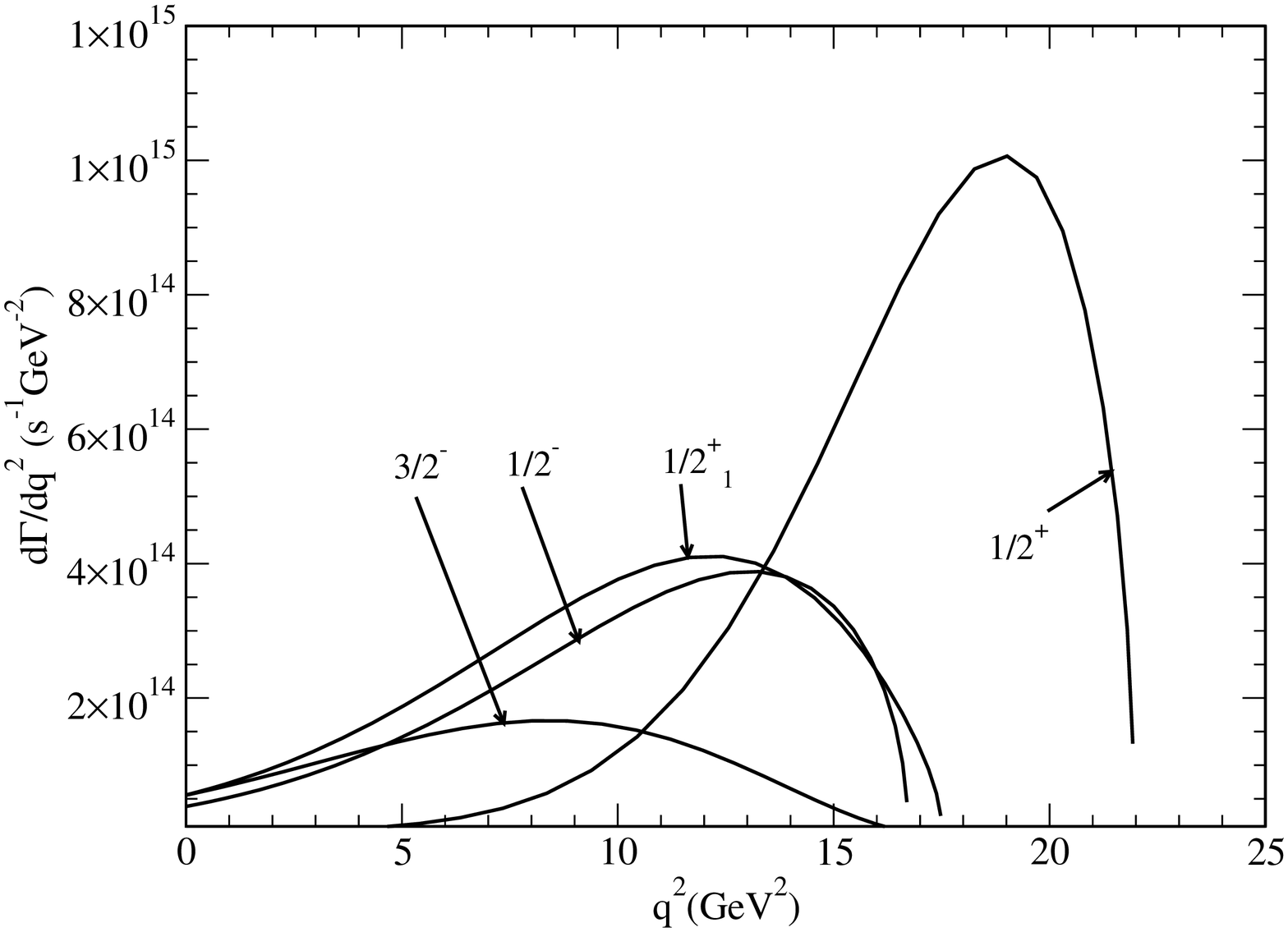,width=10cm}\hspace*{-0.3in}
\epsfig{file=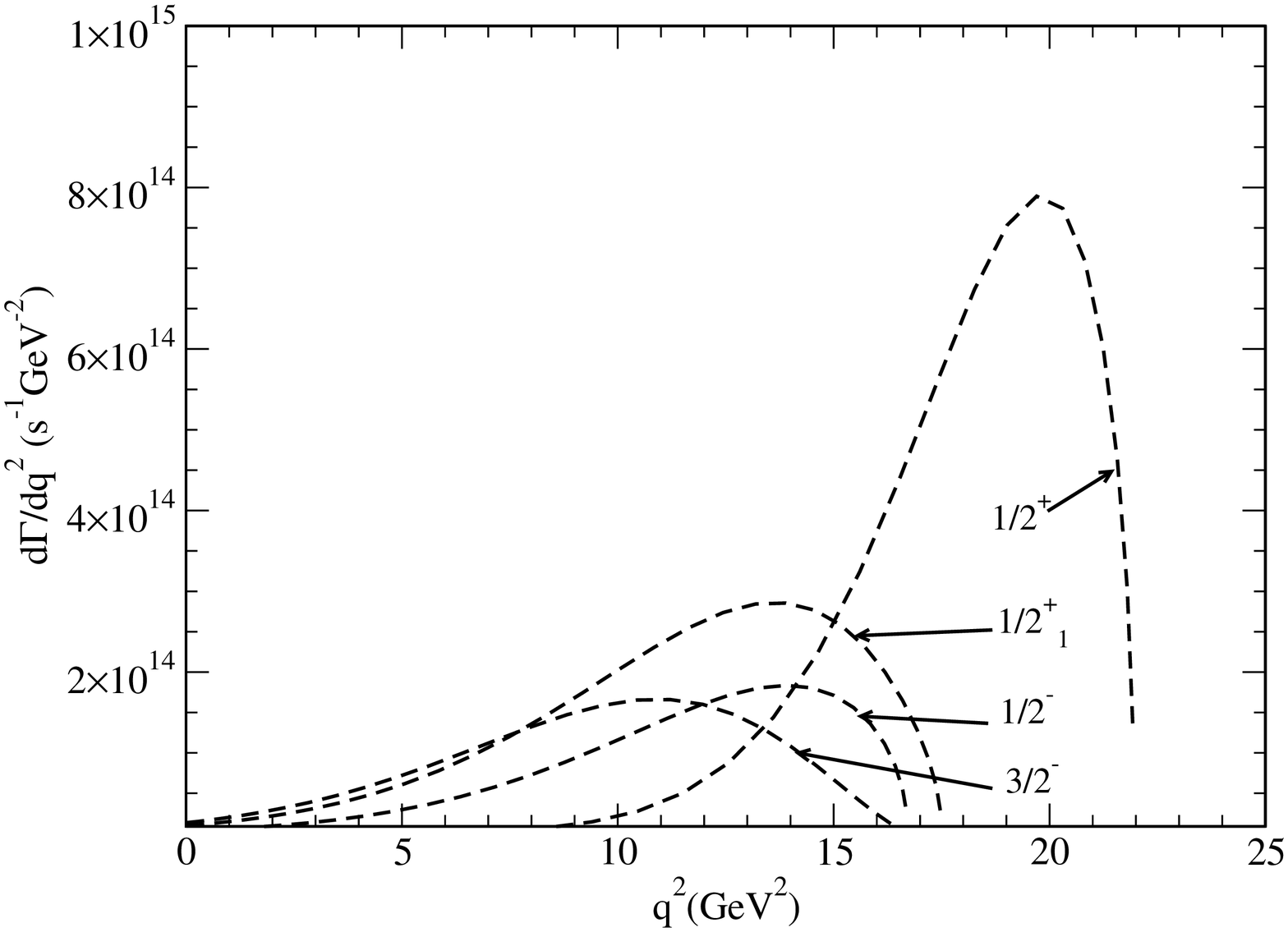,width=10cm}}
\centerline{\epsfig{file=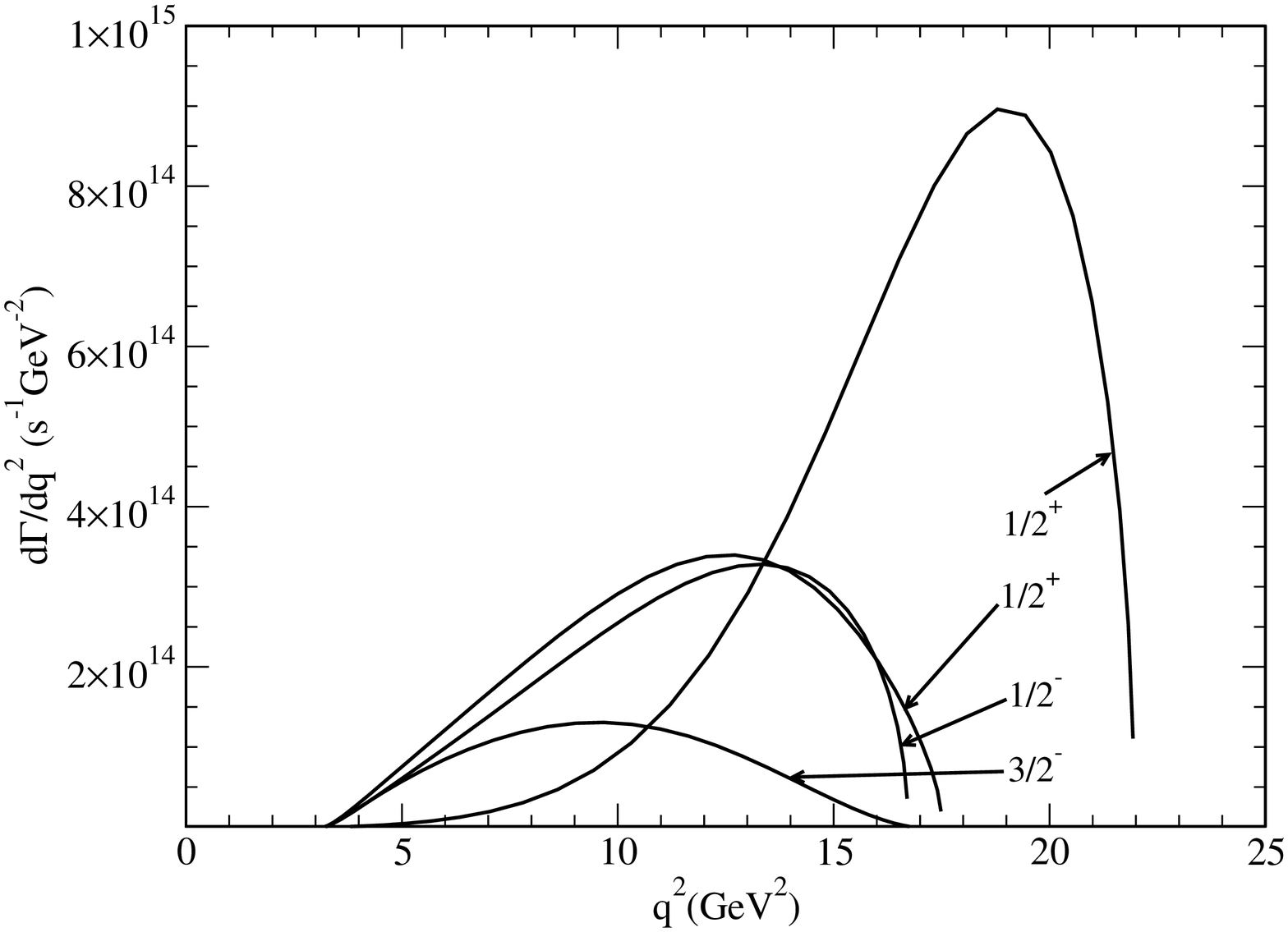,width=10cm}\hspace*{-0.3in}
\epsfig{file=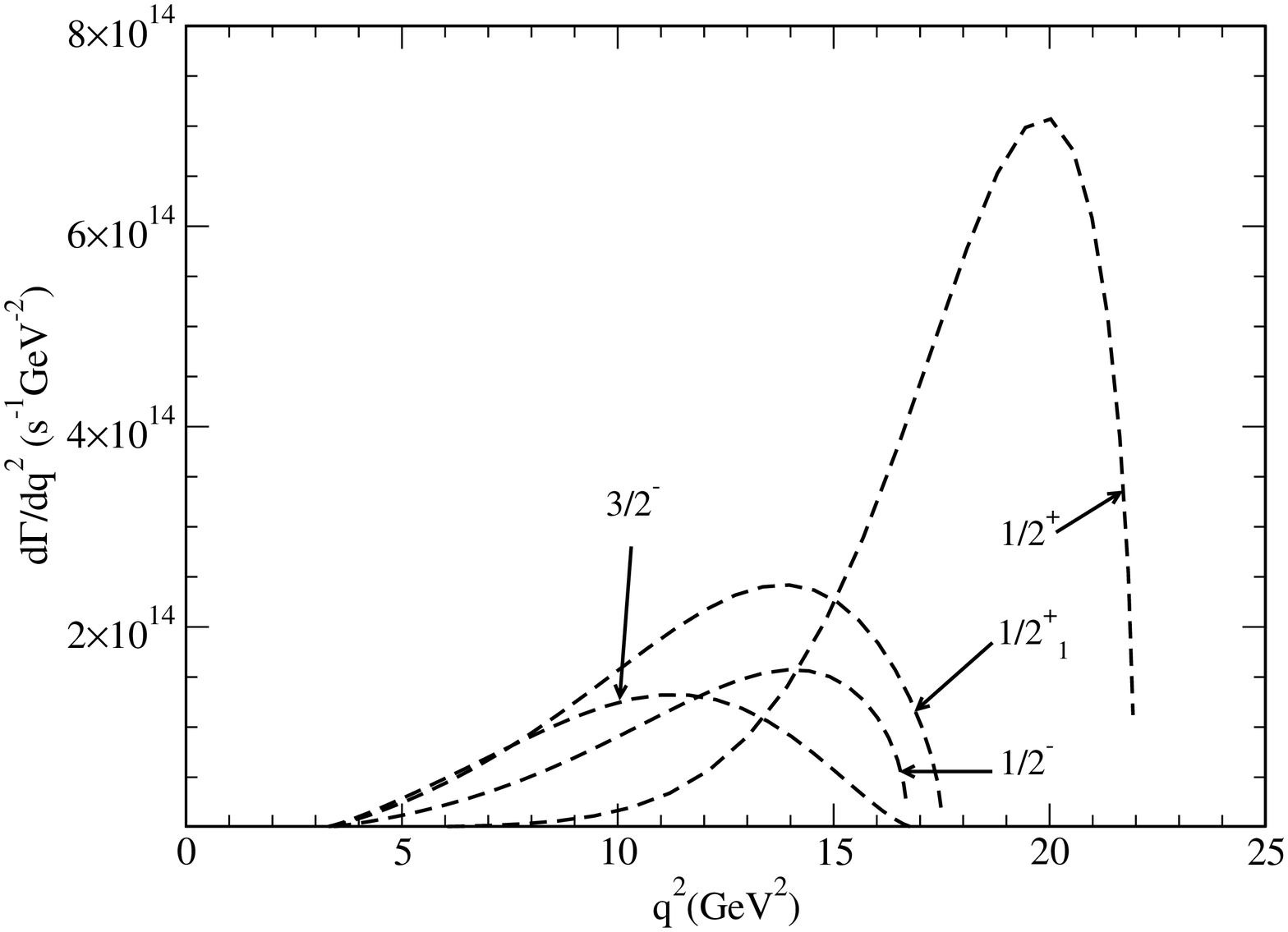,width=10cm}}
\caption{Differential decay rates for $\Lambda_b\to N^{(*)+}$ in the HONR and
HOSR models. The upper panels show the rates for $\Lambda_b\to
N^{(*)+}e^-\bar\nu_e$, while the lower panels show the rates for $\Lambda_b
\to N^{(*)+}\tau^-\bar\nu_\tau$, both in units of $|V_{ub}|^2$. The panels on
the left are from the HOSR model, while those on the right are from the HONR
model. \label{decayrate5}}
\end{figure}
In Figure~\ref{decayrate5}, we show the differential decay rates for
$\Lambda_b$ decaying semileptonically into the four lowest-lying
nucleon states, while Table~\ref{ratelbp} shows the integrated rates
into six exclusive states. Also shown in this figure and table are the
rates that we obtain when the final lepton is a $\tau$. The ground
state nucleon is the largest of the CKM suppressed decays of the
$\Lambda_b$, but it accounts for less than 50\% of these decays, in
both of the harmonic oscillator models. A large fraction (about 20\%)
goes into the first excited state, the Roper resonance, usually
treated as a radial excitation of the ground-state nucleon, as it is
in this model. As with the $\Lambda(1405)$ in the decays of the
$\Lambda_c$, this result hinges on the assumption that the Roper
resonance is a three-quark state, and that it is the first radial
excitation of the nucleon. A number of hypotheses for the internal
structure of this state have been made, such as pentaquark
partner~\cite{jaffewilczek}, dynamically generated state~\cite{krehl},
and hybrid state~\cite{kisslinger}. In each of these scenarios, the
rate at which the $\Lambda_b$ decays semileptonically into this state
is affected by its internal structure. For the three-quark,
radially-excited scenario, the prediction is that decays to this state
are about 60\% of the decays to the ground state nucleon, a rather
large fraction. If ample $\Lambda_b$'s can be produced, their
semileptonic decays may therefore provide information that can be used
in understanding the structure of the Roper resonance.

We have examined decays to other excited nucleons, and those shown in
Table~\ref{ratelbp} are by far the dominant ones. We have also
examined one additional $1/2^+$ nucleon state, two additional nucleon
states with $J^P=3/2^+$, and one additional nucleon state with
$J^P=5/2^+$, none of which are shown in Table~\ref{ratelbp}. Of these,
the rate to the additional $1/2^+$ state is less than 1\% of the
`total' rate that we have estimated, while rates to the additional
$3/2^+$ and $5/2^+$ states are similarly small or even smaller. These
small rates are a direct consequence of the structure of these states,
as their overlaps with the decaying $\Lambda_b$, in the spectator
assumption, are very small. The only other excited nucleons that may
occur with `significant' rate in the semileptonic decays of the
$\Lambda_b$ are those with higher spins, such as $7/2^+$ and
$5/2^-$. However, for such states, orbital angular momentum
centrifugal factors will lead to some suppression of the decay rate.

\begin{center}
\begin{table}
\caption{Decay rates of $\Lambda_b\to N^{(*)+}\ell\bar\nu_\ell$ in units of
$10^{12}s^{-1}\times |V_{ub}|^2$. Also shown are the rates for $\Lambda_c\to
N^{(*)0}\ell^+\nu_\ell$ in units of $10^{10}s^{-1}$, 
obtained using $|V_{cd}|=0.224$.
\label{ratelbp}}
\begin{tabular}{|l|cc|cc|}
\hline
 &\multicolumn{2}{c|}{$\Lambda_b\to N^{(*)+}\ell^-\bar\nu_\ell$}&
\multicolumn{2}{c|}
 {$\Lambda_b\to N^{(*)+}\tau^-\bar\nu_\tau$} \\ \hline
$J^P$ & $\Gamma$(HONR) & $\Gamma$(HOSR) & $\Gamma$(HONR) & $\Gamma$(HOSR)  \\
\hline
$1/2^+$ & $4.55$ &  $7.55$  & 4.01  & 6.55\\ \hline
$1/2^+_1$ & $2.92$ &  $4.44$  & 2.20  & 3.05\\ \hline
$1/2^-$ & $1.42$ &  3.85 & 1.10 & 2.73    \\ \hline
$3/2^-$ & $1.54$ &  $1.82$  & 1.03 & 1.07 \\ \hline
$3/2^+$ & $1.03$ &  2.16 & 0.28 & 0.58    \\ \hline
$5/2^+$ & $0.79$ &  $1.49$  & 0.38 & 0.55 \\ \hline
Total & 12.25 & 21.31  & 9.00 &15.53\\ \hline\hline
 &\multicolumn{2}{c|}{$\Lambda_c\to N^{(*)0}\ell^+\nu_\ell$} &-&-\\\hline
 $1/2^+$ & $1.02$ &  $1.35$ &- &-\\ \hline
$1/2^-$ & $0.02$ &  $0.04$  &-& - \\ \hline
\end{tabular}
\end{table}
\end{center}

\begin{figure}[h]
\centerline{\epsfig{file=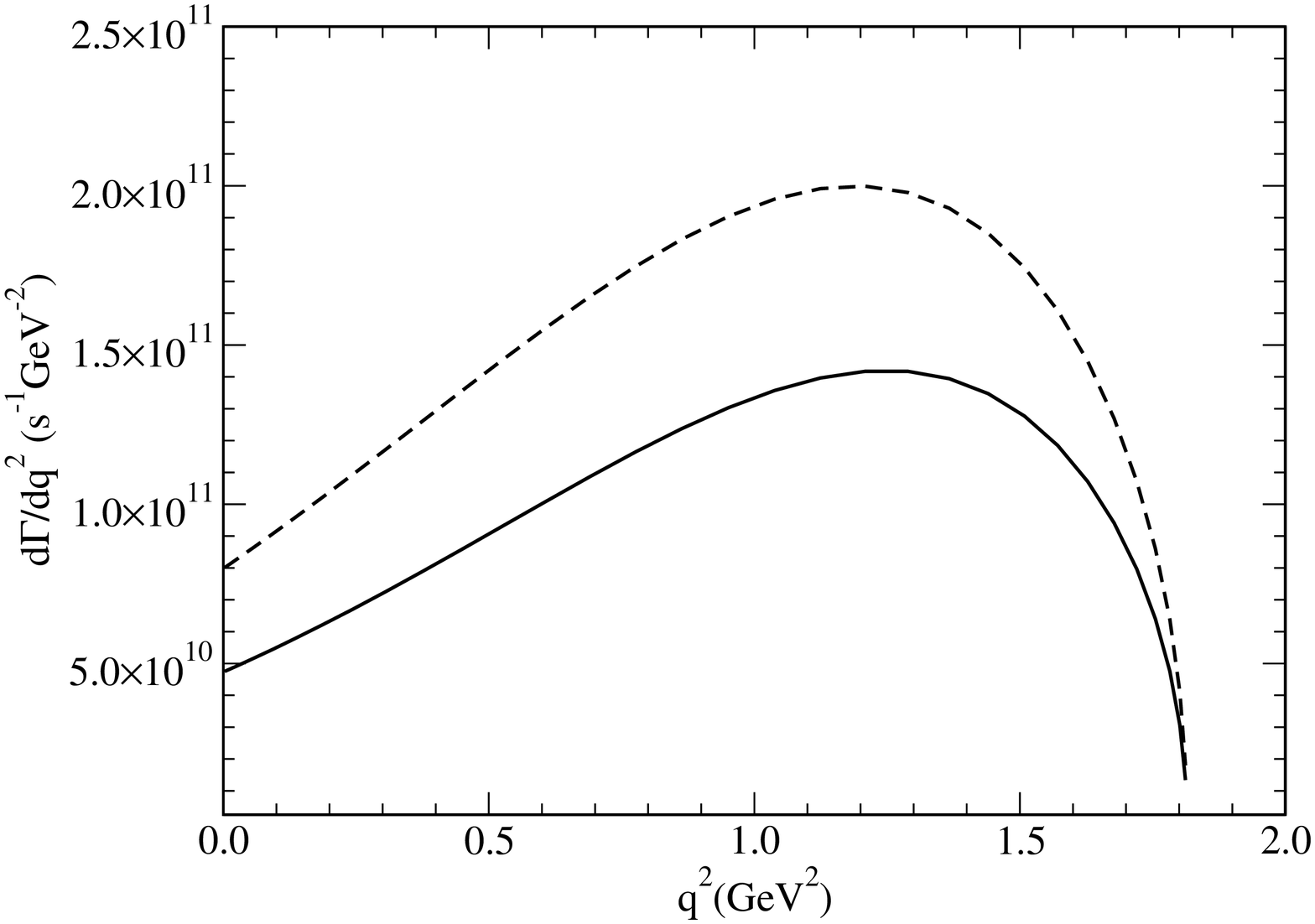,width=10cm}}
\caption{Differential decay rate for $\Lambda_c\to n$ in the HONR and HOSR models.
\label{decayrate6}}
\end{figure}
Figure~\ref{decayrate6} shows the differential decay rate for
$\Lambda_c\to n$, while the integrated decay rates for two exclusive
modes $\Lambda_c\to N^{(*)0}$, obtained using $|V_{cd}|=0.224$, are
shown in Table~\ref{ratelbp}. It is clear from this table that decays
of the $\Lambda_c$ to excited states of the nucleon are strongly
suppressed, due in part to the reduced phase space.

\section{Conclusions and Outlook}

A constituent quark model calculation of semileptonic decays of $\Lambda_b$ and
$\Lambda_c$ baryons, which has several novel features, is described here.
Analytic results for the form factors for the decays to $J^P=1/2^+$ ground
states and excited states with different quantum numbers are evaluated, and
compared to HQET predictions.  For $\Lambda_b\to\Lambda_c$ transitions, the
relations among the form factors, predicted by HQET, are satisfied by the form
factors obtained in the model, independent of the basis used to describe the
baryon wave functions. For the elastic form factors, as well as for the form
factors for decays to the $(1/2^-,\,3/2^-)$ doublet, the HQET relationships
among the form factors are found to hold up to the order we have examined,
namely $1/m_b$ and $1/m_c$. For states of higher spin, we have compared our
model form factors to the HQET predictions at leading order, and the expected
relations hold at that order. 

These form factors depend on the size parameters of the initial and final
baryon wave functions, and so a fit to the spectrum of the states treated here
is performed. Two model Hamiltonians are used, with either a nonrelativistic or
semi-relativistic kinetic energy term, and with Coulomb and spin-spin contact
interactions. The wave functions are expanded in either a harmonic oscillator
or Sturmian basis, up to second-order polynomials, and our numerical results
for form factors and rates are calculated using the resulting mixed wave
functions. Four sets of predictions are made for form factors and rates, with
wave functions, size parameters and mixing coefficients arising from fits using
both the non-relativistic and semi-relativistic Hamiltonians, and using the two
different bases. These predictions can be used to assess the model dependence
in the results we obtain.

Interestingly, the form factors for decays to ground state daughter
baryons evaluated using the Sturmian basis for the wave functions have
slopes at the non-recoil point that are significantly larger than
those evaluated using the harmonic oscillator basis. As a result, the
corresponding integrated decay rates for $\Lambda_c\to \Lambda$
elastic decays, calculated using the Sturmian wave functions, are
smaller than those obtained using the harmonic oscillator basis wave
functions. The Sturmian rates are both consistent within errors with
the experimentally reported rate of $1.05\pm 0.35\times 10^{11}$
s$^{-1}$, while those calculated using the harmonic oscillator basis
are significantly larger. As pointed out by Keister and
Polyzou~\cite{KP}, although calculations using the Sturmian basis are
not as simple as those using the harmonic oscillator basis, the
resulting form factors have shapes which are expansions in inverse
powers of $1+k^2/\Lambda^2$, with $k$ the decay three-momentum (in a
non-relativistic decay calculation like ours), and $\Lambda$ a
constant which is calculated in terms of quark masses and wave
function size parameters. This is closer to the form expected from
experimental studies of hadron decay form factors, and so the use of
Sturmian basis functions produces realistic results for decay
calculations even with the inevitable truncations of the basis
required for tractability. Larger scale numerical calculations using
the Sturmian basis require fewer basis states than those using the
harmonic oscillator basis to yield accurate energies and decay form
factors for excited states.

Although the use of a semi-relativistic Hamiltonian does not
necessarily lead to a better fit to the spectrum, in calculations
using both bases it results in an integrated decay rate for
$\Lambda_c\to \Lambda$ elastic channel that is closer to the central
value that has been experimentally reported. However, the rate
obtained in the nonrelativistic version of the Sturmian model is also
consistent (within 1 $\sigma$) with the experimentally reported value.

Decay form factors and rates to all available excited state daughter
baryons are evaluated using these four models. Significant branching
fractions are found for $\Lambda_c$ inelastic semileptonic decays in
all four calculations, with the total to all excited states ranging
from 11 to 19\%. This has important consequences for the absolute
normalization of the branching fractions to the many observed final
states in $\Lambda_c^+$ decay, most of which are measured relative to
the decay mode $\Lambda_c^+\to pK^-\pi^+$. The extraction of the
absolute branching fraction of this mode, from measurements of the
cross section for $\Lambda_c^+X$ production in $e^+e^-$ annihilation,
requires knowledge of the fraction $f$ of semi-leptonic decays
$\Lambda_c^+\to X_s\ell^+\nu_\ell$ to the elastic channel. This
contradicts the available CLEO analyses, in which it is assumed that
the elastic decay of the $\Lambda_c$ saturates its semileptonic
decays. A larger fraction, from 23\ to 38\%, of $\Lambda_b\to
\Lambda_c$ semileptonic decays, are found to be inelastic. Elastic
decays of the $\Lambda_b$ involving tau leptons in the final state are
suppressed by roughly a factor of two because of the reduction in the
final-state phase space, and those to excited baryon states are
suppressed more strongly.

HQET predicts that the form factors $\xi_1$ and $\xi_2$, defined earlier,
should be the same for the decays $\Lambda_b\to N^0$ and $\Lambda_c\to N^+$, up
to $1/m_b$ or $1/m_c$ corrections. Within our models, we find that the two form
factors are very similar, but not identical, with the differences arising from
differences in the size parameters for the $\Lambda_b$ and $\Lambda_c$. In the
case of the decay of the $\Lambda_b$ to nucleons, we find that the `elastic'
fraction is quite small, of the order of 35\% when the leptons produced in the
decay are light. A number of excited nucleons contribute to the total rate,
with the radially excited Roper having the next largest branching fraction.
This may be used as a test of the structure of this resonance, if ample
$\Lambda_b$'s can be produced.

The work presented in this manuscript can be extended in a number of
directions. We plan to examine the semileptonic decays of heavy $\Omega_Q$
baryons, both to ground states and to a number of excited states, in a
calculation similar to the one outlined here. Since the description of these
states using both the quark model and HQET is more complex, it will be
interesting to see if the correspondence  between quark-model results and the
predictions of HQET still holds. We can also apply our model to the description
of the semileptonic decays of the light baryons, although these are already
successfully described by Cabbibo theory. Essentially all experimentally
accessible observables for these decays have been measured, and it will be
interesting to see if our model, constructed with no special reference to
chiral symmetry or current algebra, can describe the results of these
measurements.

We have not examined the predictions of our model for the many polarization
observables which can, in principle, be measured in semileptonic decays. One
example is the asymmetry parameter $\alpha_{\Lambda_c}$ in the decays of the
$\Lambda_c$, which has already been extracted by the CLEO collaboration. The
predictions of our model for this and similar quantities are therefore of some
interest. In addition, the rare decays of heavy baryons, such as
$\Lambda_b\to\Lambda$ can easily be treated in the framework that we have
developed. Such processes, along with their meson analogs, are used in searches
for physics beyond the standard model. However, the interpretation of the
measured rates depend strongly on estimates of the form factors involved (in
much the same way that extraction of CKM matrix elements depend on the form
factors that describe semileptonic decays). Finally, if  factorization, in some
form, is valid, the semileptonic form factors calculated in the manuscript  may
also be useful in the description of nonleptonic weak decays.

It may also be possible to systematically improve the quark model used in the
present calculation. An obvious first step is the implementation of full
symmetrization of the spatial wave functions in the Sturmian basis, which would
allow calculation of results for decays to final state nucleons in this basis.

\section{Acknowledgments}
Helpful discussions with Dr.~J.~Piekarewicz and Dr.~L.~Reina are
gratefully acknowledged.  This research is supported by the
U.S.~Department of Energy under contracts DE-FG02-92ER40750 (M.P.~and
S.C.) and DE-FG05-94ER40832 (W.R.).

\appendix

\section{Wave Functions}
\label{wavefunctions}

As mentioned in the text, our wave functions are expanded in two  different bases.
For the states of different spins and parities considered here, the expansions 
are given in this Appendix. For $\Lambda_Q$ states with $J^P=1/2^+$, the expansion
reads

\begin{eqnarray} \label{lhfpl}
\Psi_{\Lambda_Q,1/2^+M}&=&\phi_{\Lambda_Q}\left(\vphantom{\sum_i}
\left[\eta_1^{\Lambda_Q}\psi_{000000}({\bf p}_\rho, {\bf 
p}_\lambda)
+\eta_2^{\Lambda_Q}\psi_{001000}({\bf p}_\rho, {\bf 
p}_\lambda)
+\eta_3^{\Lambda_Q}\psi_{000010}({\bf p}_\rho, 
{\bf p}_\lambda)\right]\chi_{1/2}^\rho(M)\right.\nonumber \\
&+&\eta_4^{\Lambda_Q}\psi_{000101}({\bf p}_\rho, 
{\bf p}_\lambda)\chi_{1/2}^\lambda(M)
+\eta_5^{\Lambda_Q}\left[\psi_{1M_L0101}({\bf p}_\rho, {\bf 
p}_\lambda)\chi_{3/2}^S(M-M_L)\right]_{1/2,M}\\
&+&\left.\eta_6^{\Lambda_Q}\left[\psi_{1M_L0101}({\bf p}_\rho, {\bf 
p}_\lambda)\chi_{1/2}^\lambda(M-M_L)\right]_{1/2,M}
+\eta_7^{\Lambda_Q}\left[\psi_{2M_L0101}({\bf p}_\rho, {\bf p}_\lambda)
\chi_{3/2}^S(M-M_L)\right]_{1/2,M}\right),\nonumber 
\end{eqnarray}
where $\left[\psi_{LM_Ln_\rho\ell_\rho n_\lambda\ell_\lambda}({\bf p}_\rho,
{\bf  p}_\lambda)\chi_{S}(M-M_L)\right]_{J,M}$ is a shorthand notation that
denotes the Clebsch-Gordan sum $\sum_{M_L}\< JM|LM_L, SM-M_L\>
\psi_{LM_Ln_\rho\ell_\rho n_\lambda\ell_\lambda}({\bf p}_\rho, {\bf 
p}_\lambda)\chi_{S}(M-M_L)$.

For $\Lambda_Q$ states with $J^P=1/2^-$ and $3/2^-$, the expansion is
\begin{eqnarray} \label{lthm}
\Psi_{\Lambda_Q,J^-M}&=&\phi_{\Lambda_Q}\left(
\eta_1^{\Lambda_Q}\left[\psi_{1M_L0100}({\bf p}_\rho, {\bf
p}_\lambda)\chi_{3/2}^S(M-M_L)\right]_{JM}\right.\nonumber\\
&&+\eta_2^{\Lambda_Q}\left[\psi_{1M_L0100}({\bf p}_\rho, 
{\bf p}_\lambda)
\chi_{1/2}^\lambda(M-M_L)\right]_{JM}\nonumber\\
&&+\left.\eta_3^{\Lambda_Q}\left[\psi_{1M_L0001}({\bf 
p}_\rho, 
{\bf p}_\lambda)\chi_{1/2}^\rho(M-M_L)\right]_{JM}\right),
\end{eqnarray}
where $J$ can take the values 1/2 or 3/2.

For $\Lambda_Q$ states with $J^P=3/2^+$, the expansion is
\begin{eqnarray}
\Psi_{\Lambda_Q,3/2^+M}&=&\phi_{\Lambda_Q}\left(\vphantom{\sum_i}
\eta_1^{\Lambda_Q}\psi_{000101}({\bf p}_\rho, {\bf 
p}_\lambda)\chi_{3/2}^S(M)
+\eta_2^{\Lambda_Q}\left[\psi_{1M_L0101}
({\bf p}_\rho, {\bf p}_\lambda)\chi_{3/2}^S(M-M_L)\right]_{3/2,M}\right.\nonumber\\
&+&\eta_3^{\Lambda_Q}\left[\psi_{1M_L0101}
({\bf p}_\rho, {\bf p}_\lambda)\chi_{1/2}^\lambda(M-M_L)\right]_{3/2,M}
+\eta_4^{\Lambda_Q}\left[\psi_{2M_L0200}
({\bf p}_\rho, {\bf p}_\lambda)\chi_{1/2}^\rho(M-M_L)\right]_{3/2,M}\nonumber\\
&+&\eta_5^{\Lambda_Q}\left[\psi_{2M_L0101}({\bf p}_\rho, {\bf 
p}_\lambda)\chi_{3/2}^S(M-M_L)\right]_{3/2,M}
+\eta_6^{\Lambda_Q}\left[\psi_{2M_L0101}({\bf p}_\rho, {\bf 
p}_\lambda)\chi_{1/2}^\lambda(M-M_L)\right]_{3/2,M}\nonumber \\
&+&\left.\eta_7^{\Lambda_Q}\left[\psi_{2M_L0002}({\bf p}_\rho, 
{\bf p}_\lambda)\chi_{1/2}^\rho(M-M_L)\right]_{3/2,M}\right)
\end{eqnarray}

For $J^P=5/2^+$, the expansion is 
\begin{eqnarray}
\Psi_{\Lambda_Q,5/2^+M}&=&\phi_{\Lambda_Q}\left(\vphantom{\sum_i}
\eta_1^{\Lambda_Q}\psi_{1M_L0101}({\bf p}_\rho, {\bf 
p}_\lambda)\chi_{3/2}^S(M)
+\eta_2^{\Lambda_Q}\left[\psi_{2M_L0101}
({\bf p}_\rho, {\bf p}_\lambda)\chi_{3/2}^S(M-M_L)\right]_{5/2,M}\right.\nonumber\\
&+&\eta_3^{\Lambda_Q}\left[\psi_{2M_L0101}
({\bf p}_\rho, {\bf p}_\lambda)\chi_{1/2}^\lambda(M-M_L)\right]_{5/2,M}
+\eta_4^{\Lambda_Q}\left[\psi_{2M_L0200}
({\bf p}_\rho, {\bf p}_\lambda)\chi_{1/2}^\rho(M-M_L)\right]_{5/2,M}\nonumber\\
&+&\left.\eta_5^{\Lambda_Q}\left[\psi_{2M_L0002}({\bf p}_\rho, 
{\bf p}_\lambda)\chi_{1/2}^\rho(M-M_L)\right]_{5/2,M}\right)
\end{eqnarray}
No other states are expected to have significant overlap with the decaying 
ground-state $\Lambda_Q$ in the spectator approximation that we use.

The wave function components for nucleons are different from those shown
above, due to the different (12) symmetry in the wave functions, and are shown
below. For $J^P=1/2^+$, nucleon wave functions are expanded as 
\begin{eqnarray}
\Psi_{N,1/2^+M}&=&\phi_{N}\left(\vphantom{\sum_i}
\left[\eta_1^{N}\psi_{000000}({\bf p}_\rho, {\bf 
p}_\lambda)
+\eta_2^{N}\psi_{001000}({\bf p}_\rho, {\bf 
p}_\lambda)
+\eta_3^{N}\psi_{000010}({\bf p}_\rho, 
{\bf p}_\lambda)\right]\chi_{1/2}^\lambda(M)\right.\nonumber \\
&+&\eta_4^{N}\psi_{000101}({\bf p}_\rho, 
{\bf p}_\lambda)\chi_{1/2}^\rho(M)
+\eta_5^{N}\left[\psi_{1M_L0101}({\bf p}_\rho, {\bf 
p}_\lambda)\chi_{1/2}^\rho(M-M_L)\right]_{1/2,M} \\
&+&\left.\eta_6^{N}\left[\psi_{2M_L0200}({\bf p}_\rho, {\bf 
p}_\lambda)\chi_{3/2}^S(M-M_L)\right]_{1/2,M}
+\eta_7^{N}\left[\psi_{2M_L0002}({\bf p}_\rho, {\bf p}_\lambda)
\chi_{3/2}^S(M-M_L)\right]_{1/2,M}\right),\nonumber
\end{eqnarray}

For $J^P=1/2^-$ and $3/2^-$, the expansion is
\begin{eqnarray} \label{nthm}
\Psi_{N,J^-M}&=&\phi_{N}\left(
\eta_1^{N}\left[\psi_{1M_L0100}({\bf p}_\rho, {\bf
p}_\lambda)\chi_{1/2}^\rho(M-M_L)\right]_{JM}\right.\nonumber\\
&&+\eta_2^{N}\left[\psi_{1M_L0001}({\bf p}_\rho, 
{\bf p}_\lambda)
\chi_{3/2}^S(M-M_L)\right]_{JM}\nonumber\\
&&+\left.\eta_3^{N}\left[\psi_{1M_L0001}({\bf 
p}_\rho, 
{\bf p}_\lambda)\chi_{1/2}^\lambda(M-M_L)\right]_{JM}\right),
\end{eqnarray}
where $J$ can take the values 1/2 or 3/2.

\section{Integrals in the Sturmian Basis}

Wave functions expanded in the Sturmian basis have been used by other authors in exploring aspects of heavy meson
phenomenology~\cite{olsson}. However, to the best of our knowledge, there have been no prior
applications to baryon phenomenology. We therefore believe that it is useful to
outline some of the steps needed in using this basis for calculations of the kind that we
present.

\subsection{Integrals for Hamiltonian Matrix Elements}

We begin by reminding the reader that, in coordinate space, say, the spatial wave 
function components are written as 
\begin{eqnarray}
\psi_{LMn_{\rho}\ell_{\rho}
n_{\lambda}\ell_\lambda}(\lpmb{\rho}, \lpmb{\lambda}) = 
\sum_m\langle LM|\ell_{\rho}m,\ell_\lambda M-m\rangle\psi_{n_\rho \ell_\rho m}
(\lpmb{\rho}) \psi_{n_\lambda \ell_\lambda M-m}(\lpmb{\lambda}),\nonumber
\end{eqnarray}
with $\lpmb{\rho}$ and $\lpmb{\lambda}$ as defined in the main text.

In the Sturmian basis, evaluation of the matrix elements of the
non-relativistic kinetic energy operator, as well as those of the
parts of the potential that depend only on 
$r_{12}\equiv\left|{\bf r}_1-{\bf r}_2\right|$, 
are relatively straightforward, in the latter case because
$\rho=r_{12}/\sqrt{2}$. However, the evaluation of terms that depend
on $r_{13}$ or $r_{23}$ is not as straightforward. To illustrate the
way in which such calculations are carried out, we consider the linear
potential, and examine the term
\beq
V^{\rm lin}_{13}=b\left|{\bf r}_1-{\bf r}_3\right|=br_{13}.
\eeq
We begin by writing
\beq
r_{13}=\frac{1}{\sqrt{2}}\left|\lpmb{\rho}+\sqrt{3}\lpmb{\lambda}\right|\equiv
|\lpmb{\rho}'+\lpmb{\lambda}'|={1\over
\sqrt{2}}(\rho^2+2\sqrt{3}\rho\cdot\lambda +3\lambda^2)^{1/2}.
\eeq
In the above, $\lpmb{\rho}'\equiv \lpmb{\rho}/\sqrt{2}$ and 
$\lpmb{\lambda}'\equiv \sqrt{3/2}\lpmb{\lambda}$. 
The latter form is expanded in spherical harmonics, yielding
\beq
r_{13}=4\pi\sum_{l}
 {1\over(2l+1)}{\rho'^l\over\lambda'^{l+1}}\left({\rho'^2\over
(2l+3)}-{\lambda'^2\over (2l-1)}\right) \left(Y_l({\hat\rho})\cdot 
Y_l({\hat\lambda})\right)
\eeq
for $\rho'<\lambda'$, and a similar expression with $\rho'\leftrightarrow\lambda'$
otherwise. In this expansion, 
\beq
\left(Y_l({\hat\rho})\cdot Y_l({\hat\lambda})\right)\equiv
\sum_m(-1)^m Y_l^m({\hat\rho}) Y_l^{-m}({\hat\lambda})
\eeq

Calculation of $\left<r_{13}\right>$ then requires the evaluation of the matrix element
$\langle L' n'_\rho l'_\rho n'_\lambda l'_\lambda|{\cal Y}_l({\hat\rho})\cdot
 {\cal Y}_l({\hat\lambda})|L n_\rho l_\rho n_\lambda
 l_\lambda\rangle$, 
which symbolically denotes integrations over the angles defining
$\lpmb{\rho}$ and $\lpmb{\lambda}$. This is done with the use of 6-J
symbols, leaving integrals over the magnitudes of $\rho$ and $\lambda$
which can be done either numerically or analytically. For the
potentials we use, all terms can be handled analytically. Terms in the
potential that depend on $r_{23}$ are handled in a similar manner.

\subsection{Integrals for Current Matrix Elements}

In order to evaluate the form factors in the Sturmian basis, integrals of the form
\beq
{\cal I}^{\ell_1,\ell_2,\ell_3}_{n_1,n_2}=\int d^3p\frac{{\cal Y}_{\ell_1}\left({\bf p}\right)
{\cal Y}_{\ell_2}\left({\bf p}+a{\bf k}\right){\cal Y}_{\ell_3}\left({\bf p}\right)}{\left(p^2+\alpha^2
\vphantom{\left({\bf p}+a{\bf k}\right)^2}
\right)^{n_1}
\left[\left({\bf p}+a{\bf k}\right)^2+{\alpha^\prime}^2\right]^{n_2}}
\eeq
must be calculated. In the above, $p$ represents an internal momentum conjugate to one of
the Jacobi coordinates (for these integrals, $p_\lambda$), while ${\bf k}$ is 
the momentum of the daughter baryon in the frame in which the parent is at rest. The
constant $a= -2\sqrt{3/2}\,m_\sigma/m_{\Lambda_q}$, with $m_{\Lambda_q}$ being the mass of the daughter
baryon in the decay. The quantities ${\cal Y}_{\ell}\left({\bf p}\right)$ are the vector
harmonics, with $\ell_{1,2}$ being the orbital angular momentum in the initial or final
state, respectively, while ${\cal Y}_{\ell_3}\left({\bf p}\right)$ arises from the Pauli reduction
of the vector or axial current. For simplicity we choose $\ell_1=\ell_3=0$, but this will 
still be sufficient to illustrate the method.

With the use of Feynman parametrization, this integral is first rewritten as
\beqy
{\cal I}^{0,\ell,0}_{n_1,n_2}&=&\frac{1}{\sqrt{4\pi}}\frac{\Gamma(n_1+n_2)}{\Gamma(n_1)\Gamma(n_2)}
\int_0^1 dx\int d^3p \frac{x^{n_1-1}(1-x)^{n_2-1}{\cal Y}_{0}
\left({\bf p}\right)
{\cal Y}_{\ell}\left({\bf p}+a{\bf k}\right)}
{\left\{x \left(p^2+\vphantom{\left({\bf p}+a{\bf k}\right)^2}\alpha^2\right)+(1-x)
\left[\left({\bf p}+a{\bf k}\right)^2+\beta^2\right]\right\}^{n_1+n_2}}
\nonumber\\
&=&\frac{1}{\sqrt{4\pi}}\frac{\Gamma(n_1+n_2)}{\Gamma(n_1)\Gamma(n_2)}
\int_0^1 dx\int d^3p \frac{x^{n_1-1}(1-x)^{n_2-1}{\cal Y}_{0}
\left({\bf p}\right)
{\cal Y}_{\ell}\left({\bf p}+a{\bf k}\right)}
{\left[p^2+2a(1-x){\bf p}\cdot{\bf k}+a^2 k^2(1-x)+\beta^2(1-x)+\alpha^2
x\right]^{n_1+n_2}}\nonumber\\
&=&\frac{1}{\sqrt{4\pi}}\frac{\Gamma(n_1+n_2)}{\Gamma(n_1)\Gamma(n_2)}
\int_0^1 dx\int d^3p \frac{x^{n_1-1}(1-x)^{n_2-1}{\cal Y}_{0}
\left({\bf p}\right)
{\cal Y}_{\ell}\left({\bf p}+a{\bf k}\right)}
{\left\{\left[{\bf p}+a(1-x){\bf k}\right]^2+a^2 k^2 x(1-x)+\alpha^2 x+\beta^2(1-x)
\right\}^{n_1+n_2}},
\eeqy
where the factor of $1/\sqrt{4\pi}$ arises from one of the vector harmonics with
$\ell=0$.

Defining
\beq
{\bf u}={\bf p}+a(1-x){\bf k}
\eeq
and substituting into the integral gives
\beq
{\cal I}^{0,\ell,0}_{n_1,n_2}=\frac{1}{\sqrt{4\pi}}\frac{\Gamma(n_1+n_2)}{\Gamma(n_1)\Gamma(n_2)}
\int_0^1 dx\int d^3u \frac{x^{n_1-1}(1-x)^{n_2-1}{\cal Y}_{0}
\left({\bf u}\right)
{\cal Y}_{\ell}\left({\bf u}+ax{\bf k}\right)}
{\left[u^2+a^2 k^2 x(1-x)+\alpha^2 x+\beta^2(1-x)
\right]^{n_1+n_2}}.
\eeq
The angular integration can be performed after expanding the ${\cal
Y}_{\ell}({\bf u}+ax{\bf k})$ to give
\beq
{\cal I}^{\ell}_{n_1,n_2}=a^\ell {\cal Y}_{\ell}\left({\bf k}\right)
\frac{\Gamma(n_1+n_2)}{\Gamma(n_1)\Gamma(n_2)}
\int_0^1 dx\int du\, u^2 \frac{x^{n_1-1+\ell}(1-x)^{n_2-1}}
{\left[u^2+a^2 k^2 x(1-x)+\alpha^2 x+\beta^2(1-x)
\right]^{n_1+n_2}}.
\eeq
Using
\beq
\int^\infty_0 du\frac{u^{2m}}{\left(u^2+{\cal A}\right)^n}=\frac{1}{2
{\cal A}^{n-m-1/2}}\frac{\Gamma(m+1/2)\Gamma(n-m-1/2)}{\Gamma(n)}
\eeq
in the above equation gives
\beqy
{\cal I}^{\ell}_{n_1,n_2}&=&a^\ell {\cal Y}_{\ell}\left({\bf k}\right)
\frac{\Gamma(n_1+n_2)}{\Gamma(n_1)\Gamma(n_2)}\frac{\Gamma(3/2)\Gamma(n_1+n_2-3/2)}
{\Gamma(n_1+n_2)}
\int_0^1 dx\frac{x^{n_1-1+\ell}(1-x)^{n_2-1}}
{2\left[a^2 k^2 x(1-x)+\alpha^2 x+\beta^2(1-x)
\right]^{n_1+n_2-3/2}}\nonumber\\
&=&
a^\ell {\cal Y}_{\ell}\left({\bf k}\right)
\frac{\Gamma(3/2)\Gamma(n_1+n_2-3/2)}{\Gamma(n_1)\Gamma(n_2)}
\int_0^1 dx\frac{x^{n_1-1+\ell}(1-x)^{n_2-1}}
{2\left[a^2 k^2 x(1-x)+\alpha^2 x+\beta^2(1-x)
\right]^{n_1+n_2-3/2}}.
\eeqy
This integral can now be written as a sum of terms ${\cal J}^m_n$, with
\beq
{\cal J}^m_n\equiv\int^1_0 dx \frac{x^m}{\left(c_0+c_1x+c_2x^2\right)^{n+1/2}},
\eeq
where
\beq
c_0=\beta^2, \,\,\,\, c_1=a^2 k^2+\alpha^2-\beta^2, \,\,\,\, c_2=-a^2 k^2.
\eeq
Each of these terms can be then be integrated analytically to give the required matrix 
element.

This procedure works as long as $2n>m$. When $2n\le m$, the last integration leads to logarithms.
Such terms are expanded around $k=0$ before the form factors are extracted.

\section{Expressions for the Form Factors}

\label{formfactors}

The analytic expressions that we obtain for the form factors are shown in the
following subsections. The results shown are valid for single-component wave functions. 
We separate the results obtained using the harmonic
oscillator basis from those obtained using the Sturmian basis.

\subsection{Harmonic Oscillator Basis}

\subsubsection{$1/2^+$}
\begin{eqnarray}
F_1 &=&  I_H \left[1 +\frac{m_\sigma}{\alpha^2_{\lambda\lambda'}}\left(\frac{\alpha^2_
     {\lambda'}}{m_q}+\frac{\alpha^2_{\lambda}}{m_Q}\right)\right],\nonumber\\
F_2 &=&  - I_H  \left[\frac{m_\sigma}{m_q}\frac{\alpha^2_{\lambda'}}{\alpha^2_{
\lambda\lambda'}}
   -\frac{\alpha^2_{\lambda} \alpha_{\lambda'}^2}{4\alpha^2_{\lambda\lambda'}m_
q m_Q}\right],\nonumber\\
F_3 &=&  - I_H\frac{m_\sigma}{m_Q}\frac{\alpha^2_{\lambda}}{\alpha^2_{\lambda\lambda'}},\nonumber\\
G_1 &=&  I_H\left[1-\frac{\alpha^2_{\lambda}\alpha^2_{\lambda'}}{12\alpha^2_{\lambda\lambda'}m_q
m_Q}\right],\nonumber\\
G_2 &=& -I_H\left[\frac{m_\sigma}{m_q}\frac{\alpha^2_{\lambda'}}{\alpha^2_{\lambda\lambda'}}
+\frac{\alpha^2_{\lambda}\alpha^2_{\lambda'}}{12m_qm_Q\alpha^2_{\lambda\lambda'}}
\left(1+\frac{12m^2_\sigma}{\alpha^2_{\lambda\lambda'}}\right)\right],\nonumber\\
G_3 &=& I_H \left[\frac{m_\sigma}{m_Q}\frac{\alpha^2_{\lambda}}{\alpha^2_{\lambda\lambda'}}
+\frac{m^2_\sigma\alpha^2_{\lambda}\alpha^2_{\lambda'}}{m_qm_Q\alpha^4_{\lambda
\lambda'}}\right]\nonumber
\end{eqnarray}
where
\begin{eqnarray} \label{elasticff}
I_H =\left(\frac{\alpha_\lambda\alpha_{\lambda'}}{\alpha_{\lambda\lambda'}^2}
\right)^{3/2}\exp\left( -\frac{3 m^2_\sigma}{2m^2_{\Lambda_q}}\frac{p^2}{\alpha_{\lambda\lambda'}^2}
\right),\nonumber
\end{eqnarray}
$\alpha_{\lambda\lambda'}^2 =\frac{1}{2}(\alpha_\lambda^2
+\alpha_\lambda'^2)$, and $m_\sigma$ is the mass of the light quark.

\subsubsection{$1/2^+_1$}
\begin{eqnarray}
F_1 &=&  I_H \frac{1}{2\alpha^2_{\lambda\lambda'}}\left[(\alpha^2_\lambda-\alpha^2_{\lambda'})
-\frac{m_\sigma}{3\alpha^2_{\lambda\lambda'}}\left(\frac{\alpha^2_{\lambda'}}{m
_q}
(7\alpha^2_{\lambda}-3\alpha^2_{\lambda'})+\frac{\alpha^2_\lambda}{m_Q}(7\alpha^
2_{\lambda'}
-3\alpha^2_\lambda)\right)\right],\nonumber\\
F_2 &=&-I_H\frac{\alpha^2_{\lambda'}}{6m_q\alpha^4_{\lambda\lambda'}}\left(7\alpha^2_{\lambda
}-3\alpha^2_{\lambda'}\right)\left[m_\sigma-\frac{\alpha^2_\lambda}{4m_Q}\right
],\nonumber\\
F_3 &=&   I_H\frac{\alpha^2_{\lambda}m_\sigma}{6m_Q\alpha^4_{\lambda\lambda'}}
\left(7\alpha^2_{\lambda'}-3\alpha^2_{\lambda}\right),\nonumber\\
G_1 &=&  I_H \left[\frac{(\alpha^2_\lambda -\alpha^2_{\lambda'})}{2\alpha^2_{\lambda\lambda'}}
-\frac{\alpha^2_\lambda\alpha^2_{\lambda'}}{72\alpha^4_{\lambda\lambda'}m_qm_Q}
(7\alpha^2_{\lambda}-3\alpha^2_{\lambda'})\right],\nonumber\\
G_2 &=&  -I_H\frac{\alpha^2_{\lambda'}}{6m_q\alpha^4_{\lambda\lambda'}}
\left[(7\alpha^2_{\lambda}-3\alpha^2_{\lambda'})\left(m_\sigma+\frac{\alpha^2_\lambda}{6m_Q}\right)
+\frac{7m^2_\sigma\alpha^2_\lambda}{m_Q\alpha^2_{\lambda\lambda'}}
(\alpha^2_\lambda-\alpha^2_{\lambda'})\right],\nonumber\\
G_3 &=&  -I_H\frac{\alpha^2_{\lambda}m_\sigma}{6m_Q\alpha^4_{\lambda\lambda'}}\left[
(7\alpha^2_{\lambda'}-3\alpha^2_{\lambda})-\frac{7m_\sigma\alpha^2_{\lambda'}}{m_q
\alpha^2_{\lambda\lambda'}}(\alpha^2_{\lambda}-\alpha^2_{\lambda'})\right],\nonumber
\end{eqnarray}
where
\begin{eqnarray}
I_H =\sqrt{\frac{3}{2}}\left(\frac{\alpha_\lambda\alpha_{\lambda'}}
{\alpha_{\lambda\lambda'}^2}
\right)^{3/2}\exp\left( -\frac{3 m^2_\sigma}{2m^2_{\Lambda_q}}\frac{p^2}{\alpha_{\lambda\lambda'}^2}\right).\nonumber
\end{eqnarray}

\subsubsection{$1/2^-$}
\begin{eqnarray}
F_1 &=&I_H \frac{\alpha_{\lambda}}{6}\left[\frac{3}{m_q}-\frac{1}{m_Q}\right],\nonumber\\
F_2 &=&-I_H\left[\frac{2m_\sigma}{\alpha_{\lambda}}-\frac{\alpha_{\lambda}}{2m_q}+
\frac{2m^2_\sigma\alpha_{\lambda}}{m_Q\alpha^2_{\lambda\lambda'}}-\frac{m_\sigma\alpha_{\lambda}}
{6m_qm_Q\alpha^2_{\lambda\lambda'}}(3\alpha^2_{\lambda}-2\alpha^2_{\lambda'})\right],\nonumber\\
F_3 &=&  I_H \frac{2m^2_\sigma\alpha_{\lambda}}{m_Q\alpha^2_{\lambda\lambda'}},\nonumber\\
G_1 &=&  I_H\left[\frac{2m_\sigma}{\alpha_{\lambda}}-\frac{\alpha_{\lambda}}{6m_Q}
+ \frac{m_\sigma\alpha_{\lambda}}{6m_qm_Q\alpha^2_{\lambda\lambda'}}
(3\alpha^2_{\lambda} -2\alpha^2_{\lambda'})\right],\nonumber\\
G_2 &=&  I_H \left[-\frac{2m_\sigma}{\alpha_{\lambda}}+\frac{\alpha_\lambda}{2m_q}
 +\frac{\alpha_\lambda}{3m_Q} \right],\nonumber\\
G_3 &=& I_H \frac{\alpha_{\lambda}}{3m_Q}\left[1- \frac{m_\sigma}{2m_q\alpha^2_
{\lambda\lambda'}}(3\alpha^2_{\lambda} -2\alpha^2_{\lambda'})\right],\nonumber
\end{eqnarray}
where
\begin{eqnarray}
I_H = \left(\frac{\alpha_\lambda\alpha_{\lambda'}}{\alpha_{\lambda\lambda'}^2}\right)^{5/2}
\exp\left( -\frac{3 m^2_\sigma}{2m^2_{\Lambda_q}}\frac{p^2}{\alpha_{\lambda\lambda'}^2}\right),\nonumber
\end{eqnarray}

\subsubsection{$3/2^-$}
\begin{eqnarray}
F_1 &=&  I_H \frac{3m_\sigma}{\alpha_\lambda} \left[1
+\frac{m_\sigma}{\alpha^2_{\lambda\lambda'}}
\left(\frac{\alpha^2_{\lambda'}}{m_q}+\frac{\alpha^2_{\lambda}}{m_Q}\right)\right],\nonumber\\
F_2 &=&  - I_H \left[\frac{3m_\sigma^2}{m_q}\frac{\alpha^2_{\lambda'}}{\alpha^2
_{\lambda\lambda'}\alpha_\lambda}-\frac{5\alpha_\lambda\alpha^2_{\lambda'}m_\sigma}
{4\alpha^2_{\lambda\lambda'}m_qm_Q}\right],\nonumber\\
F_3 &=&  - I_H \left[\frac{3m^2_\sigma}{m_Q}\frac{\alpha_{\lambda}}{\alpha^2_{\lambda\lambda'}}+
\frac{\alpha_{\lambda}}{2m_Q}\right],\nonumber\\
F_4 &=& I_H \frac{\alpha_\lambda}{m_Q},\nonumber\\
G_1 &=&I_H\left[\frac{3m_\sigma}{\alpha_\lambda}-\frac{\alpha_\lambda}{2m_Q}\left(1
+\frac{3m_\sigma\alpha^2_{\lambda'}}{2m_q\alpha^2_{\lambda\lambda'}}\right)\right],\nonumber\\
G_2 &=& -I_H \left[\frac{3m_\sigma^2}{m_q}\frac{\alpha^2_{\lambda'}}{\alpha^2
_{\lambda\lambda'}\alpha_\lambda}+\frac{m_\sigma\alpha_\lambda\alpha^2_{\lambda'}}
{4m_qm_Q\alpha^4_{\lambda\lambda'}}(\alpha^2_{\lambda\lambda'} +12 m_\sigma^2)\right],\nonumber\\
G_3 &=& I_H\frac{\alpha_{\lambda}}{m_Q\alpha^2_{\lambda\lambda'}}\left[\frac
{\alpha^2_{\lambda\lambda'}}{2} + 3 m^2_\sigma +\frac{\alpha^2_{\lambda'}m_\sigma}
{m_q\alpha^2_{\lambda\lambda'}}(\alpha^2_{\lambda\lambda'}+6 m_\sigma^2)\right],\nonumber\\
G_4 &=& - I_H \left[\frac{\alpha_\lambda}{m_Q}+\frac{m_\sigma}{m_qm_Q}\frac{\alpha^2_{\lambda'}
\alpha_\lambda}{\alpha^2_{\lambda\lambda'}}\right],\nonumber
\end{eqnarray}
where
\begin{eqnarray}
I_H =-\frac{1}{\sqrt{3}}\left(\frac{\alpha_\lambda\alpha_{\lambda'}}
{\alpha_{\lambda\lambda'}^2}\right)^{5/2}\exp\left( -\frac{3m^2_\sigma}{2m^2_{\Lambda_q}}
\frac{p^2}{\alpha_{\lambda\lambda'}^2}\right),\nonumber
\end{eqnarray}

\subsubsection{$3/2^+$}
\begin{eqnarray}
F_1 &=&  -I_H \frac{m_\sigma}{2}\left[\frac{5}{m_q}-\frac{3}{m_Q}\right],\nonumber\\
F_2 &=&  I_H \frac{m_\sigma}{\alpha_{\lambda}}\left[\frac{6m_\sigma}{\alpha_{\lambda}}-\frac{5\alpha_{\lambda}}{2m_q}
+\frac{6m^2_\sigma\alpha_{\lambda}}{\alpha^2_{\lambda\lambda'}m_Q}-\frac{m_\sigma\alpha_{\lambda}}{2\alpha^2_{\lambda\lambda'}
m_qm_Q}(\alpha^2_{\lambda}-2\alpha^2_{\lambda'})\right],\nonumber\\
F_3 &=&  -I_H \frac{m_\sigma}{m_Q}\left[1+\frac{6m^2_\sigma}{\alpha^2_{\lambda\lambda'}}\right],\nonumber\\
F4 &=&   I_H \frac{2m_\sigma}{m_Q},\nonumber\\
G_1 &=& - I_H\left[\frac{6m^2_\sigma}{\alpha^2_\lambda}-\frac{m_\sigma}{2m_Q}+\frac{m^2_\sigma}{6
\alpha^2_{\lambda\lambda'}m_qm_Q}(11\alpha^2_{\lambda}-6\alpha^2_{\lambda'})\right],\nonumber\\
G_2 &=&  I_H\left[\frac{6m^2_\sigma}{\alpha^2_\lambda}-\frac{5m_\sigma}{2m_q}-\frac{2m_\sigma}{m_Q}+
\frac{5\alpha^2_\lambda}{12m_qm_Q}-\frac{2m^2_\sigma\alpha^2_\lambda}{3\alpha^2
_{\lambda\lambda'}m_qm_Q}\right],\nonumber\\
G_3 &=& -I_H\left[\frac{m_\sigma}{2m_Q} -\frac{5\alpha^2_\lambda}{24m_qm_Q}-\frac
{m^2_\sigma}{4m_qm_Q\alpha^2_{\lambda\lambda'}}(5\alpha^2_{\lambda}-2\alpha^2_{\lambda'})\right],\nonumber\\
G4 &=& - I_H\frac{5\alpha^2_\lambda}{6m_qm_Q},\nonumber
\end{eqnarray}
where
\begin{eqnarray}
I_H=\frac{1}{\sqrt{5}}\left(\frac{\alpha_\lambda\alpha_{\lambda'}}
{\alpha_{\lambda\lambda'}^2}\right)^{7/2}
\exp\left(-\frac{3 m^2_\sigma}{2m^2_{\Lambda_q}}\frac{p^2}{
\alpha^2_{\lambda\lambda'}}\right),\nonumber
\end{eqnarray}

\subsubsection{$5/2^+$}
\begin{eqnarray}
F_1 &=&  I_H \frac{3m^2_\sigma}{\alpha^2_\lambda} \left[1
+\frac{m_\sigma}{\alpha^2_{\lambda\lambda'}
}\left(\frac{\alpha^2_{\lambda'}}{m_q}+\frac{\alpha^2_{\lambda}}{m_Q}\right)\right],\nonumber\\
F_2 &=&  - I_H\frac{m^2_\sigma}{m_q\alpha^2_{\lambda\lambda'}}\left[\frac{3m_\sigma\alpha^2_{\lambda'}}
{\alpha^2_\lambda}-\frac{1}{4m_Q}(8\alpha^2_\lambda+7\alpha^2_{\lambda'})\right],\nonumber\\
F_3 &=& -I_H \frac{m_\sigma}{m_Q}\left[1+\frac{3m^2_\sigma}{\alpha^2_{\lambda\lambda'}}\right],\nonumber\\
F_4 &=& I_H \frac{2m_\sigma}{m_Q},\nonumber\\
G_1 &=&  I_H\left[\frac{3m^2_\sigma}{\alpha^2_{\lambda}}-\frac{m_\sigma}{m_Q}-\frac{m^2_\sigma}
{12m_qm_Q\alpha^2_{\lambda\lambda'}}(8\alpha^2_\lambda+15\alpha^2_{\lambda'})\right],\nonumber\\
G_2 &=& - I_H\frac{m^2_\sigma}{m_q\alpha^2_{\lambda\lambda'}}\left[\frac{3m_\sigma\alpha^2_{\lambda'}}{\alpha^2_\lambda}+
\frac{1}{12m_Q}(8\alpha^2_\lambda+3\alpha^2_{\lambda'})+\frac{3m^2_\sigma\alpha
^2_{\lambda'}}
{m_Q\alpha^2_{\lambda\lambda'}}\right],\nonumber\\
G_3 &=& I_H \frac{m_\sigma}{m_Q}\left[1 +\frac{3m^2_\sigma}{\alpha^2_{\lambda\lambda'}}
+\frac{m_\sigma\alpha^2_{\lambda'}}{m_q\alpha^2_{\lambda\lambda'}}\left(1+\frac
{6m_\sigma^2}
{\alpha^2_{\lambda\lambda'}}\right)\right],\nonumber\\
G_4 &=&  - I_H \frac{2m_\sigma}{m_Q}\left[1+\frac{m_\sigma}{m_q}\frac{\alpha^2_
{\lambda'}}
{\alpha^2_{\lambda\lambda'}}\right],\nonumber
\end{eqnarray}
where
\begin{eqnarray}
I_H =\frac{1}{\sqrt{2}}\left(\frac{\alpha_\lambda\alpha_{\lambda'}}
{\alpha_{\lambda\lambda'}^2}\right)^{7/2}\exp\left( -\frac{3 m^2_\sigma}{2m^2_{\Lambda_q}}\
\frac{p^2}{\alpha_{\lambda\lambda'}^2}\right),\nonumber
\end{eqnarray}

\subsection{Sturmian Basis}
\subsubsection{$1/2^+$}
\begin{eqnarray}
F_1 &=&  I_S \left[1 +\frac{m_\sigma}{\beta_{\lambda\lambda'}}\left(\frac{\beta
_{\lambda'}}{m_q}+ \frac{\beta_\lambda}{m_Q}\right)\right],\nonumber\\
F_2 &=&  - I_S \left[\frac{m_\sigma}{m_q}\frac{\beta_{\lambda'}}{\beta_{\lambda
\lambda'}}-\frac{\beta_{\lambda}\beta_{\lambda'}}{6m_qm_Q}\right],\nonumber\\
F_3 &=&  - I_S\frac{m_\sigma}{m_Q}\frac{\beta_{\lambda}}{\beta_{\lambda\lambda'
}},\nonumber\\
G_1 &=&  I_S\left[1-\frac{\beta_{\lambda}\beta_{\lambda'}}{18m_qm_Q}\right],\nonumber\\
G_2 &=&  -I_S\left[\frac{m_\sigma\beta_{\lambda'}}{m_q\beta_{\lambda\lambda'}}+
\frac{4m^2_\sigma\beta_{\lambda}\beta_{\lambda'}}{5m_qm_Q\beta^2_{\lambda\lambda'}}+
\frac{\beta_{\lambda}\beta_{\lambda'}}{18m_q m_Q}\right],\nonumber\\
G_3 &=& I_S\left[\frac{m_\sigma\beta_{\lambda}}{m_Q\beta_{\lambda\lambda'}}+
\frac{4m^2_\sigma\beta_{\lambda}\beta_{\lambda'}}{5m_qm_Q\beta^2_{\lambda\lambda'}}\right],\nonumber
\end{eqnarray}
where
\begin{eqnarray}
I_S = \frac{\left(\frac{\beta_\lambda\beta_{\lambda'}}
{\beta_{\lambda\lambda'}}\right)^{3/2}}{\left[1+  \frac{3}{2}\frac{m^2_\sigma}{m^2_{\Lambda_q}}
\frac{p^2}{\beta_{\lambda\lambda'}^2}\right]^2},\nonumber
\end{eqnarray}
and $\beta_{\lambda\lambda'} =\frac{1}{2}(\beta_\lambda + \beta_\lambda')$.

\subsubsection{$1/2^+_1$}
\begin{eqnarray}
F_1 &=&  I_S\frac{1}{2\beta_{\lambda'}\beta_\lambda}\left[(\beta^2_\lambda-\beta^2_{\lambda'})
-\frac{2m_\sigma}{3}\left(\frac{\beta_{\lambda}}{m_Q}(5\beta_{\lambda'}-3\beta_
\lambda)
-\frac{\beta_{\lambda'}}{m_q}(5\beta_{\lambda}-3\beta_{\lambda'})\right)\right]
,\nonumber\\
F_2 &=&  -I_S\frac{(5\beta_{\lambda}-3\beta_{\lambda'})}{3m_q}
\left[\frac{m_\sigma}{\beta_{\lambda}}-\frac{\beta_{\lambda\lambda'}}{3m_Q}\right],\nonumber\\
F_3 &=&   I_S\frac{m_\sigma}{6m_Q\beta_{\lambda'}}\left(5\beta_{\lambda'}
-3\beta_{\lambda}\right),\nonumber\\
G_1 &=&  I_S\left[\frac{(\beta^2_\lambda -\beta^2_{\lambda'})}{2\beta_{\lambda'
}\beta_\lambda}-\frac{\beta_{\lambda\lambda'}}{54m_qm_Q}(5\beta_{\lambda}-3\beta_{\lambda'})\right],\nonumber\\
G_2 &=& -I_S\frac{m_\sigma}{3m_q\beta_{\lambda}}\left[(5\beta_{\lambda}-3\beta_
{\lambda'})+\frac{4m_\sigma\beta_{\lambda}}{m_Q\beta_{\lambda\lambda'}}(\beta_{\lambda}-\beta_{\lambda'})
+\frac{\beta_{\lambda\lambda'}}{18m_Q}(5\beta_{\lambda}-\beta_{\lambda'})\right
],\nonumber\\
G_3 &=&-I_S\frac{m_\sigma}{3m_Q\beta_{\lambda'}}\left[(5\beta_{\lambda'}-3\beta
_{\lambda})-\frac{4m_\sigma\beta_{\lambda'}}{m_Q\beta_{\lambda\lambda'}}(\beta_{\lambda}-\
\beta_{\lambda'})\right],\nonumber
\end{eqnarray}
where
\begin{eqnarray}
I_S = \frac{\sqrt{3}}{2}\frac{\left(\frac{\beta_\lambda\beta_{\lambda'}}
{\beta_{\lambda\lambda'}}\right)^{5/2}}{\left[1+  \frac{3}{2}\frac{m^2_\sigma}{m^2_{\Lambda_q}}
\frac{p^2}{\beta_{\lambda\lambda'}^2}\right]^3}.\nonumber
\end{eqnarray}

\subsubsection{$1/2^-$}
\begin{eqnarray}
F_1 &=& I_S\frac{\beta_{\lambda\lambda'}}{12}\left[\frac{3}{m_q}
-\frac{1}{m_Q}\right],\nonumber\\
F_2 &=& -I_S \left[\frac{2m_\sigma}{\beta_{\lambda}}-\frac{\beta_{\lambda\lambda'}}{4m_q}
+\frac{2m_\sigma^2}{\beta_{\lambda\lambda'}m_Q} -\frac{m_\sigma}{12m_qm_Q}
(5\beta_{\lambda}-3\beta_{\lambda'})\right],\nonumber\\
F_3 &=& I_S\frac{2m^2_\sigma}{m_Q\beta_{\lambda\lambda'}},\nonumber\\
G_1 &=& I_S\left[\frac{2m_\sigma}{\beta_{\lambda}}-\frac{\beta_{\lambda\lambda'
}}{12m_Q}+\frac{m_\sigma}{36m_qm_Q}(11\beta_{\lambda}-5\beta_{\lambda'})\right],\nonumber\\
G_2 &=& -I_S\left[\frac{2m_\sigma}{\beta_{\lambda}}-\frac{\beta_{\lambda\lambda
'}}{4m_q}-\frac{\beta_{\lambda\lambda'}}{6m_Q}+\frac{m_\sigma}{18m_qm_Q}
(\beta_{\lambda}-\beta_{\lambda'})\right],\nonumber\\
G_3 &=& I_S \frac{\beta_{\lambda\lambda'}}{6m_Q}\left[1 +\frac{m_\sigma}
{2m_q\beta_{\lambda\lambda'}}(\beta_{\lambda'}-3\beta_{\lambda})\right],\nonumber
\end{eqnarray}
where
\begin{eqnarray}
I_S = \sqrt{2}\frac{\left(\frac{\beta_\lambda\beta_{\lambda'}}
{\beta_{\lambda\lambda'}}\right)^{5/2}}{\left[1+  \frac{3}{2}\frac{m^2_\sigma}{m^2_{\Lambda_q}}
\frac{p^2}{\beta_{\lambda\lambda'}^2}\right]^3}.\nonumber
\end{eqnarray}

\subsubsection{$3/2^-$}
\begin{eqnarray}
F_1 &=&  I_S \frac{3m_\sigma}{\beta_\lambda} \left[1+\frac{m_\sigma}{\beta_
{\lambda\lambda'}}\left(\frac{\beta_{\lambda'}}{m_q}+\frac{\beta_{\lambda}}
{m_Q}\right)\right],\nonumber\\
F_2 &=&  - I_S\left[\frac{3m_\sigma^2}{m_q}\frac{\beta_{\lambda'}}{\beta_
{\lambda\lambda'}\beta_\lambda}-\frac{m_\sigma}{4m_qm_Q}(\beta_\lambda-3\beta_{\lambda'})\right],\nonumber\\
F_3 &=&  - I_S\left[\frac{3m_\sigma^2}{m_Q\beta_{\lambda\lambda'}}+\frac{\beta_
{\lambda\lambda'}}{4m_Q}\right],\nonumber\\
F_4 &=& I_S \frac{\beta_{\lambda\lambda'}}{2m_Q},\nonumber\\
G_1 &=&  I_S\left[\frac{3m_\sigma}{\beta_\lambda}-\frac{\beta_{\lambda\lambda'}}
{4m_Q}+\frac{m_\sigma}{60m_qm_Q} (5\beta_\lambda-23\beta_{\lambda'})\right],\nonumber\\
G_2 &=& -I_S\left[\frac{3m^2_\sigma}{m_q}\frac{\beta_{\lambda'}}{\beta_\lambda
\beta_{\lambda\lambda'}}-\frac{m_\sigma}{60m_qm_Q}(5\beta_\lambda-11\beta_{\lambda'})
+\frac{18m_\sigma^3\beta_{\lambda'}}{7\beta^2_{\lambda\lambda'}m_qm_Q}\right],\nonumber\\
G_3 &=& I_S\frac{1}{m_Q}\left[\frac{3m^2_\sigma}{\beta_{\lambda\lambda'}}+\frac{\beta_{\lambda
\lambda'}}{4} +\frac{m_\sigma\beta_{\lambda'}}{5m_q}+\frac{18m_\sigma^3\beta_
{\lambda'}}{7\beta^2_{\lambda\lambda'}m_q}\right],\nonumber\\
G_4 &=&  - I_S\frac{1}{m_Q} \left[\frac{\beta_{\lambda\lambda'}}{2} +\frac{2m_\sigma\beta_
{\lambda'}}{5m_q}\right],\nonumber
\end{eqnarray}
where
\begin{eqnarray}
I_S = -\frac{\sqrt{2}}{3}\frac{\left(\frac{\beta_\lambda\beta_{\lambda'}}
{\beta_{\lambda\lambda'}}\right)^{5/2}}{\left[1+  \frac{3}{2}\frac{m^2_\sigma}{m^2_{\Lambda_q}}
\frac{p^2}{\beta_{\lambda\lambda'}^2}\right]^3}.\nonumber
\end{eqnarray}

\subsubsection{$3/2^+$}
\begin{eqnarray}
F_1 &=&  I_S \frac{m_\sigma\beta_{\lambda\lambda'}}{2\beta_\lambda}\left[\frac{
1}{m_Q}-\frac{5}{3m_q}\right],\nonumber\\
F_2 &=&  I_S \frac{m_\sigma}{\beta_{\lambda}}\left[\frac{6m_\sigma}{\beta_{\lambda}}
-\frac{5\beta_{\lambda\lambda'}}{6m_q}+\frac{6m^2_\sigma}{\beta_{\lambda\lambda
'}m_Q} -
\frac{m_\sigma}{6m_qm_Q}(5\beta_\lambda-\beta_{\lambda'})\right],\nonumber\\
F_3 &=& - I_S \frac{m_\sigma}{3\beta_{\lambda}m_Q}\left[\beta_{\lambda\lambda'}
+\frac{18m^2_\sigma}
{\beta_{\lambda\lambda'}}\right],\nonumber\\
F4 &=&   I_S \frac{2m_\sigma\beta_{\lambda\lambda'}}{3m_Q\beta_\lambda},\nonumber\\
G_1 &=& - I_S\frac{m_\sigma}{\beta_{\lambda}}\left[\frac{6m_\sigma}{\beta_{\lambda}}-
\frac{\beta_{\lambda\lambda'}}{6m_Q}+\frac{m_\sigma}{6m_qm_Q}(5\beta_\lambda-\beta_{\lambda'})\right],\nonumber\\
G_2 &=&  I_S\frac{\beta_{\lambda\lambda'}}{\beta_{\lambda}}\left[\frac{6m^2_\sigma}{\beta_{\lambda}
\beta_{\lambda\lambda'}}-\frac{5m_\sigma}{6m_q}-\frac{2m_\sigma}{3m_Q}+\frac{\beta_{\lambda\lambda'}}{72m_qm_Q}
(5\beta_\lambda+\beta_{\lambda'})\right],\nonumber\\
G_3 &=&- I_S\frac{\beta_{\lambda\lambda'}}{3\beta_{\lambda}m_Q}\left[m_\sigma-\frac{m^2_\sigma}
{2m_q\beta_{\lambda\lambda'}}(5\beta_\lambda-\beta_{\lambda'})+\frac{\beta_{\lambda\lambda'}}
{24m_q}(5\beta_\lambda+\beta_{\lambda'})\right],\nonumber\\
G4&=&  -I_S\frac{\beta^2_{\lambda\lambda'}}{36m_qm_Q\beta_\lambda}(\beta_{\lambda'}+5\beta_\lambda),\nonumber
\end{eqnarray}
where
\begin{eqnarray}
I_S = \frac{\sqrt{6}}{5}\frac{\left(\frac{\beta_\lambda\beta_{\lambda'}}
{\beta_{\lambda\lambda'}}\right)^{7/2}}{\left[1+  \frac{3}{2}\frac{m^2_\sigma}{m^2_{\Lambda_q}}
\frac{p^2}{\beta_{\lambda\lambda'}^2}\right]^4}.\nonumber
\end{eqnarray}

\subsubsection{$5/2^+$}
\begin{eqnarray}
F_1 &=&  I_S \frac{3m^2_\sigma}{\beta^2_\lambda} \left[1
+\frac{m_\sigma}{\beta_{\lambda\lambda'}}
\left(\frac{\beta_{\lambda'}}{m_q}+\frac{\beta_{\lambda}}{m_Q}\right)\right],\nonumber\\
F_2 &=&  -I_S\frac{m^2_\sigma\beta_{\lambda'}}{m_q\beta^2_\lambda}\left[\frac{3
m_\sigma}
{\beta_{\lambda\lambda'}}-\frac{\beta_{\lambda}}{2m_Q}\right],\nonumber\\
F_3 &=&  - I_S \frac{m_\sigma}{3m_Q\beta_\lambda}\left[\beta_{\lambda\lambda'}+
\frac{9m^2_\sigma}
{\beta_{\lambda\lambda'}}\right],\nonumber\\
F_4 &=& I_S \frac{2m_\sigma}{3m_Q}\frac{\beta_{\lambda\lambda'}}{\beta_\lambda}
,\nonumber\\
G_1 &=&  I_S \left[\frac{3m^2_\sigma}{\beta^2_{\lambda}}-\frac{m_\sigma}{m_Q\beta_\lambda}\left(
\frac{\beta_{\lambda\lambda'}}{3}+\frac{5m_\sigma\beta_{\lambda'}}{14m_q}\right
)\right],\nonumber\\
G_2 &=& - I_S\frac{m^2_\sigma\beta_{\lambda'}}{m_q\beta_{\lambda\lambda'}\beta_
{\lambda}}\left[
\frac{3m_\sigma}{\beta_{\lambda}}+\frac{\beta_{\lambda\lambda'}}{14m_Q}+\frac{8
m^2_\sigma}
{3m_Q\beta_{\lambda\lambda'}}\right],\nonumber\\
G_3 &=& I_S \frac{m_\sigma}{m_Q\beta_\lambda}\left[\frac{\beta_{\lambda\lambda'
}}{3}+\frac{3m^2_\sigma}
{\beta_{\lambda\lambda'}} +\frac{m_\sigma\beta_{\lambda'}}{m_q}\left(\frac{2}{7
}+\frac{8m^2_\sigma}
{3\beta^2_{\lambda\lambda'}}\right)\right],\nonumber\\
G_4 &=&  - I_S \frac{2m_\sigma}{m_Q}\left[\frac{\beta_{\lambda\lambda'}}{3\beta
_\lambda}+\frac{2m_\sigma}{7m_q}
\frac{\beta_{\lambda'}}{\beta_{\lambda}}\right],\nonumber
\end{eqnarray}
where
\begin{eqnarray}
I_S = -\sqrt{3}\frac{\left(\frac{\beta_\lambda\beta_{\lambda'}}
{\beta_{\lambda\lambda'}}\right)^{7/2}}{\left[1+  \frac{3}{2}\frac{m^2_\sigma}{m^2_{\Lambda_q}}
\frac{p^2}{\beta_{\lambda\lambda'}^2}\right]^4}.\nonumber
\end{eqnarray}

\section{Hadronic Tensor}
\label{hadrontensor}

The hadronic tensor for these semileptonic decays takes the form
\begin{eqnarray}
H_{\mu\nu} &=& -\alpha G_{\mu\nu} + \beta_{++}(p+p')_\mu (p+p')_\nu + 
\beta_{+-}(p+p')_\mu (p-p')_\nu\nonumber \\
&+& \beta_{-+}(p-p')_\mu (p+p')_\nu + \beta_{--}(p-p')_\mu (p-p')_\nu\nonumber\\
 &+&  i \gamma \epsilon_{\mu\nu\rho\sigma}(p+p')^\rho (p-p')^\sigma.\nonumber
\end{eqnarray}
The forms of the terms $\alpha$, $\beta_{\pm\pm}$ and $\gamma$ for the
different final states we consider are given in the subsections below.

\subsection{$1/2^+$}
\beq
\alpha(1/2^+)=2\left\{[(\ma-\mb)^2-q^2]F_1^2+[(\ma+\mb)^2-q^2]G_1^2\right\},
\eeq
\beq
\beta_{++}(1/2^+)=\sum_{i=1,j=1}^{i=3,j=3}(A_{ij} F_i F_j + A^\prime_{ij} G_i G_j),
\eeq
with 
\beqy
A_{11} &=& A^\prime_{11} = 2,\nonumber\\
A_{22} &=& {1\over2\ma^2}[(\ma+\mb)^2-q^2],\nonumber\\
A_{33} &=& {1\over2\mb^2}[(\ma+\mb)^2-q^2],\nonumber\\
A_{12} &=& {1\over\ma}(\ma+\mb),\nonumber\\
A_{23} &=& {1\over\ma\mb}[(\ma+\mb)^2-q^2],\nonumber\\
A_{31} &=& {2\over \mb}(\ma+\mb),\nonumber\\
A^\prime_{22} &=& {1\over2\ma^2}[(\ma-\mb)^2-q^2],\nonumber\\
A^\prime_{33} &=& {1\over2\mb^2}[(\ma-\mb)^2-q^2],\nonumber\\
A^\prime_{12} &=& {1\over\ma}(\ma-\mb),\nonumber\\
A^\prime_{23} &=& {1\over\ma\mb} [(\ma-\mb)^2-q^2],\nonumber\\
A^\prime_{31} &=& {2\over \mb}(\ma-\mb),\nonumber
\eeqy
\beq
\gamma(1/2^+)= 4 F_1 G_1.
\eeq
\subsection{$1/2^-$}
\beq
\alpha(1/2^-)=2\{[(\ma+\mb)^2-q^2]F_1^2+[(\ma-\mb)^2-q^2]G_1^2\},
\eeq

\beq
\beta_{++}(1/2^-)=\sum_{i=1,j=1}^{i=3,j=3}(A_{ij} F_i F_j + A^\prime_{ij} G_i G_j), 
\eeq
with 
\beqy
A_{11} &=& A^\prime_{11} = 2,\nonumber\\
A_{22} &=& {1\over2\ma^2}[(\ma-\mb)^2-q^2],\nonumber\\
A_{33} &=& {1\over2\mb^2}[(\ma-\mb)^2-q^2],\nonumber\\
A_{12} &=& {1\over\ma}(\ma-\mb),\nonumber\\
A_{23} &=& {1\over\ma\mb}[(\ma\mb)^2-q^2],\nonumber\\
A_{31} &=& {2\over \mb}(\ma-\mb),\nonumber\\
A^\prime_{22} &=& {1\over2\ma^2}[(\ma+\mb)^2-q^2],\nonumber\\
A^\prime_{33} &=& {1\over2\mb^2}[(\ma+\mb)^2-q^2],\nonumber\\
A^\prime_{12} &=& {1\over\ma}(\ma+\mb),\nonumber\\
A^\prime_{23} &=& {1\over\ma\mb} [(\ma+\mb)^2-q^2],\nonumber\\
A^\prime_{31} &=& {2\over \mb}(\ma+\mb),\nonumber
\eeqy
\beq
\gamma(1/2^-)= 4 F_1G_1.
\eeq
\subsection{$3/2^-$}
\beq
\alpha(3/2^-)=\sum_{i=1,j=1}^{i=4,j=4}{1\over Y^\prime}(B_{ij} F_i F_j + B^\prime_{ij} G_i G_j),
\eeq
where $Y^\prime=3\ma^2\mb^2$, and the non-vanishing coefficients are
\beqy
B_{11} &=& X[(\ma+\mb)^2-q^2],\nonumber\\
B_{44} &=& 4\ma^2 \mb^2[(\ma+\mb)^2-q^2],\nonumber\\
B_{14} &=& B^\prime_{14} = \ma\mb[\ma^4-2(\mb^2+q^2)\ma^2+(\mb^2-q^2)^2],\nonumber\\
B^\prime_{11} &=& X[(\ma-\mb)^2-q^2],\nonumber\\
B^\prime_{44} &=& 4\ma^2\mb^2 [(\ma-\mb)^2-q^2],\nonumber
\eeqy
\beq
\beta_{++}(3/2^-)=\sum_{i=1,j=1}^{i=4,j=4}{1\over Y}(A_{ij} F_i F_j + A^\prime_{ij} G_i G_j),
\eeq
where $Y=12\ma^4\mb^4$, $X=(\ma^2+\mb^2-q^2)^2-4\ma^2\mb^2$, and the $A_{ij}$ are
\beqy
A_{11} &=& A^\prime_{11} = 4X\ma^2\mb^2,\nonumber\\
A_{22} &=& X\mb^2[(\ma+\mb)^2-q^2],\nonumber\\
A_{33} &=& X\ma^2[(\ma+\mb)^2-q^2],\nonumber\\
A_{44} &=& 4\ma^4\mb^2[(\ma+\mb)^2-q^2],\nonumber\\
A_{12} &=&  4X\ma\mb^2(\ma+\mb),\nonumber\\
A_{23} &=& 2X\ma\mb[(\ma+\mb)^2-q^2],\nonumber\\
A_{31} &=& 4X\ma^2\mb(\ma+\mb),\nonumber\\
A_{14} &=& -8\ma^3\mb^2[(\ma+2\mb)q^2+(\ma-\mb)(\ma+\mb)^2],\nonumber\\
A_{24} &=& 4\ma^2\mb^2[(\ma+\mb)^2-q^2][\ma^2-\mb^2-q^2],\nonumber\\
A_{34} &=& 4\ma^3\mb[(\ma+\mb)^2-q^2][\ma^2-\mb^2-q^2],\nonumber\\
A^\prime_{22} &=& X\mb^2[(\ma-\mb)^2-q^2],\nonumber\\
A^\prime_{33} &=& X\ma^2[(\ma-\mb)^2-q^2],\nonumber\\
A^\prime_{44} &=& 4\ma^4\mb^2[(\ma-\mb)^2-q^2],\nonumber\\
A^\prime_{12} &=& 4X\ma\mb^2(\mb-\ma),\nonumber\\
A^\prime_{23} &=& 2X\ma\mb[(\ma-\mb)^2-q^2],\nonumber\\
A^\prime_{31} &=& 4X\ma^2\mb(\mb-\ma),\nonumber\\
A^\prime_{14} &=& 8\ma^3\mb^2[(\ma+2\mb)q^2-(\ma+\mb)(\ma-\mb)^2],\nonumber\\
A^\prime_{24} &=& 4\ma^2\mb^2[(\ma-\mb)^2-q^2][\ma^2-\mb^2-q^2],\nonumber\\
A^\prime_{34} &=& 4\ma^3\mb[(\ma-\mb)^2-q^2][\ma^2-\mb^2-q^2],\nonumber
\eeqy
\beqy
\gamma(3/2^-) &=& {2\over 3\ma^2\mb^2}\{[(\ma-\mb)^2-q^2](F_1G_4\ma\mb+F_1G_1[(\ma+\mb)^2-q^2])\nonumber\\
&+& F_4G_4\ma^2\mb^2 +F_4G_1\ma\mb[(\ma+\mb)^2-q^2]\}.
\eeqy
\subsection{$3/2^+$}
\beq
\alpha(3/2^+)=\sum_{i=1,j=1}^{i=4,j=4}{1\over Y^\prime}(B_{ij} F_i F_j + B^\prime_{ij} G_i G_j),
\eeq
where the non-vanishing coefficients are
\beqy
B_{11} &=& X[(\ma-\mb)^2-q^2],\nonumber\\
B_{44} &=& 4\ma^2 \mb^2[(\ma-\mb)^2-q^2],\nonumber\\
B_{14} &=& B^\prime_{14} = \ma\mb[\ma^4-2(\mb^2+q^2)\ma^2+(\mb^2-q^2)^2],\nonumber\\
B^\prime_{11} &=& X[(\ma+\mb)^2-q^2],\nonumber\\
B^\prime_{44} &=& 4\ma^2 \mb^2[(\ma+\mb)^2-q^2],\nonumber
\eeqy
\beq
\beta_{++}(3/2^+)=\sum_{i=1,j=1}^{i=4,j=4}{1\over Y}(A_{ij} F_i F_j + A^\prime_{ij} G_i G_j),
\eeq
where $Y=12\ma^4\mb^4$, $X=(\ma^2+\mb^2-q^2)^2-4\ma^2\mb^2$, and
\beqy
A_{11} &=& A^\prime_{11} = 4X\ma^2\mb^2,\nonumber\\
A_{22} &=& X\mb^2[(\ma-\mb)^2-q^2],\nonumber\\
A_{33} &=& X\ma^2[(\ma-\mb)^2-q^2],\nonumber\\
A_{44} &=& 4\ma^4\mb^2[(\ma-\mb)^2-q^2],\nonumber\\
A_{12} &=&  4X\ma\mb^2(\mb-\ma),\nonumber\\
A_{23} &=& 2X\ma\mb[(\ma-\mb)^2-q^2],\nonumber\\
A_{31} &=&  4X\ma^2\mb(\mb-\ma),\nonumber\\
A_{14} &=& 8\ma^3\mb^2[(\ma-2\mb)q^2-(\ma+\mb)(\ma-\mb)^2],\nonumber\\
A_{24} &=& 4\ma^2\mb^2[(\ma-\mb)^2-q^2][\ma^2-\mb^2-q^2],\nonumber\\
A_{34} &=& 4\ma^3\mb[(\ma-\mb)^2-q^2][\ma^2-\mb^2-q^2],\nonumber\\
A^\prime_{22} &=& X\mb^2[(\ma+\mb)^2-q^2],\nonumber\\
A^\prime_{33} &=& X\ma^2[(\ma+\mb)^2-q^2],\nonumber\\
A^\prime_{44} &=& 4\ma^4\mb^2[(\ma+\mb)^2-q^2],\nonumber\\
A^\prime_{12} &=& 4X\ma\mb^2(\mb+\ma),\nonumber\\
A^\prime_{23} &=& 2X\ma\mb[(\ma+\mb)^2-q^2],\nonumber\\
A^\prime_{31} &=& 4X\ma^2\mb(\mb+\ma),\nonumber\\
A^\prime_{14} &=& -8\ma^3\mb^2[(\ma-2\mb)q^2-(\ma-\mb)(\ma-\mb)^2],\nonumber\\
A^\prime_{24} &=& 4\ma^2\mb^2[(\ma+\mb)^2-q^2][\ma^2-\mb^2-q^2],\nonumber\\
A^\prime_{34} &=& 4\ma^3\mb[(\ma+\mb)^2-q^2][\ma^2-\mb^2-q^2],\nonumber
\eeqy
\beqy
\gamma(3/2^+) &=& {2\over 3\ma^2\mb^2}\{[(\ma+\mb)^2-q^2](F_1G_4\ma\mb+F_1G_1[(\ma-\mb]^2-q^2])\nonumber\\
&+& F_4G_4\ma^2\mb^2 +F_4G_1\ma\mb[(\ma-\mb)^2-q^2]\}.
\eeqy
\subsection{$5/2^+$}
\beq
\alpha(5/2^+)=\sum_{i=1,j=1}^{i=4,j=4}{1\over Y_2}X(B_{ij} F_i F_j + B^\prime_{ij} G_i G_j),
\eeq
where the non-vanishing coefficients are
\beqy
B_{11} &=& X[(\ma-\mb)^2-q^2],\nonumber\\
B_{44} &=& 3\ma^2\mb^2 [(\ma+\mb)^2-q^2],\nonumber\\
B_{14} &=& B^\prime_{14} = 2X\ma\mb,\nonumber\\
B^\prime_{11} &=& X[(\ma+\mb)^2-q^2],\nonumber\\
B^\prime_{44} &=& 3\ma^2\mb^2 [(\ma-\mb)^2-q^2],\nonumber
\eeqy
\beq
\beta_{++}(5/2^+)=\sum_{i=1,j=1}^{i=4,j=4}{1\over Y_1}(A_{ij} F_i F_j + A^\prime_{ij} G_i G_j),
\eeq
where $Y_1=80\ma^6\mb^6$, and
\beqy
A_{11} &=&A^\prime_{11} = 4X^2\ma^2\mb^2,\nonumber\\
A_{22} &=& X^2\mb^2[(\ma+\mb)^2-q^2],\nonumber\\
A_{33} &=& X^2\ma^2[(\ma+\mb)^2-q^2],\nonumber\\
A_{44} &=& 4\ma^4\mb^2[(\ma+\mb)^2-q^2][q^4-(2\ma^2+\mb^2)q^2+(\ma^2-\mb^2)^2],\nonumber\\
A_{12} &=&  4X^2\ma\mb^2(\ma+\mb),\nonumber\\
A_{23} &=& 2X^2\ma\mb[(\ma+\mb)^2-q^2],\nonumber\\
A_{31} &=&  4X^2\ma^2\mb(\ma+\mb),\nonumber\\
A_{14} &=& -8X\ma^3\mb^2[(\ma+2\mb)q^2+(\ma-\mb)(\ma+\mb)^2],\nonumber\\
A_{24} &=& 4X\ma^2\mb^2[(\ma+\mb)^2-q^2][\ma^2-\mb^2-q^2],\nonumber\\
A_{34} &=& 4X\ma^3\mb[(\ma+\mb)^2-q^2][\ma^2-\mb^2-q^2],\nonumber\\
A^\prime_{22} &=& X^2\mb^2[(\ma-\mb)^2-q^2],\nonumber\\
A^\prime_{33} &=& X^2\ma^2[(\ma-\mb)^2-q^2],\nonumber\\
A^\prime_{44} &=& 4\ma^4\mb^2[(\ma-\mb)^2-q^2][q^4-(2\ma^2+\mb^2)q^2+(\ma^2-\mb^2)^2],\nonumber\\
A^\prime_{12} &=& 4X^2\ma\mb^2(\mb-\ma),\nonumber\\
A^\prime_{23} &=& 2X^2\ma\mb[(\ma-\mb)^2-q^2],\nonumber\\
A^\prime_{31} &=& 4X^2\ma^2\mb(\mb-\ma),\nonumber\\
A^\prime_{14} &=& 8X\ma^3\mb^2[(\ma+2\mb)q^2-(\ma+\mb)(\ma-\mb)^2],\nonumber\\
A^\prime_{24} &=& 4X\ma^2\mb^2[(\ma-\mb)^2-q^2][\ma^2-\mb^2-q^2],\nonumber\\
A^\prime_{34} &=& 4X\ma^3\mb[(\ma-\mb)^2-q^2][\ma^2-\mb^2-q^2],\nonumber
\eeqy
\beqy
\gamma(5/2^+) &=& {\ma^4-2\ma^2(\mb^2+q^2)+(\mb^2+q^2)^2\over 10\ma^4\mb^4}
\left\{F_4G_4\ma^2\mb^2
+F_4G_1\ma\mb\left[\vphantom{F_4G_4\ma^2\mb^2}(\ma-\mb)^2-q^2\right]\right.\nonumber\\
&+&\left.[(\ma+\mb)^2-q^2]\left(F_1G_4\ma\mb+F_1G_1\left[\vphantom{F_4G_4\ma^2\mb^2}(\ma-\mb)^2-q^2\right]\right)
\right\}.\nonumber
\eeqy

\section{Constructing Higher Spin Representations}

It is necessary to construct explicit representations for the spin-3/2 and spin-
5/2 baryons that we treat. In the case of the former, the vector-spinor field  
$u^\alpha(p', s')$  must satisfy 
\beq
p^\prime_\alpha u^\alpha(p', s')=0, \,\,\,\,
\gamma_\alpha u^\alpha(p', s')=0,\,\,\,\, \slash{p}^\prime u^\alpha(p', s')
=m_{\Lambda_q^{(3/2)}}u^\alpha(p', s').
\eeq

A suitable representation can be constructed by using the usual Dirac spin-1/2
spinors, together with the `polarization' vectors $\epsilon_\mu(p^\prime,s_z)$.
These vectors satisfy
\beq
p^\prime_\mu \epsilon^\mu(p^\prime,s_z)=0, \,\,\,\,\epsilon_\mu^*(p^\prime,s_z)
\epsilon^\mu(p^\prime,s_z^\prime)=-\delta_{s_z,s_z^\prime}.
\eeq

Our representation of the spin-3/2 Rarita-Schwinger vector-spinor
$u_\mu(p^\prime, M)$ is given by the Clebsch-Gordan sum
\begin{eqnarray}
u_\mu(p', M)&=& \sum_m\epsilon_\mu(p', m)u(p',M-m)\langle3/2 M|1 m, 
1/2, M-m\rangle.
\end{eqnarray}
This satisfies all of the conditions required.

A representation of the spin-$5/2$ spinor $u^{\alpha\beta}(p^\prime,s)$ can be 
constructed in a similar way, but there are two additional constraints that must
be satisfied. The first is that the spinor must be symmetric in its Lorentz
indices, and the second is that it must be traceless when the two indices are
contracted, i.e.
\beq
u^\alpha_\alpha(p^\prime,s)=0.
\eeq
Such a representation can be built in one of two ways. We can use the 
previously constructed spin-3/2 spinor, and the vector $\epsilon$, to write
\begin{equation}
u_{\mu\nu}(p',M)= \sum_m\epsilon_\mu(p',m)u_\nu(p',M-m)
\langle5/2 M|1 m, 3/2, M-m\rangle.
\end{equation}
Alternatively, we can first construct a spin-2 tensor $A_{\mu\nu}$ 
from two of the $\epsilon$ vectors as
\beq
A_{\mu\nu}(p^\prime,M)=\sum_m\epsilon_\mu(p',m)\epsilon_\nu(p',M-m)
\langle2 M|1 m, 1, M-m\rangle
\eeq
The symmetry properties of the Clebsch-Gordan
coefficients guarantee that this tensor is symmetric in its indices. The
spin-5/2 spinor is then
\begin{equation}
u_{\mu\nu}(p',M)= \sum_mA_{\mu\nu}(p',m)u(p',M-m)
\langle5/2 M|2 m, 1/2, M-m\rangle.
\end{equation}
These two representations are equivalent, but the manifest symmetry of the
second representation allows us to see the symmetry in $u_{\mu\nu}$ in an
obvious way.

The conditions
\beq
p^\prime_\alpha u^{\alpha\beta}(p', s')=
p^\prime_\beta u^{\alpha\beta}(p', s')=0
\eeq 
 are clearly satisfied, since each vector $\epsilon$ satisfies
$p^\prime\cdot\epsilon$=0 (and the second equality also follows from the symmetry in
the indices). It is easy to check that the auxiliary conditions
\beq
\gamma_\alpha u^{\alpha\beta}(p', s')=\gamma_\beta u^{\alpha\beta}(p', s')
=0
\eeq 
are satisfied, as are
\beq
\slash{p}^\prime u^{\alpha\beta}(p', s')
=m_{\Lambda_q^{(5/2)}} u^{\alpha\beta}(p',s').
\eeq
The traceless condition
\beq
g_{\alpha\beta}u^{\alpha\beta}(p', s')=0,
\eeq
is less obvious, but follows from the tracelessness of $A_{\mu\nu}$. This, in
turn, follows from the symmetry properties of the Clebsch-Gordan sum in
$A_{\mu\nu}$, and the properties of the $\epsilon$ vectors.

\newif\ifmultiplepapers
\def\beginpapers{\multiplepaperstrue}
\def\endpapers{\multiplepapersfalse}  
\def\journal#1&#2(#3)#4{\rm #1~{\bf #2}\unskip, \rm  #4 (19#3)}
\def\trjrnl#1&#2(#3)#4{\rm #1~{\bf #2}\unskip, \rm #4 (19#3)}
\def\baps{\journal {Bull.} {Am.} {Phys.} {Soc.}&}
\def\jap{\journal J. {Appl.} {Phys.}&}
\def\prl{\journal {Phys.} {Rev.} {Lett.}&}
\def\pl{\journal {Phys.} {Lett.}&}
\def\pr{\journal {Phys.} {Rev.}&}
\def\np{\journal {Nucl.} {Phys.}&}
\def\rmp{\journal {Rev.} {Mod.} {Phys.}&}
\def\jmp{\journal J. {Math.} {Phys.}&}
\def\rmm{\journal {Revs.} {Mod.} {Math.}&}
\def\jetp{\journal {J.} {Exp.} {Theor.} {Phys.}&}
\def\sjetp{\trjrnl {Sov.} {Phys.} {JETP}&}
\def\dokl{\journal {Dokl.} {Akad.} Nauk USSR&}
\def\spd{\trjrnl {Sov.} {Phys.} {Dokl.}&}
\def\tmf{\journal {Theor.} {Mat.} {Fiz.}&}
\def\snp{\trjrnl {Sov.} J. {Nucl.} {Phys.}&}
\def\hpa{\journal {Helv.} {Phys.} Acta&}
\def\yf{\journal {Yad.} {Fiz.}&}
\def\zp{\journal Z. {Phys.}&}
\def\anp{\journal {Adv.} {Nucl.} {Phys.}&}
\def\ap{\journal {Ann.} {Phys.}&}
\def\am{\journal {Ann.} {Math.}&}
\def\nc{\journal {Nuo.} {Cim.}&}
\def\etal{{\sl et al.}}
\def\pre{\journal {Phys.} {Rep.}&}
\def\pca{\journal Physica (Utrecht)&}
\def\prs{\journal {Proc.} R. {Soc.} London &}
\def\jcp{\journal J. {Comp.} {Phys.}&}
\def\pna{\journal {Proc.} {Nat.} {Acad.}&}
\def\jpg{\journal J. {Phys.} G (Nuclear Physics)&}
\def\fort{\journal {Fortsch.} {Phys.}&}
\def\jfa{\journal {J.} {Func.} {Anal.}&}
\def\cmp{\journal {Comm.} {Math.} {Phys.}&}

\end{document}